\documentclass[useAMS,usenatbib]{mn2e}

\usepackage{color}
\usepackage{colortbl}
\usepackage{multirow}
\usepackage{dcolumn}
\usepackage{amsmath}
\usepackage{amssymb}
\usepackage{graphicx}
\usepackage{epsfig}
\usepackage{changebar}
\usepackage{natbib}
\usepackage{multirow}
\usepackage{subfigure}
\usepackage{txfonts}
\usepackage{rotating}

\usepackage{longtable,lscape}

\newcolumntype{d}[1]{D{.}{\cdot}{#1}}

\newcolumntype{.}{D{.}{.}{-1}}

\newcommand{\lsun}{L$_\odot$}
\newcommand{\msun}{M$_\odot$}

\newcommand{\mum}{$\mu$m}

\newcommand{\kms}{km\,s$^{-1}$}

\newcommand{\hii}{H~{\sc ii}}
\newcommand{\uchii}{UC\,H~{\sc ii}}

\newcommand{\poi}{Poisson}

\newcommand{\sex}{\texttt{SExtractor}}
\newcommand{\rms}{r.m.s.}

\newcommand{\unmatchedmmb}{43}

\title[ATLASGAL-MMB Associations]{ATLASGAL --- Environments of 6.7\,GHz methanol masers}
\author[J. S. Urquhart et al.]{
J.\,S.\,Urquhart$^{1}$\thanks{E-mail:
jurquhart@mpifr-bonn.mpg.de (MPIfR)}, T.\,J.\,T.\,Moore$^{2}$, F.\,Schuller$^{3}$, F.\,Wyrowski$^{1}$, K.\,M.\,Menten$^{1}$,  \newauthor M.\,A.\,Thompson$^{4}$, T.\,Csengeri$^{1}$, C.\,M.\,Walmsley$^{5,6}$, L.\,Bronfman$^{7}$, C.\,K\"onig$^{1}$\\
$^{1}$
 Max-Planck-Institut f\"ur Radioastronomie, Auf dem H\"ugel
  69, Bonn, Germany \\
$^{2}$Astrophysics Research Institute, Liverpool John Moores University, Twelve Quays House, Egerton Wharf, Birkenhead, CH41\,1LD, UK\\ 
$^{3}$European Southern Observatory, Alonso de Cordova 3107, Vitacura, Santiago, Chile\\
$^{4}$Science and Technology Research Institute, University of Hertfordshire, College Lane, Hatfield, AL10 9AB, UK\\
$^{5}$Osservatorio Astrofisico di Arcetri, Largo E. Fermi, 5, 50125 Firenze, Italy\\
$^{6}$Dublin Institute for Advanced Studies, Burlington Road 10, Dublin 4, Ireland\\
$^{7}$Departamento de Astronom\'{i}a, Universidad de Chile, Casilla 36-D, Santiago, Chile}
\begin{document}

\date{Accepted ??. Received ??; in original form ??}

\pagerange{\pageref{firstpage}--\pageref{lastpage}} \pubyear{2009}

\maketitle

\label{firstpage}

\begin{abstract}

Using the 870-$\mu$m APEX Telescope Large Area Survey of the Galaxy (ATLASGAL), we have identified 577 submillimetre continuum sources with masers from the methanol multibeam (MMB) survey in the region $280\degr <  \ell < 20\degr$; $|\,b\,| < 1.5\degr$.  94\,per\,cent of methanol masers in the region are associated with sub-millimetre dust emission. We estimate masses for $\sim$450 maser-associated sources and find that methanol masers are preferentially associated with massive clumps. These clumps are centrally condensed, with envelope structures that appear to be scale-free, the mean maser position being offset from the peak column density by $0 \pm 4$\arcsec.  Assuming a Kroupa initial mass function and a star-formation efficiency of $\sim$30\,per\,cent, we find that over two thirds of the clumps are likely to form clusters with masses $>$20\,\msun. Furthermore, almost all clumps satisfy the empirical mass-size criterion for massive star formation. Bolometric luminosities taken from the literature for $\sim$100 clumps range between $\sim$100 and 10$^6$\,\lsun. This confirms the link between methanol masers and massive young stars for 90\,per\,cent of our sample. The Galactic distribution of sources suggests that the star-formation efficiency is significantly reduced in the Galactic-centre region, compared to the rest of the survey area, where it is broadly constant, and shows a significant drop in the massive star-formation rate density in the outer Galaxy. We find no enhancement in source counts towards the southern Scutum-Centaurus arm tangent at $\ell \sim 315\degr$, which suggests that this arm is not actively forming stars. 

\end{abstract}
\begin{keywords}
Stars: formation -- Stars: early-type -- Galaxy: structure -- ISM: molecules -- ISM: submillimetre.
\end{keywords}

\section{Introduction}

Massive stars ($>$ 8\,\msun\ and $>$10$^3$\,\lsun) play a hugely important role in many astrophysical processes from the formation of the first solid material in the early Universe (\citealt{dunne2003}); to their substantial influence upon the evolution of their host galaxies and future generations of star formation (\citealt{kennicutt2005}). Given the profound impact massive stars have, not only on their local environment, but also on a Galactic scale, it is crucial to understand the environmental conditions and processes involved in their birth and the earliest stages of their evolution. However, massive stars form in clusters and are generally located at greater distances than regions of low-mass star formation and therefore understanding how these objects form is observationally much more challenging (see \citealt{zinnecker2007} for a review). Moreover, massive stars are rare and they evolve much more quickly than low-mass stars, reaching the main sequence while still deeply embedded in their natal environment. As a consequence of their rarity and relatively short evolution large spatial volumes need to be searched in order to identify a sufficient number of sources in each evolutionary stage. Only then can we begin to understand the processes involved in the formation and earliest stages of massive star formation. 

There have been a number of studies over the last two decades or so that have been used to identify samples of embedded young massive stars utilizing the presence of methanol masers (e.g., \citealt{walsh1997}), IRAS, MSX or GLIMPSE infrared colours (e.g., \citealt{molinari1996} and \citealt{bronfman1996}, \citealt{lumsden2002} and \citealt{robitaille2008}, respectively) and compact radio emission (e.g., \citealt{wood1989}). Although these methods have had some success, they tend to focus on a particular evolutionary type and in the case of the infrared colour selected samples are biased away from complex regions as a result of confusion in the images due to the limited angular resolution of the surveys. This is particularly acute for IRAS selected samples. Some of these surveys suffer from significant biases. For example, methanol masers require a strong mid-infrared source for the creation of high enough methanol abundances and to pump the maser transitions, while a hot ionising star must be present to ionize an \uchii\ region. Surveys such as these preclude the possibility of identifying the very earliest pre-stellar clumps that would need to be included in any complete evolutionary sequence for massive star formation. 

All of the earliest stages of massive star formation take place within massive clumps of dust and gas, which can be traced by their thermal dust continuum emission. Dust emission is generally optically thin at (sub)millimetre wavelengths and is therefore an excellent tracer of column density and total clump mass. As well as including all of the embedded stages, dust emission observations are also sensitive to the colder pre-stellar phases and so provide a means to study the whole evolutionary sequence of the massive star formation. Until recently there were no systematic and unbiased surveys of dust emission. Most studies to date that have been undertaken consisted of targeted observations of IRAS or maser selected samples (e.g., \citealt{sridharan2002,faundez2004,hill2005,thompson2006}), and so suffer from the same problems mentioned in the previous paragraph, or have concentrated on a single, often exceptionally rich region (e.g., \citealt{motte2007}), which may not be representative. However, there are now two large unbiased (sub)millimetre surveys available: the APEX Telescope Large Area Survey of the Galaxy (ATLASGAL; \citealt{schuller2009}) at 870\,\mum\ and the Bolocam Galactic Plane Survey (BGPS; \citealt{aguirre2011}) at 1.1\,mm. These surveys have identified many thousands of sources across the Galaxy with which to compile the large samples of massive clumps required to build up a comprehensive understanding of massive star formation, and test the predictions of the main two competing theoretical models (i.e., competitive accretion (\citealt{bonnell1997,bonnell2001}) and monolithic collapse (\citealt{mckee2003})). 

The complete ATLASGAL catalogue consists of approximately 12,000 compact sources (see \citealt{contreras2013} for details) distributed across the inner Galaxy. This is the first of a series of papers that will investigate the dust properties and Galactic distribution of massive star formation. Here we use the association of methanol masers, that are considered to be an excellent tracer of the early stages of massive star formation (\citealt{minier2003}), to identify a large sample of massive clumps.   

The structure of this paper is as follows: in Sect.\,2 we provide a brief summary of the two surveys used to select the sample of massive star forming clumps. In Sect.\,3 we describe the matching procedure and discuss sample statistics, while in Sect.\,4 we derive the physical properties of the clumps and their associated masers. In Sect.\,5 we evaluate the potential of the clumps to form massive stars and investigate their Galactic distribution with reference to the large scale structural features of the Milky Way. We present a summary of the results and highlight our main findings in Sect.\,6.

\section{Survey descriptions}

\subsection{ATLASGAL Survey}

The APEX Telescope Large Area Survey of the Galaxy (ATLASGAL; \citealt{schuller2009}) is the first systematic survey of the inner Galactic plane in the submillimeter wavelength range. The survey was carried out with the Large APEX Bolometer Camera (LABOCA; \citealt{siringo2009}), an array of 295 bolometers observing at 870\,$\mu$m (345 GHz). The 12-metre diameter telescope affords an angular resolution of $19\rlap{.}''2$ FWHM. The initial survey region covered a Galactic longitude region of $300\degr < \ell < 60\degr$ and $|b| < 1.5\degr$, but this was extended to include $280\degr < \ell < 300\degr$, however, the latitude range was shifted to $-2\degr < b < 1\degr$ to take account of the Galactic warp in this region of the plane and is not as sensitive as for the inner Galaxy ($\sim$60 and 100\,mJy beam$^{-1}$ for the $300\degr < \ell < 60\degr$ and $280\degr < \ell < 300\degr$ regions, respectively).

\citet{contreras2013} produced a compact source catalogue for the central part of the survey region (i.e., 330\degr\ $ <\ell <$ 21\degr) using the source extraction algorithm \sex\ \citep{bertin1996}. Signal-to-noise maps that had been filtered to remove the large scale variations due to extended diffuse emission were used by \sex\ to detect sources above a threshold of 3$\sigma$, where $\sigma$ corresponds to the background noise in the maps. Source parameters were determined for each detection from the dust emission maps. This catalogue consists of 6,639 sources and is 99\,per\,cent complete at $\sim$6$\sigma$, which corresponds to a flux sensitivity of 0.3-0.4\,Jy\,beam$^{-1}$. We have used the same source extraction algorithm and method described by \citet{contreras2013} to produce a catalogue for the currently unpublished 280\degr\ $ <\ell <$ 330\degr\ and 21\degr\ $ <\ell <$ 60\degr\ regions of the survey. When the sources identified in these regions are combined with those identified by  \citet{contreras2013} we obtain a final compact source catalogue of some 12,000 sources (full catalogue will be presented in Csengeri et al. 2013 in prep.). The telescope has an \rms\ pointing accuracy of $\sim$2\arcsec, which we adopted as the positional accuracy for the catalogue. This catalogue provides a complete census of dense dust clumps located in the inner Galaxy and includes all potential massive star forming clumps with masses greater than 1,000\,\msun\ out to 20\,kpc.

\subsection{MMB Survey}
\label{sect:mmb_description}

Methanol masers are well-known indicators of the early phases of high-mass star formation, in particular sources showing emission in the strong 6.7\,GHz Class-II maser transition \citep{menten1991}. The Methanol Multibeam (MMB) Survey mapped the Galactic plane for this maser transition using a 7-beam receiver on the Parkes telescope with a sensitivity of 0.17\,Jy\,beam$^{-1}$ and a half-power beamwidth of 3.2\arcmin\ (\citealt{green2009}). All of these initial maser detections were followed up at high-resolution ($\sim$2\arcsec) with the Australia Telescope Compact Array (ATCA) to obtain sub-arcsec positional accuracy (0.4\arcsec\ \rms; \citealt{caswell2010b}). To date the MMB is complete between $186\degr < \ell\ < 20\degr$ and $|b| < 2\degr$ (\citealt{caswell2010b,green2010,caswell2011,green2012}) and has reported the positions of 707 methanol maser sites; these sites consist of groups of maser spots that can be spread up to 1\arcsec\ in size but are likely to be associated with a single object.

\begin{figure*}
\begin{center}

\includegraphics[width=0.33\textwidth, trim= 0 0 0 0]{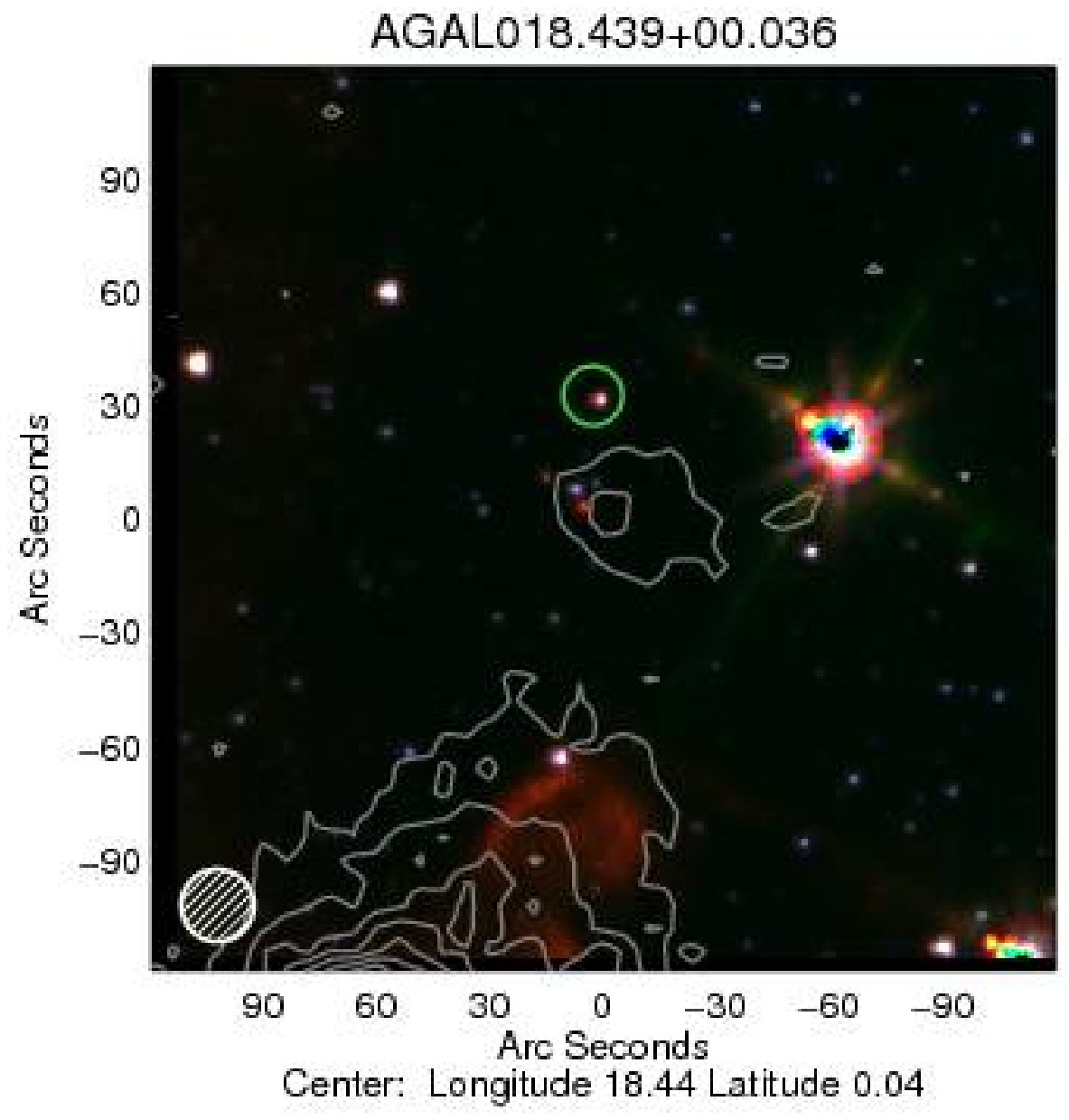}
\includegraphics[width=0.33\textwidth, trim= 0 0 0 0]{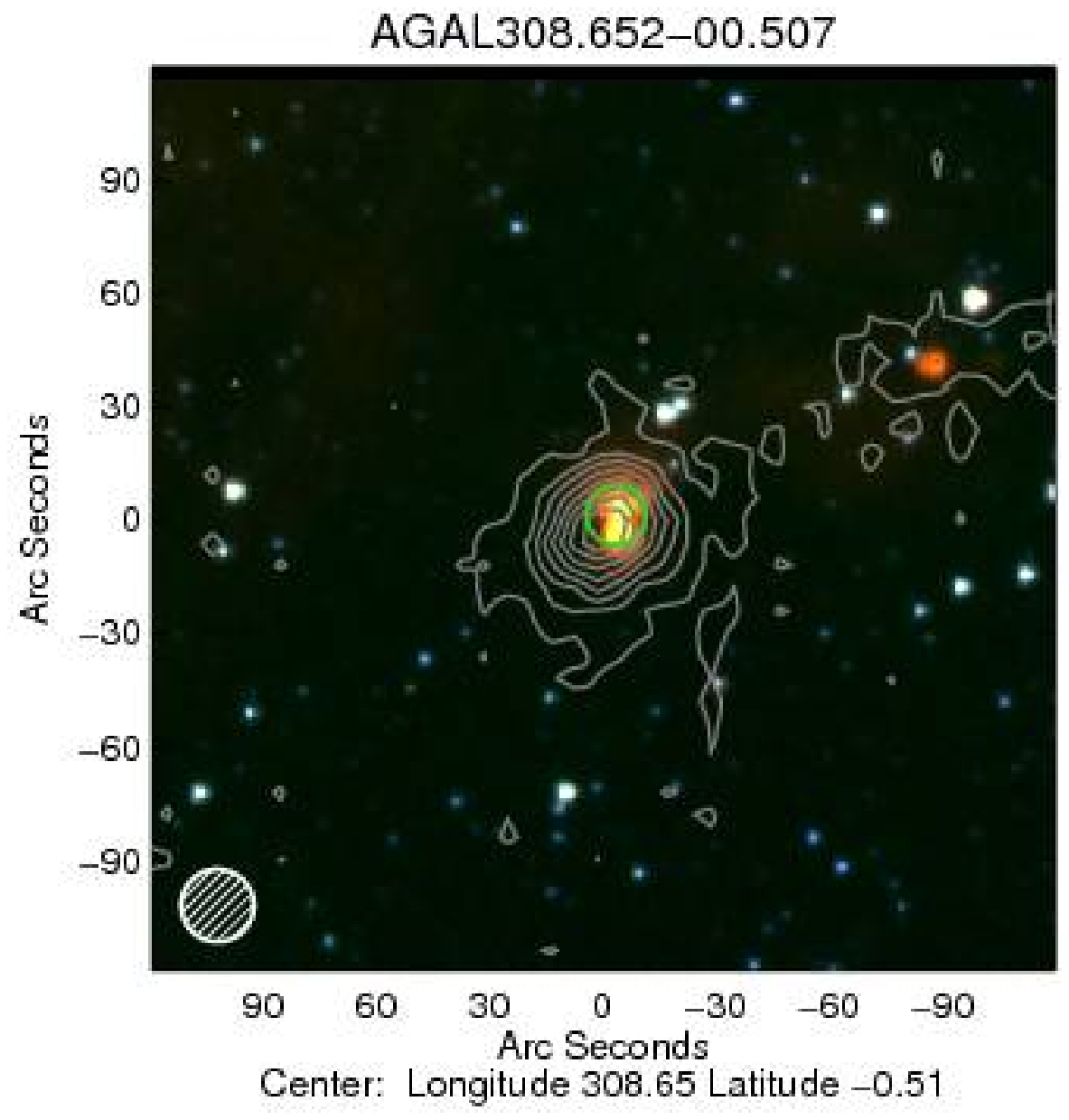}
\includegraphics[width=0.33\textwidth, trim= 0 0 0 0]{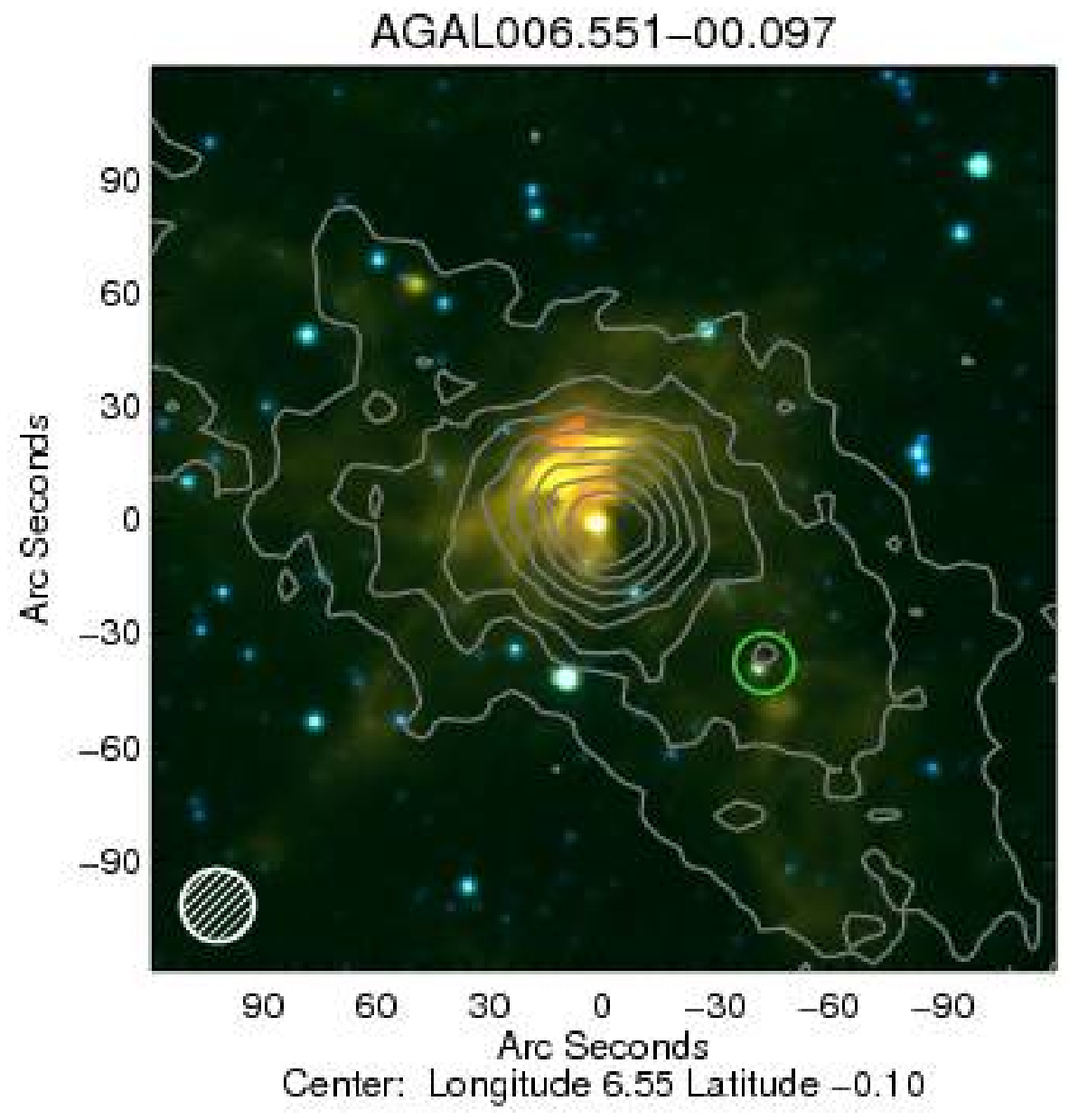}
\includegraphics[width=0.33\textwidth, trim= 0 0 0 0]{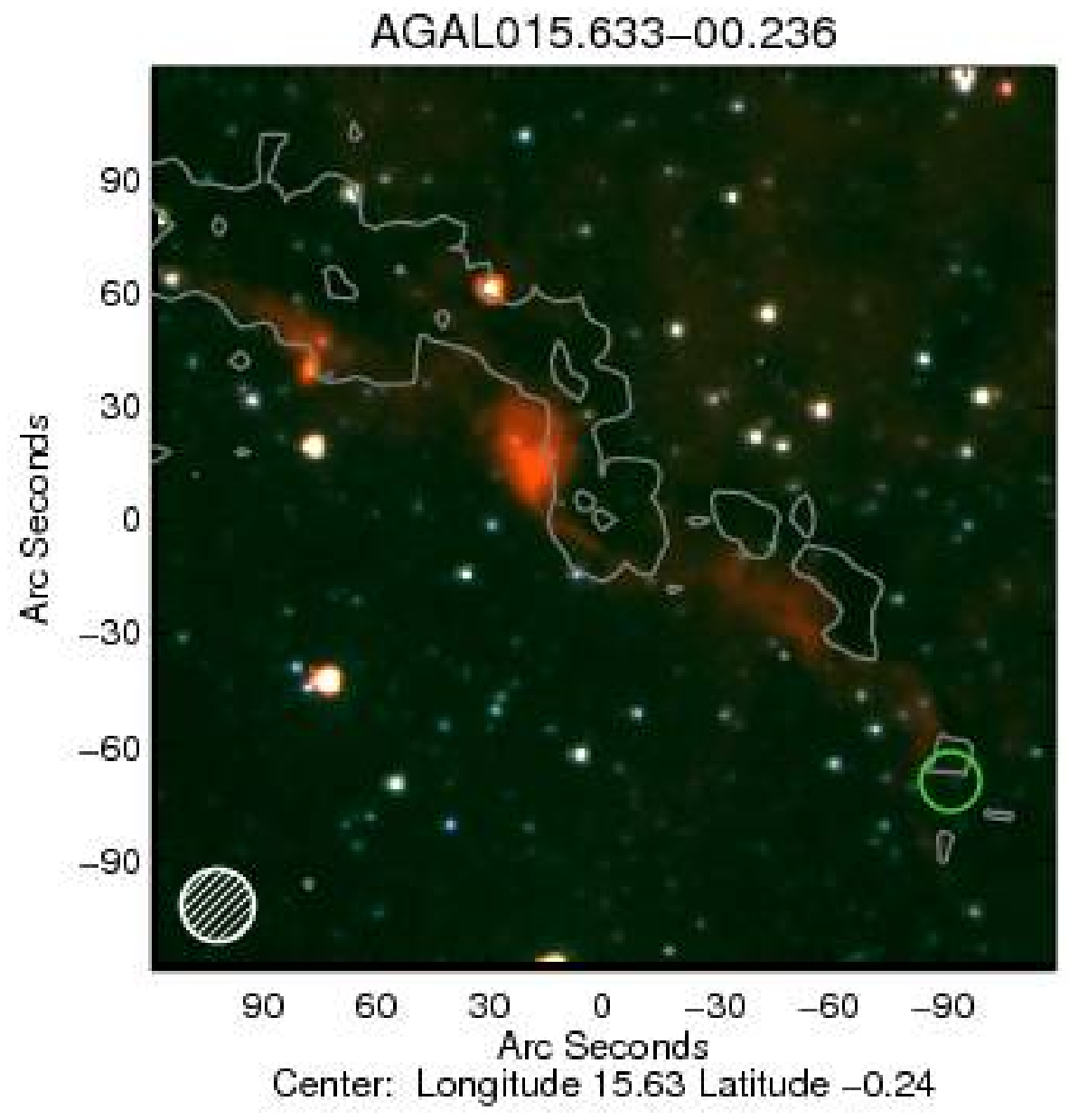}
\includegraphics[width=0.33\textwidth, trim= 0 0 0 0]{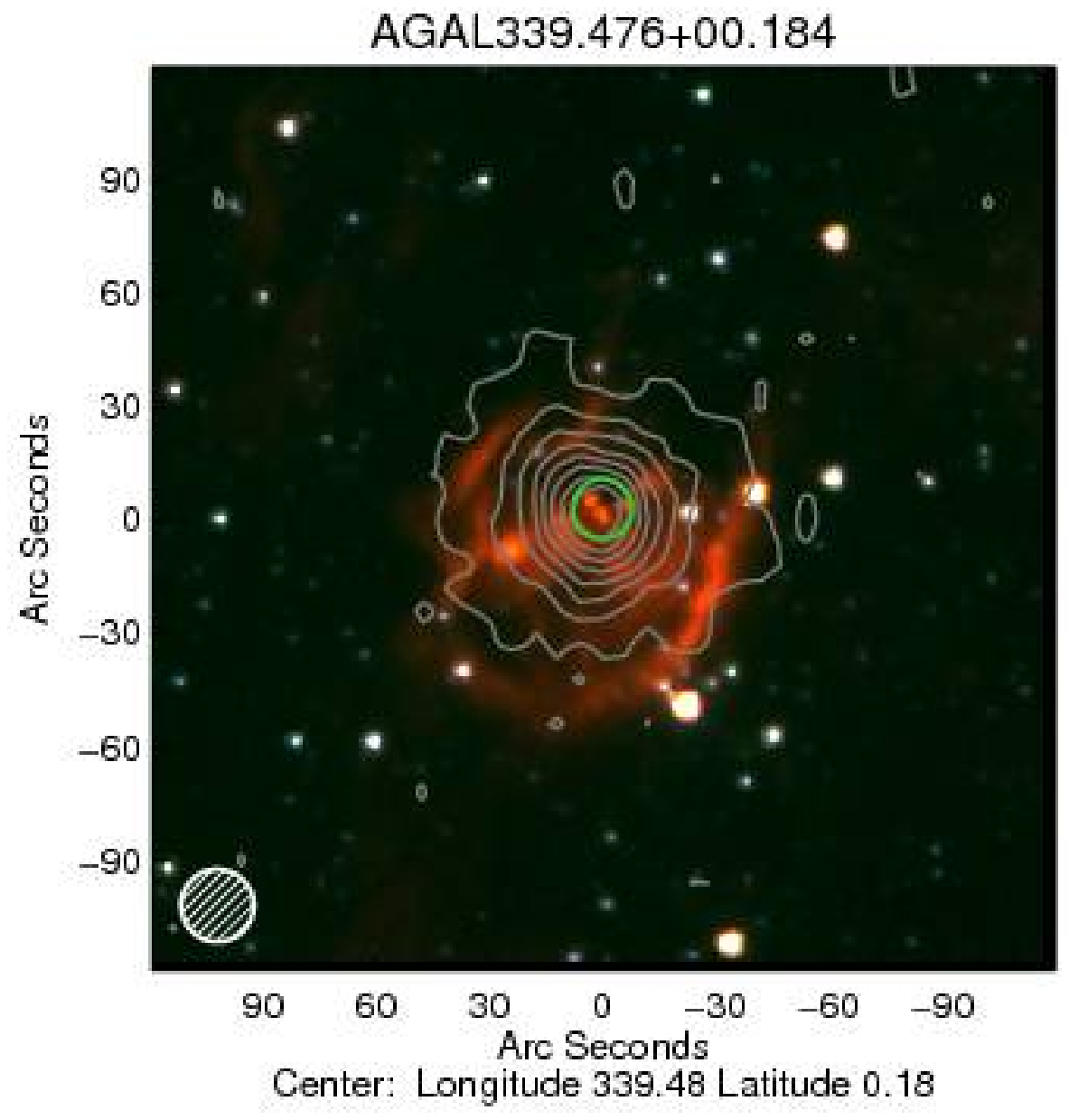}
\includegraphics[width=0.33\textwidth, trim= 0 0 0 0]{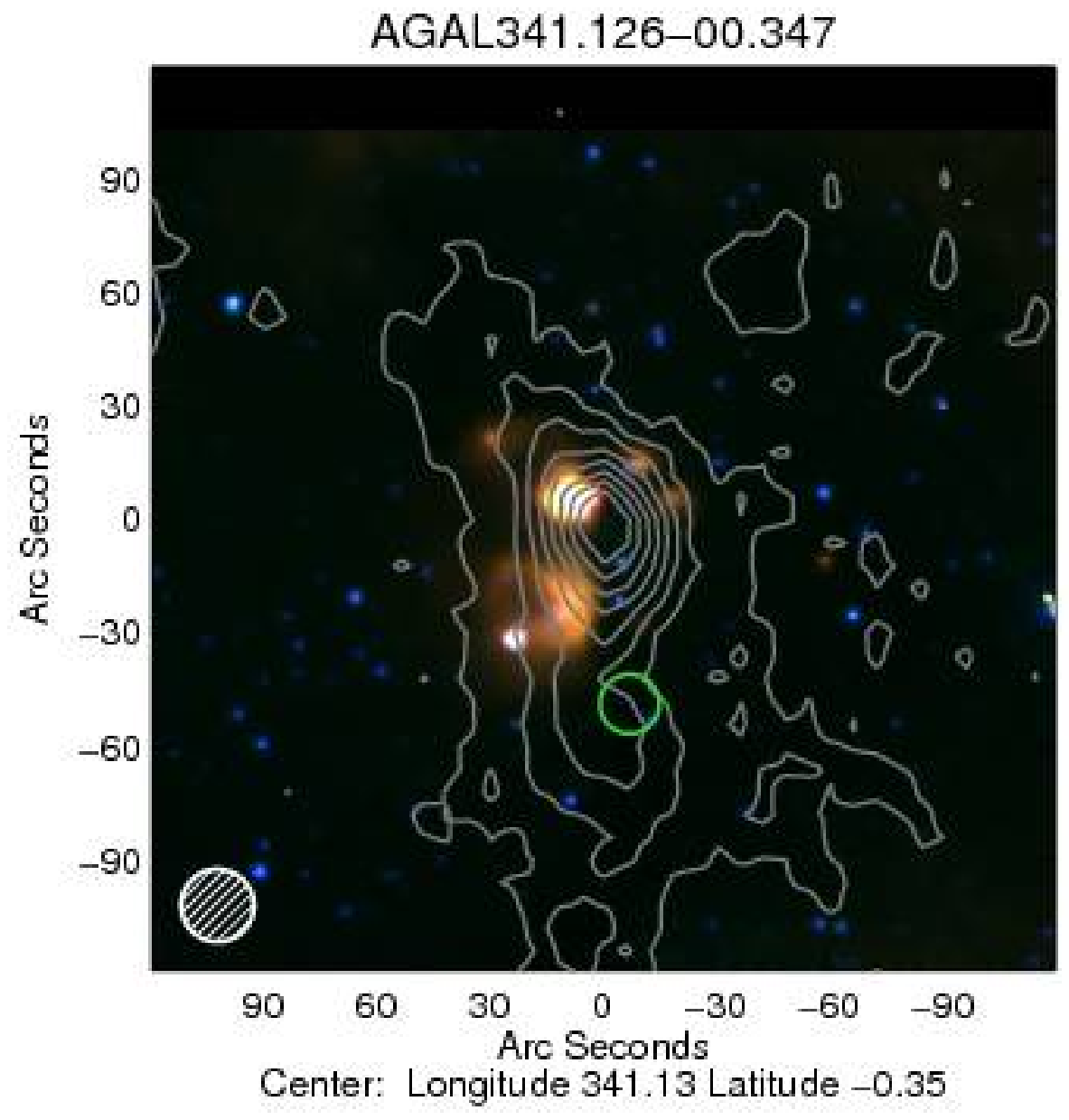}

\caption{\label{fig:irac_images_associated_atlasgal} Examples of the local mid-infrared environment found towards the ATLASGAL-MMB candidate associations. The left panels show examples of rejected matches, while the centre and right panels show genuine associations (see text for details). These images are composed of the GLIMPSE 3.6, 4.5 and 8.0\,\mum\ IRAC bands (coloured blue, green and red respectively) which are overlaid with grey contours showing the ATLASGAL 870\,$\mu$m emission. The positions of the methanol masers in each field are indicated by the green circles and the white hatched circle in the lower left corner of each image shows the size of the APEX beam at 870\,\mum. The contour levels start at 2$\sigma$ and increase in steps set by a dynamically determined power-law of the form $D=3\times N^i+2$, where $D$ is the dynamic range of the submillimetre emission map (defined as the peak brightness divided by the local r.m.s. noise), $N$ is the number of contours used (8 in this case), and $i$ is the contour power-law index. The lowest power-law index used was one, which results linearly spaced contours starting at 2$\sigma$ and increasing in steps of 3$\sigma$ (see \citet{thompson2006} for more details). The advantage of this scheme over a linear scheme is its ability to emphasize both emission from diffuse extended structures with low surface brightness and emission from bright compact sources.}

\end{center}
\end{figure*}

The MMB catalogue gives the velocity of the peak component and the flux density as measured from both the high sensitivity ATCA follow-up observations and those measured from the initial lower sensitivity Parkes observations. These observations were taken over different epochs up to two years apart and consequently the measured values can be affected by variability of the maser. We have used the velocity and peak flux densities measured from the ATCA data for all MMB sources except for MMBG321.704+01.168 as it was not detected in the ATCA observations; for this source we use the values recorded from the Parkes observations.

\section{ATLASGAL-MMB associations}
\label{sect:atlas-mmb}

\subsection{Matching statistics}
\label{sect:matching_stats}

\begin{figure*}
\begin{center}
\includegraphics[width=0.49\textwidth, trim= 0 0 0 0]{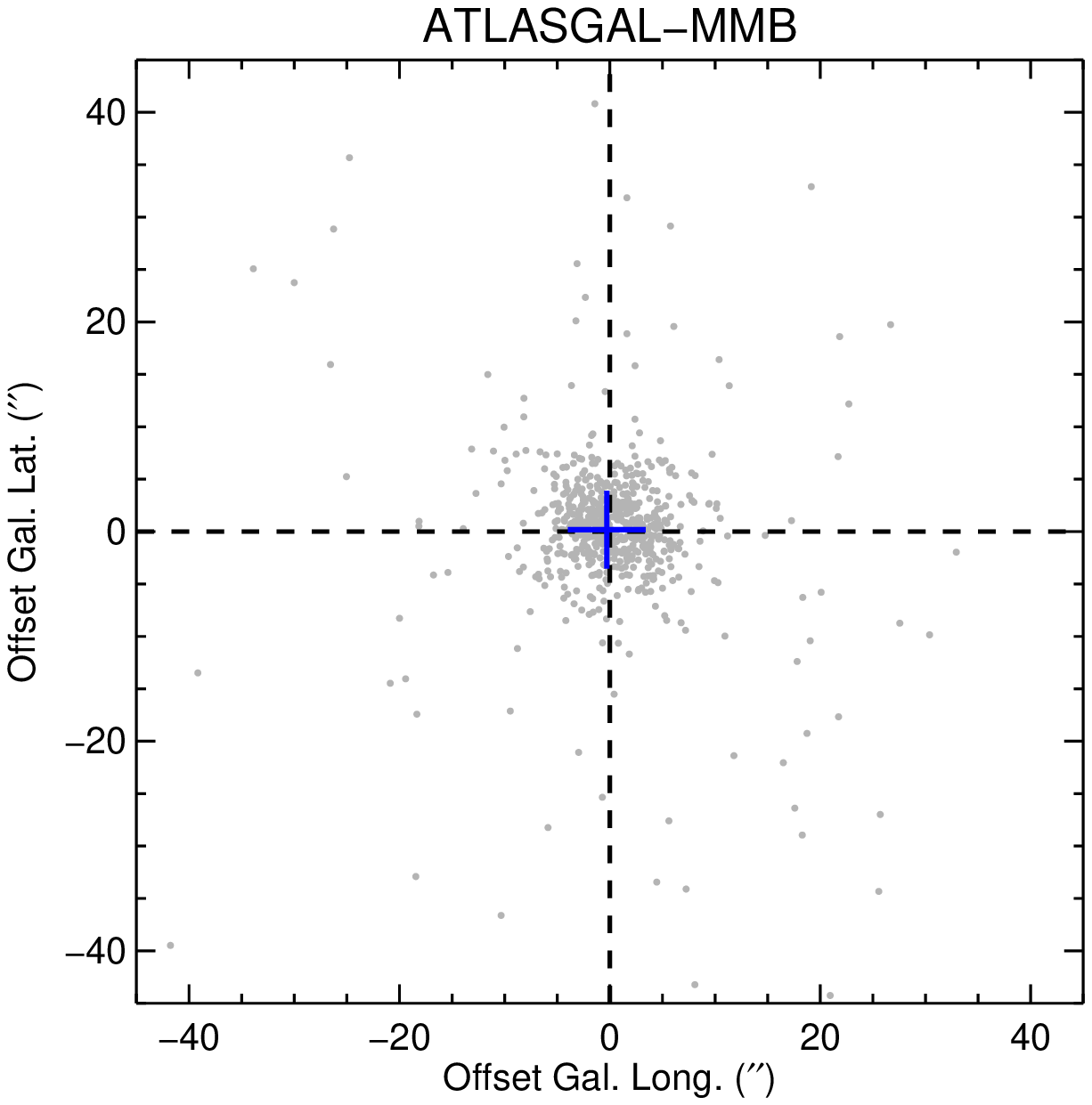}
\includegraphics[width=0.49\textwidth, trim= 0 0 0 0]{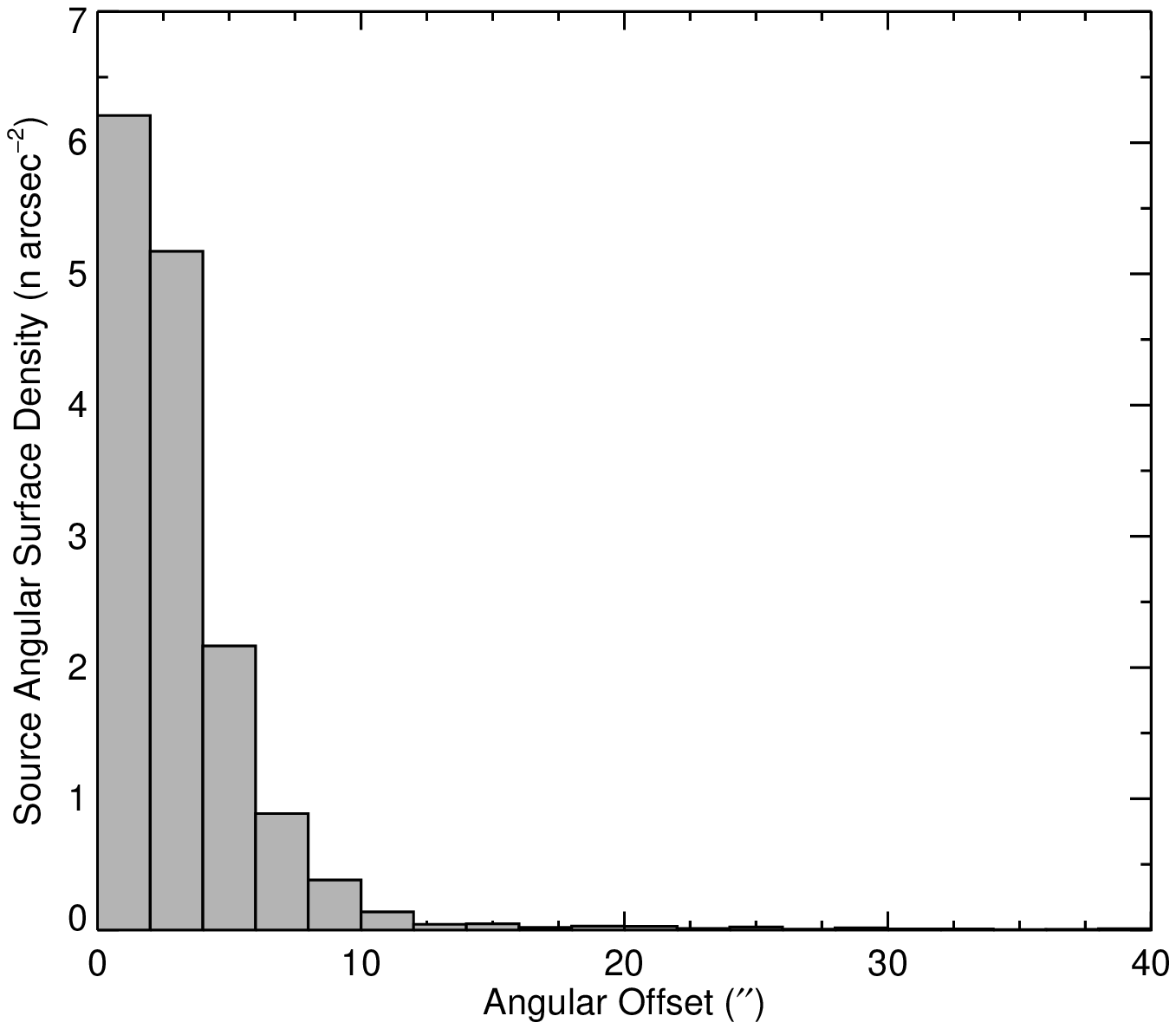}

\caption{\label{fig:mmb_atlas_offset}  Two-dimensional distribution of the angular offsets between the peak of the ATLASGAL dust emission and the matched MMB source is shown in the left panel. The dashed vertical and horizontal lines indicate the $x$ and $y=0$ axes, respectively and the blue cross shows the mean offset in longitude and latitude of the whole sample. In the right panel we present a histogram showing the surface density of methanol masers as a function of separation between them and the position of peak emission of the associated ATLASGAL source. We have truncated the $x$-axis of this plot at 40\arcsec\ as there are only 14 ATLASGAL-MMB matches with larger separations and the source density effectively falls to zero. The bin size is 2\arcsec.} 

\end{center}
\end{figure*}

Of the 707 methanol masers currently identified by the MMB survey 671 are located in the Galactic longitude and latitude range surveyed at 870\,$\mu$m by the APEX telescope as part of the ATLASGAL project. This represents $\sim$95\,per\,cent of the entire published MMB catalogue. As a first step to identify potential matches between the methanol masers and the ATLASGAL sources we used a matching radius of 120\arcsec, which is the maximum radius of sources found in the ATLASGAL catalogue \citep{contreras2013}. We used the peak 870\,$\mu$m flux as the ATLASGAL source position for these matches. In cases where a methanol maser was found to be located within this search radius of two or more ATLASGAL sources the nearest submillimetre source was selected as the most likely association. This simple radius search identified 637 potential ATLASGAL-MMB associations from the possible 671 methanol masers. To verify that these associations are genuine, we extracted $3\arcmin\times3\arcmin$ regions from the ATLASGAL emission maps and inspected these by eye to confirm the emission region and position of the methanol maser are coincident. We failed to find an ATLASGAL source at the MMB position towards nine MMB sources and therefore removed these from the associated sample; these masers form part of the sample of unassociated masers discussed in Sect.\,\ref{sect:unmatched_mmb}.

For the potential associations located in the part of the Galaxy surveyed by GLIMPSE (i.e., $\ell > 295$ and $|b| < 1$; \citealt{benjamin2003_ori}) we extracted $3\arcmin\times3\arcmin$ mid-infrared datasets from the project archive and produced false colour images of their environments. We present a sample of these images in Fig.\,\ref{fig:irac_images_associated_atlasgal} overplotted with contours of the submillimetre emission. In the left panels of Fig.\,\ref{fig:irac_images_associated_atlasgal} we present some examples of the rejected matches. The final sample consists of 628 methanol masers that are positionally coincident with 577 ATLASGAL sources (see middle and right panels of Fig.\,\ref{fig:irac_images_associated_atlasgal} for examples of these associations), with two or more methanol masers found toward 44 clumps. This sample includes $\sim$94\,per\,cent of the MMB masers located in the surveys' overlap region, however, this represents only $\sim$7\,per\,cent of the ATLASGAL sources in the same region.

For two of the ATLASGAL sources associated with multiple methanol masers we find  the maser velocities disagree by more that would be expected if the masers were associated with the same molecular complex (i.e., $|\Delta v| > 10$\,\kms). These are: AGAL355.538$-$00.104 which is associated with MMB355.538$-$00.105 (3.8\,km\,s$^{-1}$) and MMB355.545$-$00.103 ($-$28.2\,km\,s$^{-1}$); and AGAL313.766$-$00.862 which is associated with MMB313.767$-$00.863 ($-$56.3\,km\,s$^{-1}$) and MMB313.774$-$00.863 ($-$44.8\,km\,s$^{-1}$). For each of these we have adopted the velocity of the maser with the smallest angular offset from the submillimetre peak. 

In the left panel of Fig.\,\ref{fig:mmb_atlas_offset} we present a plot showing the offsets in Galactic longitude and latitude between the peak positions of the ATLASGAL and methanol maser emission. The positional correlation between the two tracers is excellent with the mean (indicated by the blue cross) centred at zero ($\Delta \ell = - 0.3\arcsec\pm0.4$\arcsec\ and  $\Delta b=+0.18\arcsec\pm0.37$\arcsec). In the right panel of Fig.\,\ref{fig:mmb_atlas_offset} we present a plot showing the MMB angular surface density as a function of angular separation between the maser position and the peak of the 870\,$\mu$m emission of the associated ATLASGAL source. This plot reveals a strong correlation between the position of the methanol maser and the peak submillimetre continuum emission. The distribution peaks at separations less than 2\arcsec\ and falls off rapidly as the angular separation increases to  $\sim$12\arcsec\ after which the distribution flattens off to an almost constant background level close to zero. We find that $\sim$87\,per\,cent of all ATLASGAL-MMB associations have an angular separation $<$\,12\arcsec\ (which corresponds to 3$\sigma$, where $\sigma$ is the standard deviation of the offsets weighted by the surface density). The high concentration and small offsets between the masers and the peak dust emission reveals that the methanol masers are embedded in the brightest emission parts of these submillimetre clumps. This would suggest that the protostars giving rise to these methanol masers are preferentially found towards the centre of their host clumps. This is in broad agreement with the predictions of the competitive accretion model (\citealt{bonnell1997,bonnell2001}) where the deeper gravitational potential at the centre of the clump is able to significantly increase the gas density by funneling material from the whole cloud toward the centre. However, high angular resolution interferometric observations (e.g., with ALMA) are required to conclusively prove this hypothesis by measuring the density distribution for the clumps on small spatial scales.

The larger angular offsets are found for approximately 80 ATLASGAL-MMB associations, which could indicate the presence of clumpy substructure that has not been properly identified by SExtractor. Alternatively, these matches with large offsets could be the result of a chance alignment of a nearby clump with an MMB source that is associated with more distant dust clump that falls below the ATLASGAL sensitivity limit. In the right panels of Fig.\,\ref{fig:irac_images_associated_atlasgal} we present mid-infrared images towards two sources where the methanol maser is offset from the peak position of the dust emission by more than 12\arcsec. Inspection of the dust emission (shown by the contours) does reveal the presence of weak localised peaks close to that of the methanol maser position, which would suggest that these two particular sources do possess substructure that has not been identified independently by the ATLASGAL source extraction method. These two sources are fairly typical of the matches with offsets greater than 12\arcsec\ and therefore we would conclude that this is the most likely explanation, however, higher sensitivity observations are required to confirm this, and to properly characterise the dust properties of these methanol maser sites.

\begin{figure}
\begin{center}
\includegraphics[width=0.49\textwidth, trim= 0 0 0 0]{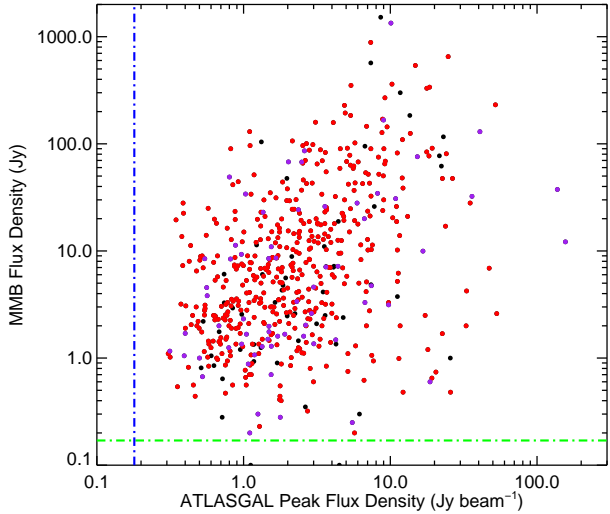}

\caption{\label{fig:flux_density_mmb_atlasgal} Comparison of the methanol maser flux density and the peak 870\,$\mu$m flux density of the associated ATLASGAL source. The colours indicate whether or not a particular maser has been associated with mid-infrared emission (\citealt{gallaway2013}); red, purple and black colours correspond to infrared bright, infrared dark, and sources for which no data is available, respectively. The dashed-dotted blue and green lines indicate the 3$\sigma$ sensitivities of the ATLASGAL and MMB surveys, respectively. The results of a Pearson correlation test returns a coefficient value of 0.19.}

\end{center}
\end{figure}

In Fig.\,\ref{fig:flux_density_mmb_atlasgal} we present a plot comparing the methanol flux density and the 870\,$\mu$m peak flux density of the ATLASGAL-MMB associations. There is no apparent  correlation between these parameters from a visual inspection of this plot, however, the  correlation coefficient is 0.19 with a significance value of 5$\times$10$^{-6}$ and so there is a weak correlation. Methanol masers are found to be associated with a range of evolutionary stages of massive star formation (i.e., from the hot molecular core (HMC) through to the ultra-compact (UC) HII region stage). Therefore the low level of correlation could simply be a reflection of the spread in evolutionary stages covered by the methanol masers. It is also important to bear in mind that these methanol masers are likely to be associated with the circumstellar envelope/disk of a single embedded source, and given the resolution of the ATLASGAL survey it is almost certain that the measured submillimetre flux is related to the mass of the whole clump, which is likely to go on to form an entire cluster. It is therefore not surprising that we find the flux densities of the masers and the dust clumps to be only weakly correlated.

To investigate whether there is a correlation between the submillimetre and methanol maser fluxes and the mid-infrared properties of the source we have cross-correlated the matched sources with the \citet{gallaway2013} catalogue. Using this catalogue we separate our matched sample into three groups, those associated with mid-infrared emission, infrared dark sources and those located outside the region covered by the GLIMPSE Legacy project, upon which the work of \citet{gallaway2013} is based. The distribution of these three groups are shown in Fig.\,\ref{fig:flux_density_mmb_atlasgal} as red, purple and black symbols, respectively. Comparing the flux distributions of the infrared bright and dark samples with a Kolmogorov-Smirnov (KS) test we do not find them to be significantly different.

\subsection{ATLASGAL 870\,$\mu$m flux distribution}
\label{sect:flux_dist}

\begin{figure}
\begin{center}
\includegraphics[width=0.49\textwidth, trim= 0 0 0 0]{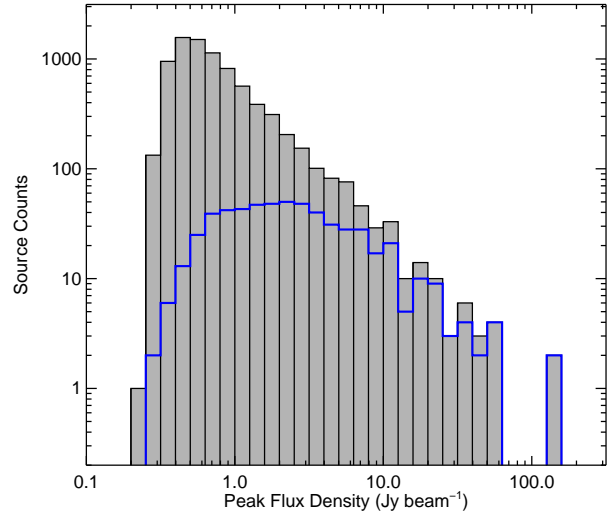}
\includegraphics[width=0.49\textwidth, trim= 0 0 0 0]{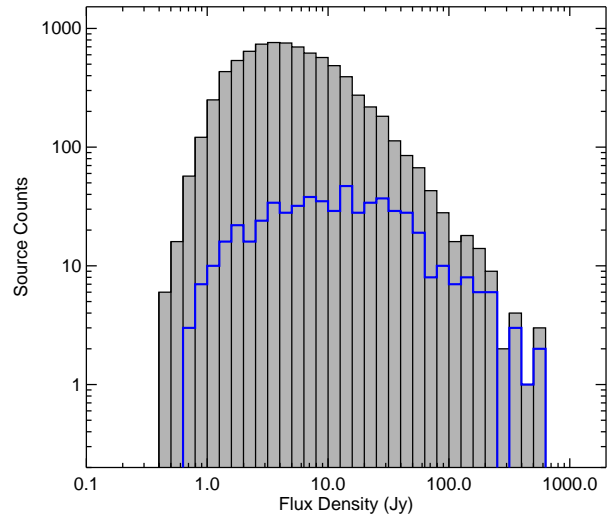}

\caption{\label{fig:flux_density} Flux density distribution for ATLASGAL sources in the overlap region (grey filled histogram) and the ATLASGAL-MMB associated sources (blue histogram). In the upper and lower panels we present histograms of the peak and integrated flux densities measured for each source, respectively.}

\end{center}
\end{figure}

In Fig.\,\ref{fig:flux_density} we present plots of the 870\,$\mu$m peak and integrated flux distribution (upper and lower panels, respectively) of the ATLASGAL catalogue (grey histogram) and ATLASGAL-MMB associations (blue histogram). It is clear from these plots that the methanol masers are preferentially associated with the brighter ATLASGAL sources in the sense that the probability of an association with a maser approaches 100\,per\,cent for brighter clumps. This is particularly evident in the peak flux distribution, which reveals that only a relatively small number of submillimetre sources brighter than $\sim$7\,Jy\,beam$^{-1}$ are not associated with a methanol maser. It is also clear from these plots that there is a stronger correlation between the brightest peak flux density ATLASGAL sources and the presence of a methanol maser than between the integrated flux and the presence of a maser. The integrated flux is a property of the whole clump/cloud, whereas the peak flux is more likely to be associated with the highest column density and/or warmest regions of the clump where star formation is taking place. Furthermore, some of the ATLASGAL sources with the highest integrated fluxes are large extended sources, that can have relatively low peak flux densities.

\begin{figure}
\begin{center}
\includegraphics[width=0.49\textwidth, trim= 0 0 0 0]{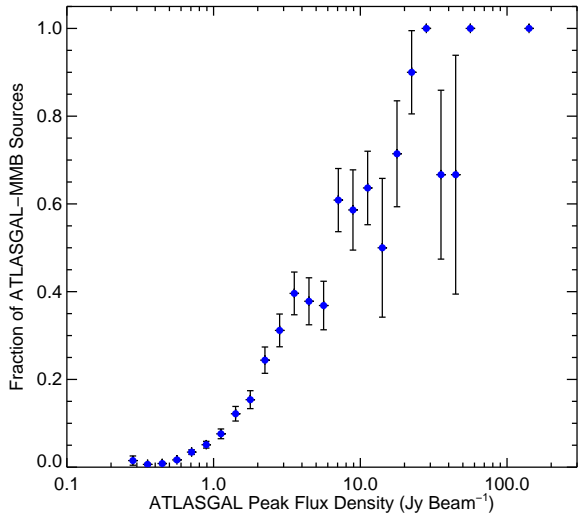}

\caption{\label{fig:peak_flux_density_association_ratio} Fraction of ATLASGAL sources associated with a methanol maser as a function of peak 870\,$\mu$m flux density. The errors are estimated using binomial statistics.}

\end{center}
\end{figure}

In Fig.\,\ref{fig:peak_flux_density_association_ratio} we present a plot showing the ratio of ATLASGAL sources found to be associated with a methanol maser as a function of peak 870\,$\mu$m flux density. Although the errors in the ratios for the higher flux bins are relatively large, due to the smaller numbers of sources they contain, there is still clearly a strong correlation between bright submillimetre emission and the presence of a methanol maser. Given that the association rate of ATLASGAL sources with methanol masers increases rapidly with peak flux density ($\sim$100\,per\,cent for sources above 20\,Jy\,beam$^{-1}$) an argument can be made for the maser emission being effectively isotropic. Although the radiation beamed from individual maser spots is highly directional, there are many very high resolution studies that have revealed significant numbers of maser spots to be associated with a single source (e.g., \citealt{goddi2011}). These individual maser spots are distributed around the central embedded protostar and are therefore spatially distinct. 

\subsection{Unassociated MMB Sources}
\label{sect:unmatched_mmb}

\begin{figure}
\begin{center}
\includegraphics[width=0.49\textwidth, trim= 0 0 0 0]{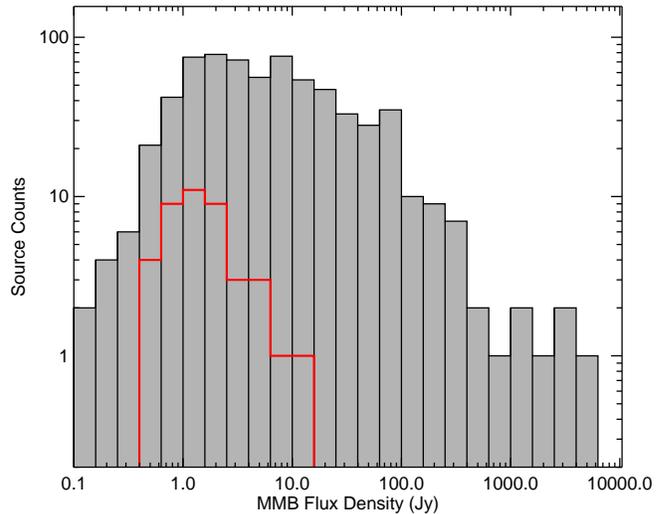}

\caption{\label{fig:mmb_flux_distribution} MMB flux distribution for the whole sample (grey filled histogram) and those not associated with an ATLASGAL source (red histogram). The bin size is 0.2 dex.} 

\end{center}
\end{figure}

\unmatchedmmb\ MMB masers could not be matched with an ATLASGAL source. These masers have integrated flux densities between 0.43 and 15.65\,Jy, with a mean and median value of 2.2 and 1.5\,Jy, respectively. In Fig.\,\ref{fig:mmb_flux_distribution} we present a histogram showing the distribution of maser fluxes for the whole MMB sample (grey filled histogram) and of the unassociated masers (red histogram). The flux densities of the unmatched masers are significantly above the MMB survey's sensitivity of 0.17\,Jy\,beam$^{-1}$, and although they are principally found towards the lower flux end of the distribution, they are not the weakest masers detected. For comparison the mean and median fluxes for the whole maser sample are 47.8 and 5.1\,Jy, respectively. Using a KS test to compare the distribution of the dustless MMB masers with that of the whole MMB catalogue we are able to reject the null hypothesis that these are drawn from the same population with greater than three sigma confidence. 

It is widely accepted that methanol masers are almost exclusively associated with high-mass star forming regions (e.g., \citealt{minier2003,pandian2010}). If this is the case then we might expect all of these unmatched sources to be located at the far side of the Galaxy where their dust emission falls below the ATLASGAL detection sensitivity, however, it is possible that these are associated with more evolved stars (e.g., \citealt{walsh2003}). We will investigate the nature of these unassociated methanol masers in more detail in Sect.\,5.4.

\section{Physical properties}

In the previous section we identified two subsamples based on the possible combinations of ATLASGAL and MMB associations: 1) methanol masers associated with the thermal continuum emission from cold dust and likely tracing star formation; 2) apparently dustless methanol masers which could be a combination of masers associated with more distant dust clumps not currently detected. These two subsamples consist of approximately 94\,per\,cent and 6\,per\,cent of the MMB sources in the overlap region, respectively. 

In this section we will concentrate on the ATLASGAL-MMB associations to determine the physical properties of their environments; these are given for each clump in Table\,\ref{tbl:derived_clump_para} while in Table\,\ref{tbl:derived_para} we summarise the global properties.

\setlength{\tabcolsep}{4pt}

\begin{table*}

\begin{center}\caption{Derived clump parameters. The columns are as follows: (1) and (2) ATLASGAL and MMB names; (3) angular offset between the peak of the submillimetre emission and the masers position; (4) ratio of the semi-major to semi-minor sizes of the ATLASGAL source; (5) ratio of the integrated and peak submillimetre emission ($Y$-factor); (6) heliocentric distance; (7) Galactocentric distance; (8) effective physical radius; (9) column density; (10) clump mass derived from the integrated 870\,$\mu$m emission assuming a dust temperature of 20\,K; (11) isotropic methanol maser luminosity. }
\label{tbl:derived_clump_para}
\begin{minipage}{\linewidth}
\small
\begin{tabular}{ll.........}
\hline \hline
  \multicolumn{1}{c}{ATLASGAL name}&  \multicolumn{1}{c}{MMB Name$^{\rm{a}}$}&	\multicolumn{1}{c}{Offset}  &\multicolumn{1}{c}{Aspect}  &\multicolumn{1}{c}{$Y$-factor}  &	\multicolumn{1}{c}{Distance} &\multicolumn{1}{c}{R$_{\rm{GC}}$}&\multicolumn{1}{c}{Radius}&\multicolumn{1}{c}{Log(N(H$_2$))} &	\multicolumn{1}{c}{Log(Mass$_{\rm{Clump}}$)}&	\multicolumn{1}{c}{Log(L$_{\rm{MMB}}$)}\\
    \multicolumn{1}{c}{ }&  \multicolumn{1}{c}{ }&	\multicolumn{1}{c}{($^{\prime\prime}$)}  &\multicolumn{1}{c}{Ratio}  &\multicolumn{1}{c}{ }  &	\multicolumn{1}{c}{(kpc)} &\multicolumn{1}{c}{(kpc)}&\multicolumn{1}{c}{(pc)}&\multicolumn{1}{c}{(cm$^{-2}$)} &	\multicolumn{1}{c}{(\msun)}&	\multicolumn{1}{c}{(Jy\,kpc$^2$)}\\
        \multicolumn{1}{c}{(1)}&  \multicolumn{1}{c}{(2)}&	\multicolumn{1}{c}{(3)}  &\multicolumn{1}{c}{(4)}  &\multicolumn{1}{c}{(5)}  &	\multicolumn{1}{c}{(6)} &\multicolumn{1}{c}{(7)}&\multicolumn{1}{c}{(8)}&\multicolumn{1}{c}{(9)} &	\multicolumn{1}{c}{(10)}&	\multicolumn{1}{c}{(11)}\\

\hline
AGAL281.709$-$01.104	&	MMB281.710$-$01.104	&	2.3	&	1.6	&	3.44	&	4.2	&	8.7	&	0.50	&	22.97	&	2.94	&	1.88	\\
AGAL284.352$-$00.417	&	MMB284.352$-$00.419	&	6.5	&	1.5	&	10.44	&	5.2	&	8.8	&	1.05	&	22.84	&	3.49	&	2.93	\\
AGAL284.694$-$00.359	&	MMB284.694$-$00.361	&	6.2	&	1.0	&	2.00	&	6.3	&	9.2	&	\multicolumn{1}{c}{$\cdots$}	&	22.41	&	2.51	&	3.22	\\
AGAL285.339$-$00.001	&	MMB285.337$-$00.002	&	8.3	&	1.7	&	5.29	&	5.1	&	8.7	&	0.53	&	22.63	&	2.96	&	3.56	\\
AGAL287.372+00.646	&	MMB287.371+00.644	&	6.8	&	1.1	&	2.59	&	5.2	&	8.6	&	0.19	&	22.67	&	2.72	&	4.56	\\
AGAL291.272$-$00.714	&	MMB291.270$-$00.719	&	19.6	&	1.4	&	11.25	&	1.0	&	8.2	&	0.36	&	23.88	&	3.13	&	2.00	\\
AGAL291.272$-$00.714	&	MMB291.274$-$00.709	&	20.5	&	1.4	&	11.25	&	1.0	&	8.2	&	0.36	&	23.88	&	3.13	&	2.94	\\
AGAL291.579$-$00.432	&	MMB291.579$-$00.431	&	4.3	&	1.4	&	5.05	&	8.1	&	9.4	&	2.02	&	23.59	&	4.31	&	2.92	\\
AGAL291.579$-$00.432	&	MMB291.582$-$00.435	&	14.8	&	1.4	&	5.05	&	7.7	&	9.1	&	1.90	&	23.59	&	4.26	&	3.31	\\
AGAL291.636$-$00.541	&	MMB291.642$-$00.546	&	28.0	&	1.5	&	14.72	&	7.8	&	9.2	&	2.94	&	23.33	&	4.49	&	2.37	\\
AGAL291.879$-$00.809	&	MMB291.879$-$00.810	&	2.7	&	1.2	&	2.89	&	9.9	&	10.4	&	\multicolumn{1}{c}{$\cdots$}	&	22.61	&	3.27	&	3.22	\\
AGAL292.074$-$01.129	&	MMB292.074$-$01.131	&	8.1	&	2.1	&	3.18	&	3.2	&	7.9	&	\multicolumn{1}{c}{$\cdots$}	&	22.34	&	2.05	&	2.04	\\
AGAL293.828$-$00.746	&	MMB293.827$-$00.746	&	1.4	&	1.1	&	2.48	&	10.7	&	10.7	&	0.94	&	23.10	&	3.75	&	3.56	\\
AGAL293.941$-$00.874	&	MMB293.942$-$00.874	&	3.7	&	1.2	&	3.10	&	11.2	&	10.9	&	0.86	&	22.81	&	3.59	&	3.83	\\
AGAL294.336$-$01.705	&	MMB294.337$-$01.706	&	5.6	&	1.1	&	1.72	&	1.0	&	8.1	&	\multicolumn{1}{c}{$\cdots$}	&	22.59	&	1.03	&	0.11	\\
AGAL294.511$-$01.622	&	MMB294.511$-$01.621	&	3.3	&	1.1	&	3.60	&	1.0	&	8.1	&	0.14	&	23.11	&	1.89	&	1.98	\\
AGAL294.976$-$01.734	&	MMB294.977$-$01.734	&	3.1	&	1.3	&	6.73	&	0.2	&	8.4	&	0.05	&	23.20	&	1.02	&	-1.10	\\
AGAL294.989$-$01.719	&	MMB294.990$-$01.719	&	2.2	&	1.1	&	4.87	&	1.1	&	8.1	&	0.17	&	23.10	&	2.03	&	2.19	\\
AGAL296.893$-$01.306	&	MMB296.893$-$01.305	&	2.4	&	1.3	&	0.98	&	10.0	&	9.8	&	\multicolumn{1}{c}{$\cdots$}	&	22.52	&	2.71	&	3.18	\\
AGAL297.391$-$00.634	&	MMB297.406$-$00.622$^\dagger$	&	68.2	&	1.6	&	10.82	&	10.7	&	10.1	&	0.50	&	22.38	&	3.66	&	3.31	\\
AGAL298.182$-$00.786	&	MMB298.177$-$00.795$^\dagger$	&	37.7	&	1.0	&	3.60	&	10.4	&	9.9	&	1.49	&	23.15	&	3.94	&	3.56	\\
AGAL298.224$-$00.339	&	MMB298.213$-$00.343$^\dagger$	&	41.4	&	1.3	&	9.89	&	11.4	&	10.5	&	3.89	&	23.31	&	4.61	&	3.33	\\
AGAL298.263+00.739	&	MMB298.262+00.739$^\dagger$	&	2.5	&	1.3	&	3.23	&	4.0	&	7.5	&	0.37	&	22.92	&	2.83	&	3.47	\\
AGAL298.631$-$00.362	&	MMB298.632$-$00.362$^\dagger$	&	6.0	&	1.6	&	1.31	&	11.9	&	10.8	&	\multicolumn{1}{c}{$\cdots$}	&	22.31	&	2.77	&	3.38	\\
AGAL298.724$-$00.086	&	MMB298.723$-$00.086$^\dagger$	&	5.0	&	1.1	&	1.84	&	10.6	&	9.9	&	0.65	&	22.94	&	3.45	&	3.20	\\
AGAL299.012+00.127	&	MMB299.013+00.128$^\dagger$	&	3.5	&	1.9	&	5.29	&	10.2	&	9.6	&	1.66	&	22.72	&	3.66	&	4.03	\\
AGAL300.504$-$00.176	&	MMB300.504$-$00.176$^\dagger$	&	1.4	&	1.3	&	4.83	&	9.6	&	9.0	&	2.16	&	23.11	&	3.95	&	3.67	\\
AGAL300.969+01.146	&	MMB300.969+01.148$^\star$	&	7.0	&	1.3	&	6.49	&	4.3	&	7.3	&	1.25	&	23.41	&	3.69	&	3.05	\\
AGAL301.136$-$00.226	&	MMB301.136$-$00.226$^\dagger$	&	2.3	&	1.3	&	2.59	&	4.3	&	7.3	&	0.78	&	23.90	&	3.77	&	2.59	\\
AGAL302.032$-$00.061	&	MMB302.032$-$00.061$^\dagger$	&	1.2	&	1.2	&	3.94	&	4.5	&	7.2	&	0.70	&	23.00	&	3.10	&	3.43	\\
\hline\\
\end{tabular}\\
$^{\rm{a}}$ Sources with a superscript have been searched for mid-infrared emission by \citet{gallaway2013}:  $\dagger$ and $\ddagger$ indicate infrared bright and infrared dark sources, respectively, and $\star$ identifies the sources they were unable to classify.\\ 
Notes: Only a small portion of the data is provided here, the full table is available in electronic form at the CDS via anonymous ftp to cdsarc.u-strasbg.fr (130.79.125.5) or via http://cdsweb.u-strasbg.fr/cgi-bin/qcat?J/MNRAS/.

\end{minipage}

\end{center}
\end{table*}

\setlength{\tabcolsep}{6pt}

\begin{table*}

\begin{center}\caption{Summary of derived parameters.}
\label{tbl:derived_para}
\begin{minipage}{\linewidth}
\small
\begin{tabular}{lc.......}
\hline \hline
  \multicolumn{1}{l}{Parameter}&  \multicolumn{1}{c}{Number}&	\multicolumn{1}{c}{Mean}  &	\multicolumn{1}{c}{Standard Error} &\multicolumn{1}{c}{Standard Deviation} &	\multicolumn{1}{c}{Median} & \multicolumn{1}{c}{Min}& \multicolumn{1}{c}{Max}\\
\hline
Radius (pc) &          375&       1.27&     0.05 & 
       1.01 &      0.97 &     0.01 &       5.69\\

Aspect Ratio &          577&       1.51&     0.02 &       0.40 & 
      1.40 &       1.01  &       3.39 \\
$Y$-factor &          577&       5.63&      0.16 &        3.89 & 
      4.55 &      0.98 &       24.66\\
Log[Clump Mass (\msun)] &          442&       3.27&     0.04
 &       0.77 &       3.36 &      -2.00 &       5.43\\
 Log[Column Density (cm$^{-2}$)]&          577&       22.86&     0.04 & 
       1.06 &       22.88 &       4.16 &       24.74\\
Log[$L_{\rm{MMB}}$ (Jy\,kpc$^2$)] &          442&
       3.48  &   0.05 &       0.99 &       3.47
 &      -1.30 &       6.34\\
\hline\\
\end{tabular}\\

\end{minipage}

\end{center}
\end{table*}

\subsection{Distances}
\label{sect:distance}

Using the velocity of the peak maser component and a Galactic rotation model (e.g., \citealt{brand1993,reid2009}) it is possible to estimate a particular source's kinematic distance. However, for sources located within the solar circle  (i.e., $<$ 8.5\,kpc of the Galactic centre) there is a two fold degeneracy as the source velocity corresponds to two distances equally spaced on either side of the tangent position. In a follow-up paper to the MMB survey \citet{green2011b} used archival HI data taken from the Southern and VLA Galactic Plane Surveys (SGPS (\citealt{mcclure2005}) and VGPS (\citealt{stil2006}), respectively) to resolve the distance ambiguities to a large number of methanol masers.

\citet{green2011b} examined HI spectra for 734 methanol masers, of which 204 are located at Galactic longitudes between $\ell=20\degr$-60\degr\ and therefore are not included in the current MMB catalogue and are not considered here. However, the \citet{green2011b} study does include 525 of the 671 MMB sources located in the  overlapping ATLASGAL and MMB region. Breaking this down further we find that this includes 506 of the 628 MMB sources associated with an ATLASGAL source as discussed in Sect.\,\ref{sect:atlas-mmb}, and 31 of the \unmatchedmmb\ methanol masers not associated with submillimetre emission mentioned in Sect.\,\ref{sect:unmatched_mmb}.

\citet{green2011b} used the Galactic rotation model of \citet{reid2009} to determine the kinematic distances to their sample of methanol masers. However, this model's assumption of a flat rotation curve with a high rotational velocity leads to very noticeable differences in the fourth quadrant of the Galaxy between the model-derived tangent velocities and the empirically derived values determined by \citet{mcclure2007} from the HI termination velocities (see their Fig.\,8). This leads to the near and far distance being located farther from the tangent position than would otherwise be expected, and produces a large hole in the Galactic distribution around the tangent positions (see Fig.\,4 of \citealt{green2011b}).

In order to avoid this we have used the Galactic rotation curve of \citet{brand1993} as its model tangent velocities are a much closer match to the HI termination velocities. The use of a different rotation model in most cases results in only a slight change in the estimated kinematic distances from those given by \citet{green2011b}. In addition to the minor difference in kinematic distances imposed by the change of rotation model ($\sim$1\,kpc) we also: 1) place any source with a velocity within 10\,\kms\ of the tangent point at the tangent distance, and 2) place any sources within the solar circle at the near distance if a far-distance allocation would lead to a height above the mid-plane larger than 4 times the scale height of young massive stars (i.e., $\sim$30\,pc; \citealt{reed2000, urquhart2011}).

\citet{green2011b} provides distance solutions for 385 of the ATLASGAL-MMB associations and 16 of the unassociated MMB sources. We have adopted their distances for 378 of the ATLASGAL-MMB associations and find the values determined from the difference rotation curves, after applying the two criteria mentioned in the previous paragraph, agree within 1\,kpc in every case. We only disagree with the distance allocations given by \citet{green2011b} for seven sources. Of these, we have associated four sources with the G305 complex (AGAL305.361+00.186, AGAL305.362+00.151, AGAL305.799$-$00.244 and AGAL305.887+00.016), three located at the tangent position (AGAL309.384$-$00.134, AGAL311.627+00.266 and AGAL336.916$-$00.022). Only one source (AGAL351.774$-$00.537) is placed at the near distance of 0.4\,kpc by the \citet{reid2009} model but is placed outside the solar circle by the \citet{brand1993} model at a distance of 17.4\,kpc. 

There are 83 ATLASGAL-MMB associations that have not been assigned a distance by \citet{green2011b}. We have allocated distances to 64 of these. Fourteen have been placed at the tangent position, fourteen have been associated with a well-known complex (i.e., G305 and W31) and 19 have been placed at the near distance. Fifteen of the sources placed at the near distance are because a far-distance allocation would place than more then 120\,pc from the Galactic mid-plane. Finally, the velocities of seventeen sources places them outside the solar circle, and thus, these sources do not suffer from the kinematic distance ambiguity problem.

 \begin{figure}
\begin{center}
\includegraphics[width=0.49\textwidth, trim= 0 0 0 0]{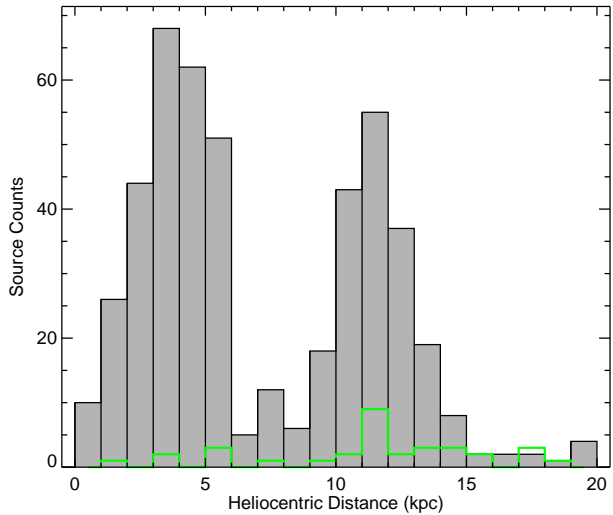}

\caption{\label{fig:atlas_mmb_distance_hist} Heliocentric distance distribution for the 473 MMB distances available (grey filled histogram) and the distance of 31 MMB sources not associated with an ATLASGAL source (green histogram). The bin size is 1\,kpc.} 

\end{center}
\end{figure}

In total we have distances to 442 ATLASGAL-MMB associated sources and their distribution is shown in Fig.\,\ref{fig:atlas_mmb_distance_hist} (grey filled histogram). The distribution is shown to be bimodal, with peaks at 3-4\,kpc and 11-12\,kpc. The low number of sources between these two peaks is due to an almost total lack of any MMB sources within 3\,kpc of the Galactic centre.

In addition to the distances we have determined for the ATLASGAL-MMB associations, we have obtained distances for 33 of the 43 MMB masers not associated with thermal dust emission. Nineteen of these were drawn from \citet{green2011b}, eleven are found to be located in the outer Galaxy using the \citet{brand1993} model, two others are located at the tangent position and one is placed at the near distance as a far distance would place more than 120\,pc from the Galactic mid-plane. It is clear from the distribution of this sample of sources (green histogram shown in Fig.\,\ref{fig:atlas_mmb_distance_hist}) that the majority have distances larger than 9\,kpc. In the previous section (i.e., Sect.\,3.3) we suggested that one possible explanation for the non-detection of submillimetre dust emission from these sources could be that they are located at larger distances; this distribution offers some support for that hypothesis. The median value for the distance for these unassociated MMB sources is $\sim$13\,kpc compared with a median value of $\sim$5\,kpc for the ATLASGAL-MMB associations. A KS test on the two samples shows their distance distributions are significantly different and we are able to reject the null hypothesis that they are drawn from the same population.

\subsection{Sizes and morphology}
\label{sect:size}

\begin{figure}
\begin{center}
\includegraphics[width=0.49\textwidth, trim= 0 0 0 0]{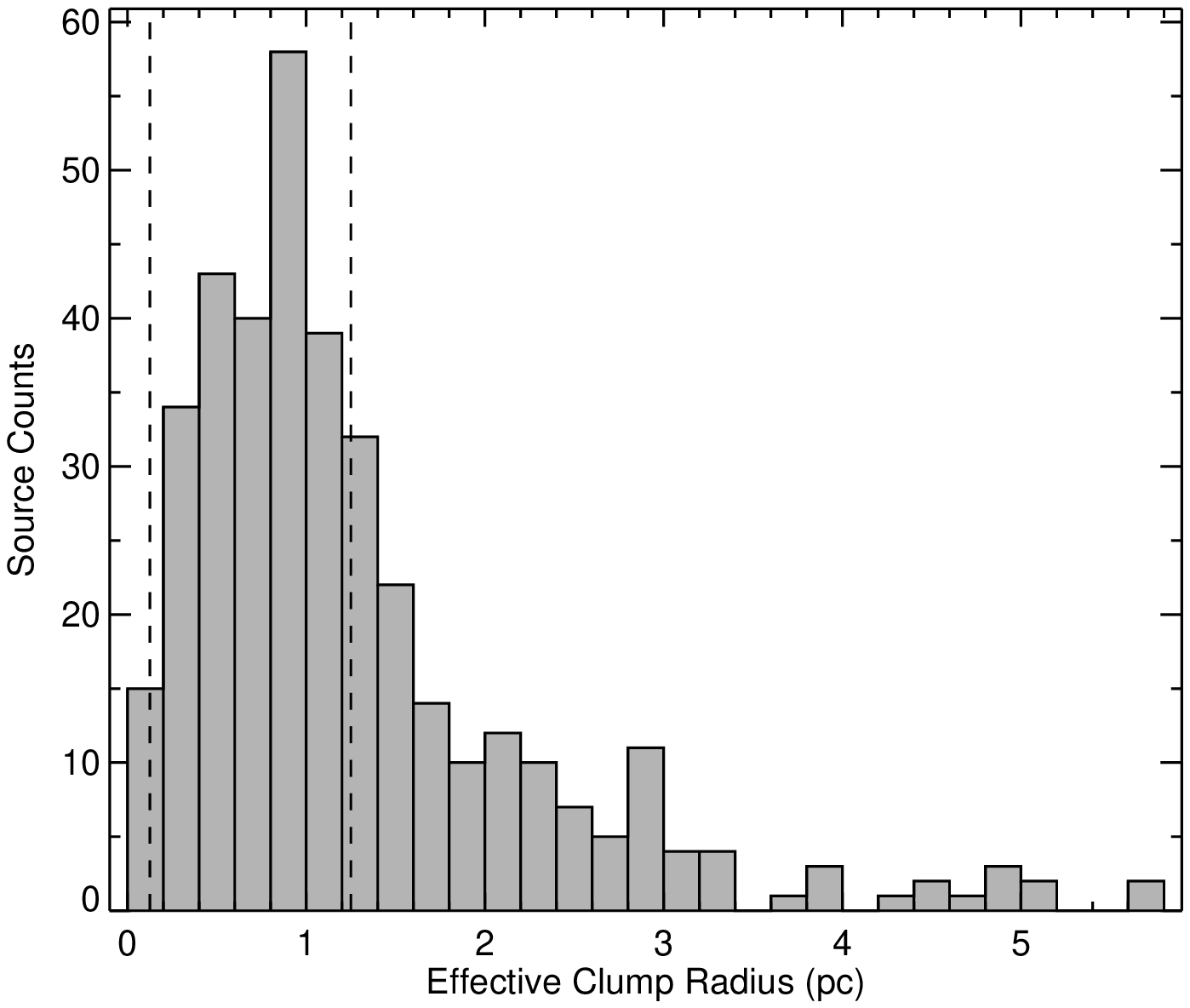}
\includegraphics[width=0.49\textwidth, trim= 0 0 0 0]{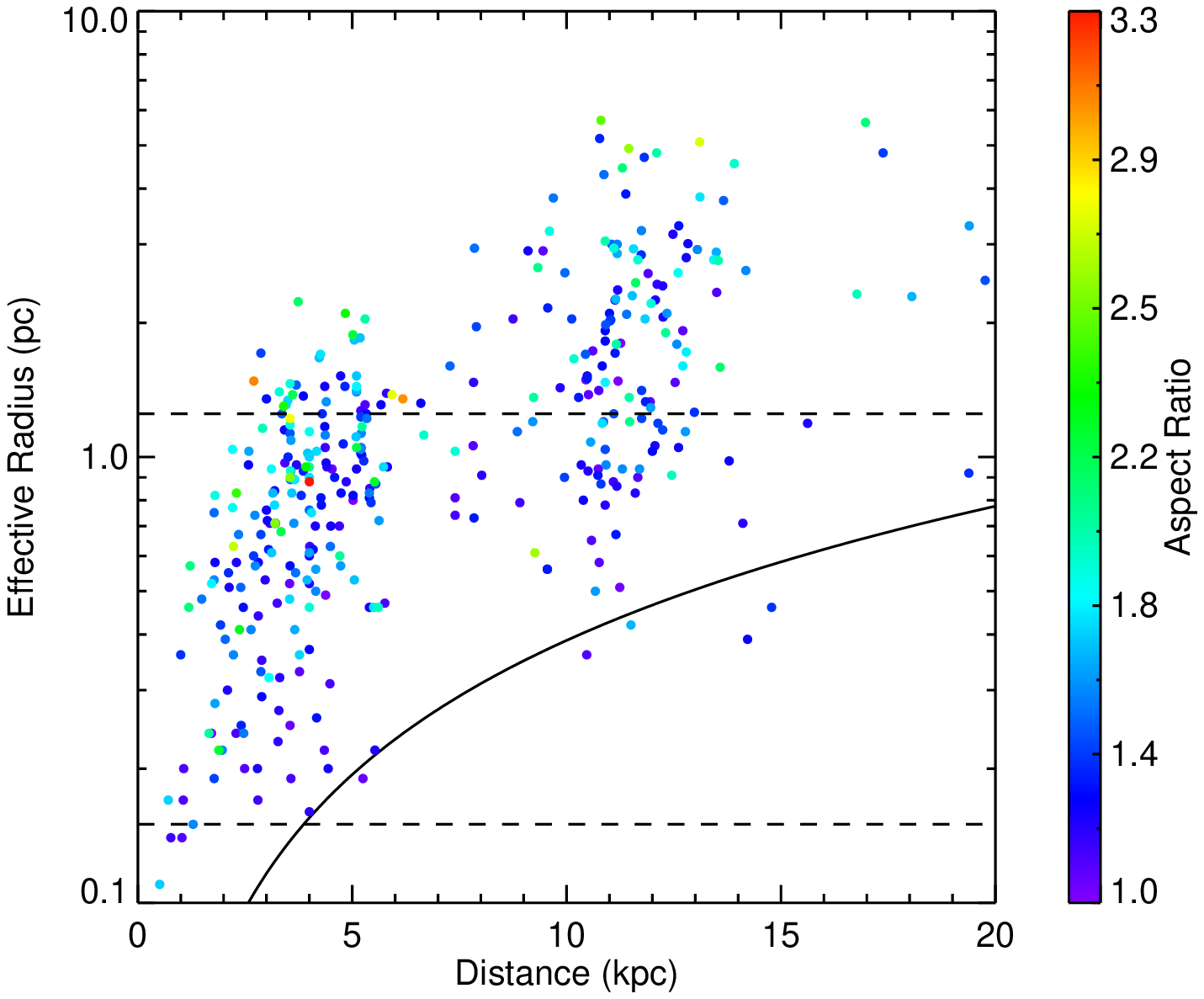}

\caption{\label{fig:physical_size} Upper panel: the distribution of effective radius for the whole ATLASGAL-MMB associated sample (grey filled histogram). Lower panel: the radius and aspect ratio distribution as a function of heliocentric distance. The left and right (lower and upper) dashed vertical (horizontal) lines in the upper panel (lower panel) indicate the radii separating cores and clumps (0.125\,pc), and clumps and clouds (1.25\,pc), respectively. The solid curved line shown in the lower panel shows the physical resolution of the survey based on the APEX 19.2\arcsec\ beam at 870\,\mum\ while the colours give an indication of the aspect ratio of each sources (see colour bar on the right for values). } 

\end{center}
\end{figure}

Using the kinematic distances discussed in the previous section and the effective angular radii derived for the clumps by \sex\ we are able to estimate their physical sizes (see \citealt{rosolowsky2010} for definition of effective radius). The distribution of sizes is presented in the upper panel of Fig.\,\ref{fig:physical_size} and ranges from 0.1 to several pc, with a peak at $\sim$1\,pc. The dashed vertical lines shown on this plot at radii of 0.15 and 1.25\,pc indicate the boundary between cores and clumps, and clumps and clouds, respectively (as adopted by \citealt{dunham2011} from Table\,1 of \citealt{bergin2007}). However, we note that the distribution is continuous with no features at these sizes so the definitions are probably somewhat arbitrary.

As also reported by \cite{tackenberg2012}, from a study of ATLASGAL candidate starless clumps we find no correlation between angular size and distance, which results in an approximately linear correlation between physical sizes and distance. It is important to bear in mind that at distances of a kpc or so we are primarily resolving structures on the size scale of cores, at intermediate distances, clumps, and at greater distances, entire cloud structures. This will have an effect on some of the derived parameters such as column and volume densities. However, although the sample covers a large range of sizes, the majority falls into the size range suggested for clumps, which are more likely to be in the process of forming clusters rather than a single massive star. For simplicity we refer to our sample as clumps, with the caveat that it includes a large range of physical sizes.

In the lower panel of Fig.\,\ref{fig:physical_size} we plot the source radius as a function of heliocentric distance. The colours of the symbols used in this plot give an indication of the aspect ratio of each source, the values of which can be read off from the colour bar to the right of the plot. This plot clearly illustrates that at larger distances we are probing larger physical structures, however, the aspect ratio of this sample of objects does not appear to have a significant distance dependence. This may suggest that even at larger distance the molecular clouds associated with MMB sources are still single structures. 

\begin{figure}
\begin{center}
\includegraphics[width=0.49\textwidth, trim= 0 0 0 0]{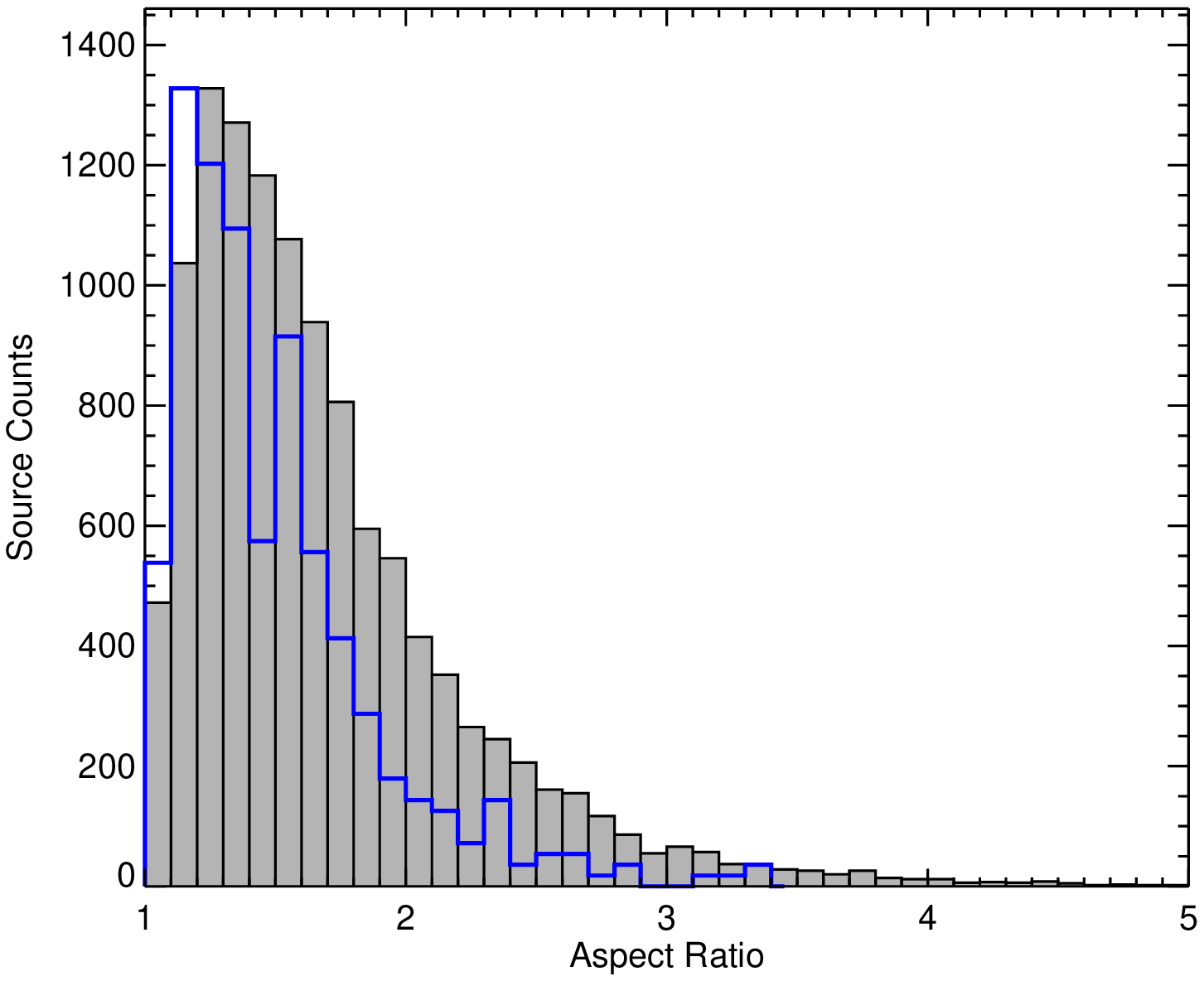}
\includegraphics[width=0.49\textwidth, trim= 0 0 0 0]{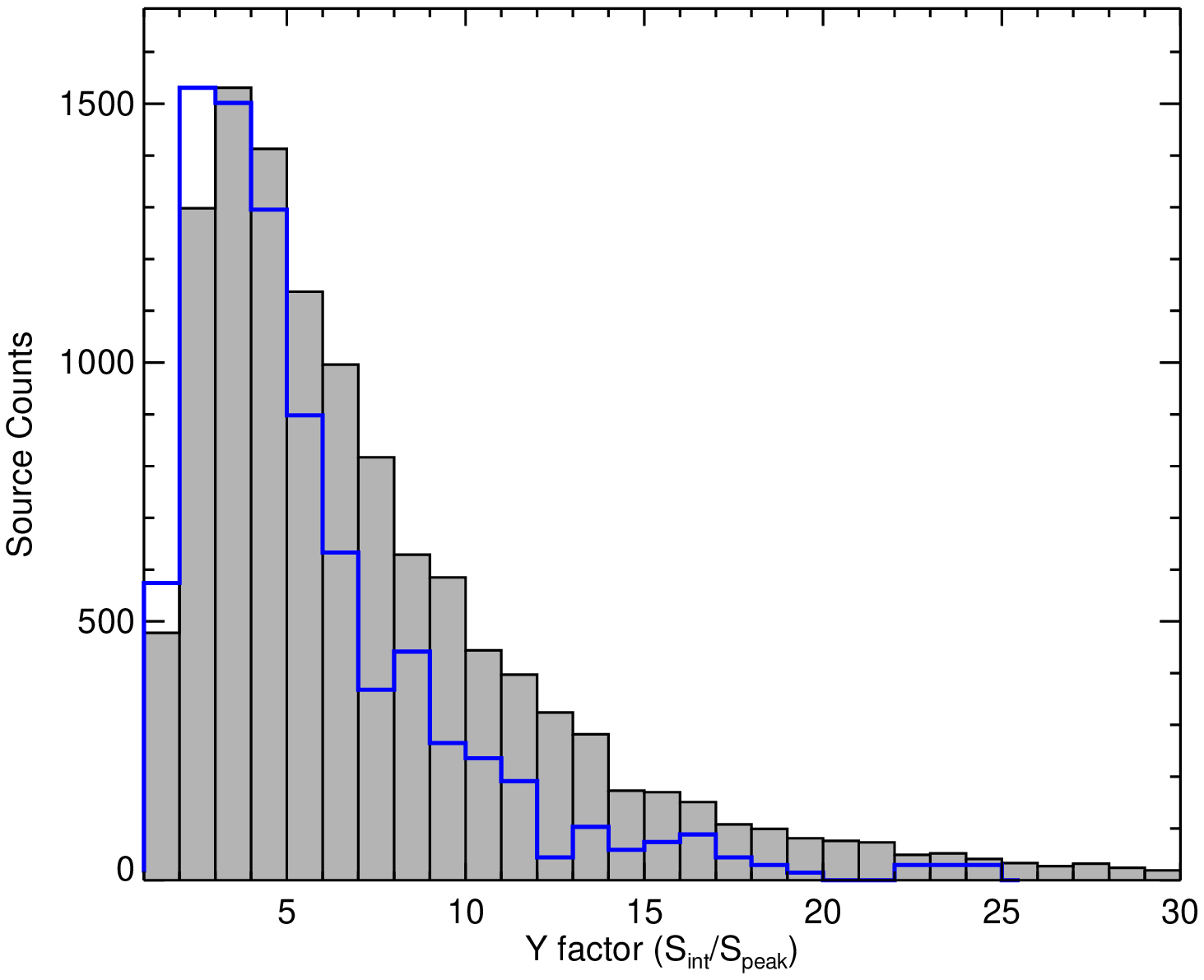}

\caption{\label{fig:morphology} In the upper and lower panels we present plots showing the source aspect ratio and the $Y$-factor, respectively,  of the whole ATLASGAL sample (grey histogram) and the ATLASGAL-MMB associated sources (blue histogram).  The ATLASGAL-MMB distributions have been normalised to the peak of the whole ATLASGAL sample. The bin sizes used are 0.1 and 1 for the aspect ratio and the $Y$-factor, respectively.} 

\end{center}
\end{figure}

In the upper panel of Fig.\,\ref{fig:morphology} we present the distribution of the aspect ratios of ATLASGAL-MMB sources (blue histogram), which is shown against that of the whole population of compact ATLASGAL sources (grey filled histogram). It is clear from this plot that the ATLASGAL-MMB sources have a significantly smaller aspect ratio than the general population, which would suggest they are more spherical in structure. The mean (median) values for the ATLASGAL-MMB and the whole population are 1.51$\pm$0.02 (1.40) and 1.69$\pm$0.01 (1.55), respectively. A similar median value of 1.3 was found by \citet{thompson2006} from a programme of targeted SCUBA submillimetre observations of clumps associated with \uchii\ regions.

\begin{figure*}
\begin{center}

\includegraphics[width=0.49\textwidth, trim= 0 0 0 0]{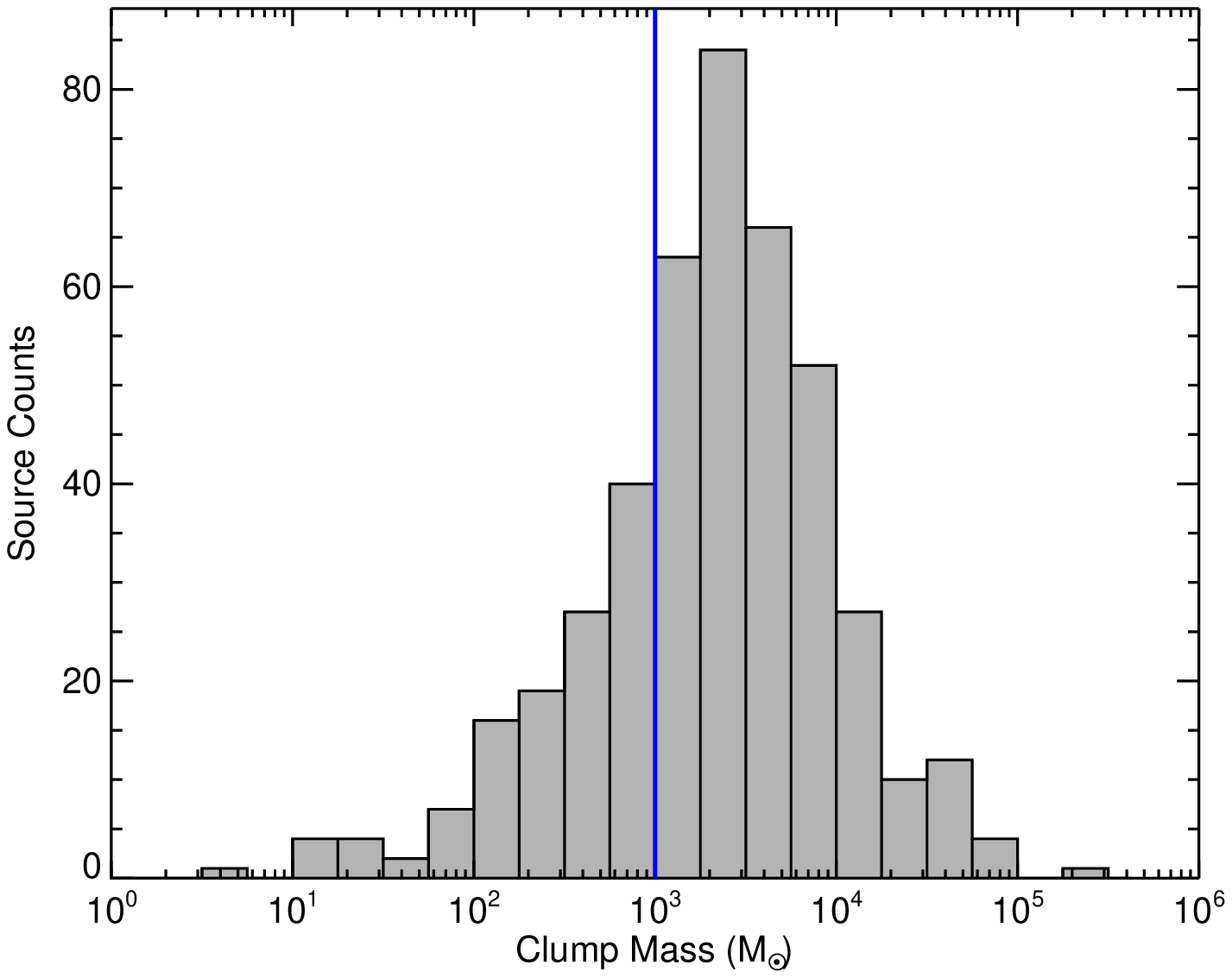}
\includegraphics[width=0.47\textwidth, trim= 0 0 0 0]{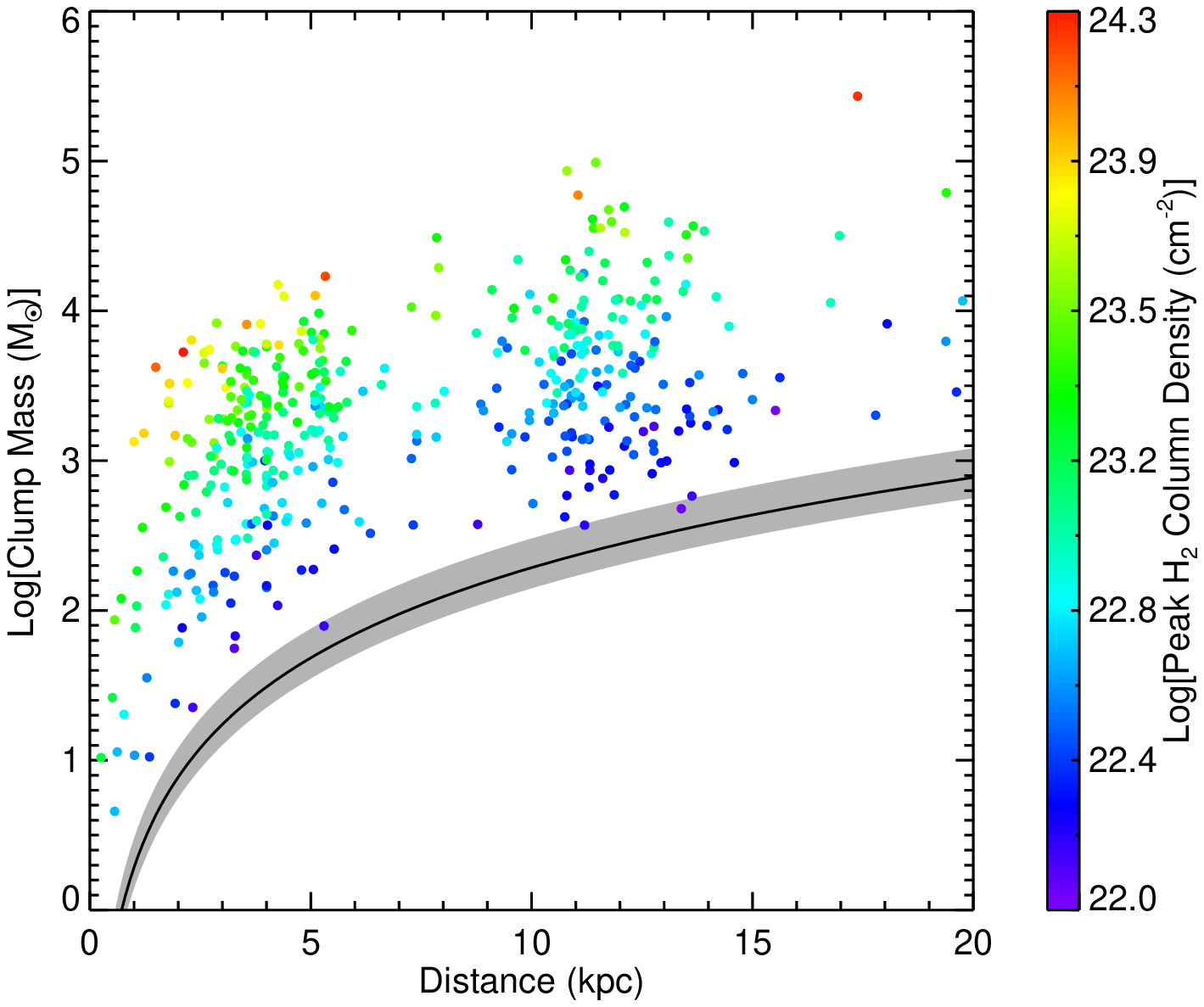}

\caption{\label{fig:clump_mass} Left panel: The isothermal dust mass distribution of the ATLASGAL-MMB associated clumps assuming a dust temperature of 20\,K. The vertical blue line indicates the completeness limit (see text for details). The bin size is 0.25. Right panel: Dust mass distribution of ATLASGAL-MMB associated clumps as a function of heliocentric distance. The colours give an indication of the peak column density of each source (see colour bar on the right for values) and the solid black line and the grey filled region indicated the mass sensitivity limit of the survey and its associated uncertainty assuming a dust temperature of 20$\pm5$\,K.} 

\end{center}
\end{figure*}

Another way to compare the morphology of the ATLASGAL-MMB associations with the general population of ATLASGAL sources is to look at the ratio of their integrated to peak fluxes; this is referred to as the $Y$-factor. This has been used to investigate the general extent of submillimetre clumps associated with high-mass protostellar cores (HMPOs) and \uchii\ regions by \citet{williams2004} and \citet{thompson2006}, respectively. In the lower panel of Fig.\,\ref{fig:morphology} we plot the $Y$-factor for the whole ATLASGAL compact source population (filled grey histogram) and the ATLASGAL-MMB associated sources (blue histogram). The $Y$-factors for both distributions peak at similar values, between 2-4, however, the ATLASGAL-MMB associated sources have significantly lower overall $Y$-factors (with a median value of 4.8 compared to 6.7 for the whole ATLASGAL compact source population). A KS test is able to reject the null hypothesis that these are drawn from the same population with greater than three sigma confidence. 

Peaks in the $Y$-factor around 3 are also seen in the HMPO and \uchii\ region samples of \citet{williams2004} and \citet{thompson2006} and appear to have a similar distribution to our sample sample. The large range of heliocentric distances over which their HMPO sample is distributed led \citet{williams2004} to suggest that the envelope structures may be scale-free. The ATLASGAL clumps are similarly located over a range of heliocentric distances and thus the ATLASGAL beam traces structure on scales from a few times 0.1 pc to a several pc. As we observe similar Y-factors for a range of spatial resolutions this suggests that the radial density distribution follows a similar power law over a range of spatial scales.  Hence the structure of the clump \emph{envelopes} (at least on pc scales) are likely to follow a scale-free power law.  The fact that all three samples (ATLASGAL-MMB, HMPO and \uchii\ regions) are likely to cover the whole range of embedded star formation and broadly show the same $Y$-factor properties would seem to support this. Similarly to \citet{williams2004} and \citet{thompson2006}, we find that a significant amount of the mass associated with the ATLASGAL-MMB sources is found in the outer regions of the clumps, from which we conclude that this situation does not change significantly over the evolution of the embedded stars.  

In summary we have determined that overall the ATLASGAL-MMB associated sources are roughly spherical, centrally condensed clumps that appear to have a scale-free envelope with a methanol maser coincident with the peak of the submillimetre emission.

\subsection{Isothermal clump masses}

In calculating masses and column densities we have assumed that all of the measured flux arises from warm dust, however, free-free emission from embedded ionised gas and/or molecular line emission could make a significant contribution for broadband bolometers such as LABOCA. \citet{schuller2009} considered these two forms of contaminating emission and concluded that even in the most extreme case of the giant HII region associated with W43, and the CO (3-2) lines associated with extreme outflows, hot cores and photon-dominated regions, the contribution from free-free emission and molecular lines is of the order 20 and 15\,per\,cent, respectively, and in the majority of cases will be almost negligible (see also \citealt{drabek2012}).

Assuming the dust is generally optically thin and can be characterised by a single temperature we are able to estimate the isothermal dust masses for the ATLASGAL-MMB associated sources. Following \citet{hildebrand1983}, the total mass in a clump is directly proportional to the total flux density integrated over the source:

\begin{equation}
M \, = \, \frac{D^2 \, S_\nu \, R}{B_\nu(T_D) \, \kappa_\nu},
\end{equation}

\noindent where $S_\nu$ is the integrated 870\,$\mu$m flux, $D$ is the heliocentric distance to the source, $R$ is the gas-to-dust mass ratio (assumed to be 100), $B_\nu$ is the Planck function for a dust temperature $T_D$, and $\kappa_\nu$ is the dust absorption coefficient taken as 1.85\,cm$^2$\,g$^{-1}$ (this value was derived by \citet{schuller2009} by interpolating to 870\,$\mu$m from Table\,1, Col.\,5 of \citet{ossenkopf1994}).

As reliable dust temperatures are not available for the majority of our sample,  we make the simplifying assumption that all of the clumps have approximately the same temperature and set this to be 20\,K. Single-dish ammonia studies have derived kinetic gas temperatures for a large of number massive star formation regions that cover the full range of evolutionary stages. These include methanol masers (\citealt{pandian2012}); 1.1\,mm thermal dust sources identified by the Bolocam Galactic Plane Survey (BGPS; \citealt{dunham2011b}); 870\,$\mu$m ATLASGAL sources (\citealt{wienen2012}) and the Red MSX Source Survey (RMS; \citealt{urquhart2011b}).\footnote{All of these observational programmes used either the Green Bank Telescope (FWHM $\sim$30\arcsec;\citealt{urquhart2011b,dunham2011b}) or the Effelsberg telescope (FWHM $\sim$40\arcsec; \citealt{pandian2012,wienen2012}) and therefore have comparable resolution and sensitivity.} For the methanol masers mean and median kinetic temperatures of 26\,K and 23.4\,K were reported by \citet{pandian2012}, and for the massive young stellar objects (MYSOs) and \uchii\ regions have mean and median kinetic temperatures of 22.1 and 21.4\,K \citet{urquhart2011b}, while \citet{wienen2012} report kinetic temperatures of $\sim$24\,K for a subsample of ATLASGAL sources associated with methanol masers. \citet{dunham2011b} report lower kinetic gas temperatures 15.6$\pm$5.0\,K, however, their sample includes a larger fraction of starless clumps and so the lower mean temperature is expected.

A kinetic gas temperature of $\sim$25\,K would seem to characterise the clumps that show evidence of star formation, however, this temperature is likely to be an upper limit to the clump-averaged kinetic temperature as these observations were pointed at the peak emission of the clumps and the kinetic temperature is likely to be significantly lower towards the edges of the clumps (\citealt{zinchenko1997}). \citet{dunham2011b} estimate that using the peak kinetic temperature for the whole clump may underestimate the isothermal mass by up to a factor of two. Therefore we have chosen to use a value of 20\,K, consistent with a number of similar studies (cf. \citealt{motte2007,hill2005}). Given that the true clump-averaged kinetic temperatures of the sample are likely to range between 15 and 25\,K the resulting  uncertainties in the derived isothermal clump masses of individual sources are $\pm$43\,per\,cent (allowing for an uncertainty in temperature of $\pm$5\,K which is is added in quadrature with the 15\,per\,cent flux measurement uncertainty). However, this is unlikely to have a significant impact on the overall mass distribution or the statistical analysis of the masses. We note that 10\,per\,cent of the ATLASGAL-MMB associations are also associated with embedded UCHII regions. However, \citet{urquhart2011b} found that the presence of an \uchii\ region only results in an increase in clump-averaged kinetic temperatures of a few Kelvin.

In the left panel of Fig.\,\ref{fig:clump_mass} we present a plot of the isothermal dust mass distribution, while in the right panel we show the mass distribution as a function of heliocentric distance. It is clear from the right panel of this figure that we are sensitive to all ATLASGAL-MMB associated clumps with masses above 1,000\,\msun\ across the Galaxy, and our statistics should be complete above this level. (This completeness limit is indicated on the left panel of Fig.\,\ref{fig:clump_mass} by the vertical blue line.) In this regard it is interesting to note that the mass distribution peaks at several thousand solar masses (see left panel of Fig.\,\ref{fig:clump_mass}), which is significantly above the completeness limit, and so the drop off in the source counts between the completeness limit and the peak is likely to be real. This is an important point as it confirms that the methanol masers are preferentially associated with very massive clumps. 

\begin{figure}
\begin{center}
\includegraphics[width=0.49\textwidth, trim= 0 0 0 0]{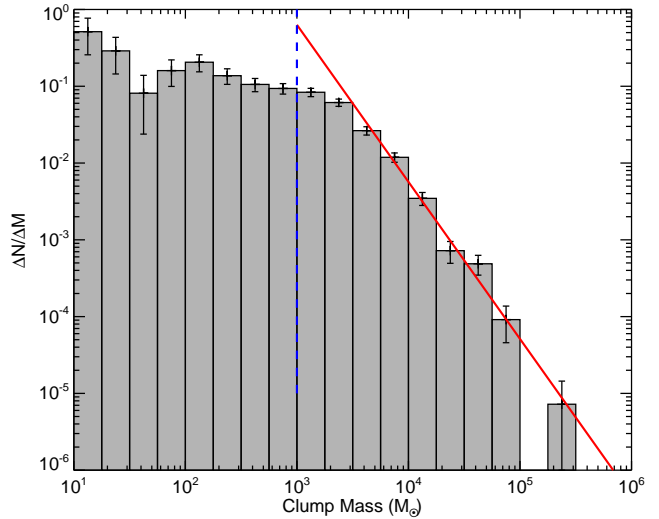}

\caption{\label{fig:mass_radius_dndm_histogram} Differential mass distribution. The red line shows a linear fit to the data and has a slope of $-$2.0$\pm$0.1 which is close to the Salpeter (1955) value and similar to that found for a number of other studies. The errors are estimated using \poi\ statistics and the bin size used is 0.25\,dex.} 

\end{center}
\end{figure}

According to \citet{lada2003} and \citet{motte2003} the radius and mass required to form stellar clusters is of the order 0.5-1\,pc and 100-1,000\,\msun, respectively. Given the sizes and masses of the ATLASGAL-MMB associated clumps  it is highly likely that the majority are in the process of forming clusters. Assuming a star formation efficiency (SFE) of 30\,per\,cent  and an initial mass function (IMF; \citealt{kroupa2001}), \citet{tackenberg2012} estimate that a clump mass of $\sim$1,000\,\msun\ is required to have the potential to form at least one 20\,\msun\ star, while a clump of $\sim$3,000\,\msun\ is required to form at least one star more massive than 40\,\msun. It is consistent with the assumption that methanol masers are associated mainly with high-mass star formation to find that the majority of ATLASGAL-MMB associations ($\sim$72\,per\,cent) have masses larger than $\sim$1,000\,\msun, and thus, satisfy the mass requirement for massive star formation.

We also note that approximately a third of the ATLASGAL-MMB associations ($\sim$28\,per\,cent) have masses below what is thought to be required to form at least one massive star. However, all of these sources also tend to be more compact objects (in most cases $\le$0.3\,pc) and may have a higher star formation efficiency and be forming smaller stellar systems where the stellar IMF does not apply (e.g., \citealt{motte2007}). 

\subsubsection{Peak column densities}

We estimate column densities from the peak flux density of the clumps using the following equation:

\begin{equation}
N_{H_2} \, = \, \frac{S_\nu \, R}{B_\nu(T_D) \, \Omega \, \kappa_\nu \, \mu
\, m_H},
\end{equation}

\noindent where $\Omega$ is the beam solid angle, $\mu$ is the mean molecular weight of the interstellar medium, which we assume to be equal to 2.8, and $m_H$ is the mass of an hydrogen atom, while $\kappa_\nu$ and $R$ are as previously defined. We again assume a dust temperature of 20\,K.  
 
The derived column densities are in the range $\sim$10$^{22-24}$\,cm$^{-2}$, peaking at 10$^{23}$\,cm$^{-2}$. This corresponds to a surface density of few times 0.1\,g\,cm$^{-2}$, which is a factor of a few lower than the value of 1\,g\,cm$^{-2}$ predicted to be the lower limit for massive star formation (i.e., \citealt{mckee2003,krumholz2008}). However, we should not read too much into this as the column densities of the more distant sources can be affected by beam dilution. This effect is nicely illustrated in the right panel of Fig.\,\ref{fig:clump_mass} where we use colours to show the column densities as a function of distance (see colour bar for values). There is clearly a dependence of column density on distance as, given the resolution of the survey, we are sampling larger scale physical structures as the distance increases, which preferentially reduces the column densities of more distant sources. It is likely that these more distant sources would fragment into smaller and denser core-like structures at higher resolution (e.g., \citealt{motte2007}). Therefore one should exercise caution when drawing conclusions from beam-averaged quantities such as the column and volume densities (cf. \citealt{tackenberg2012}).

\begin{figure*}
\begin{center}

\includegraphics[width=0.49\textwidth, trim= 0 0 0 0]{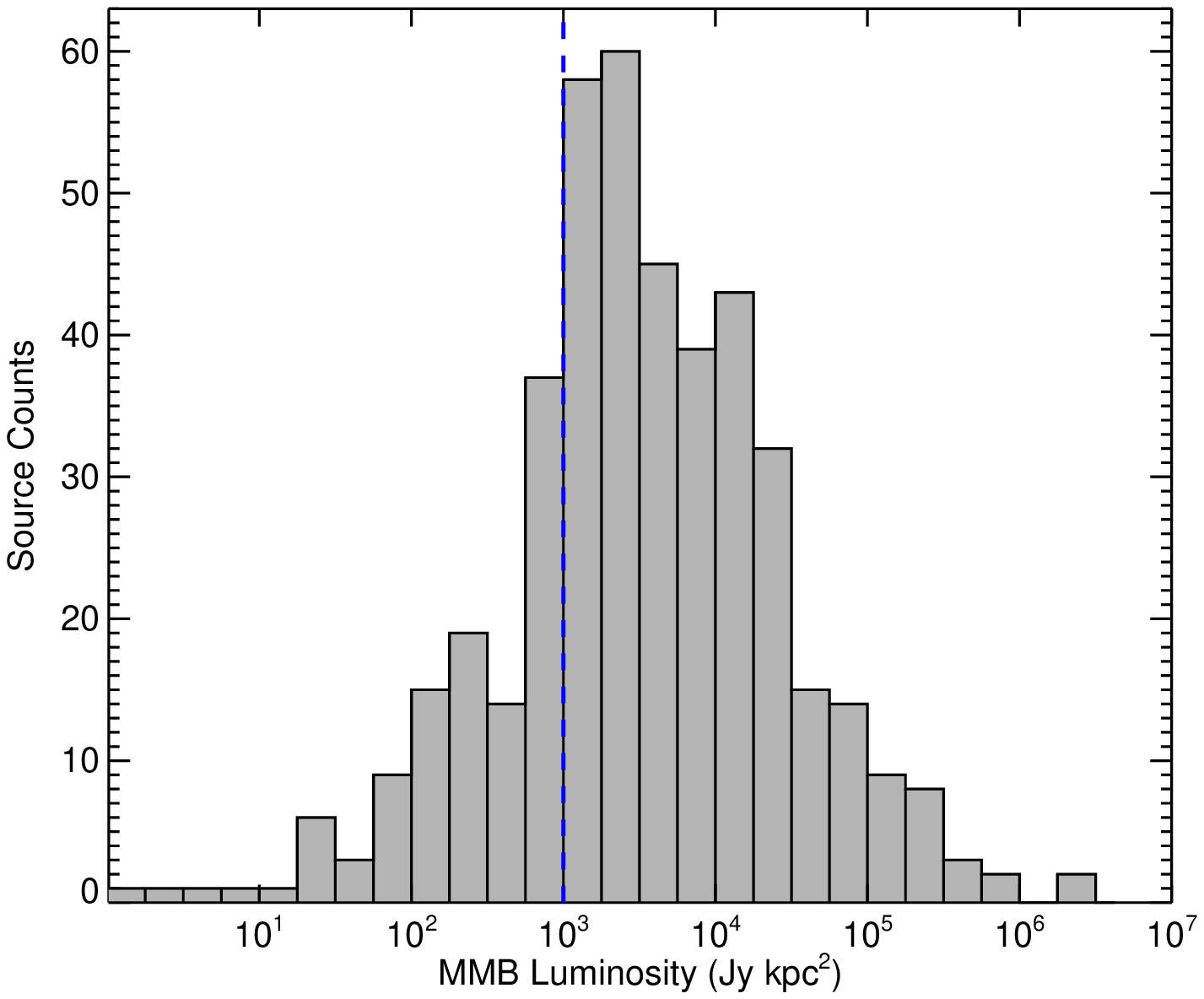}
\includegraphics[width=0.49\textwidth, trim= 0 0 0 0]{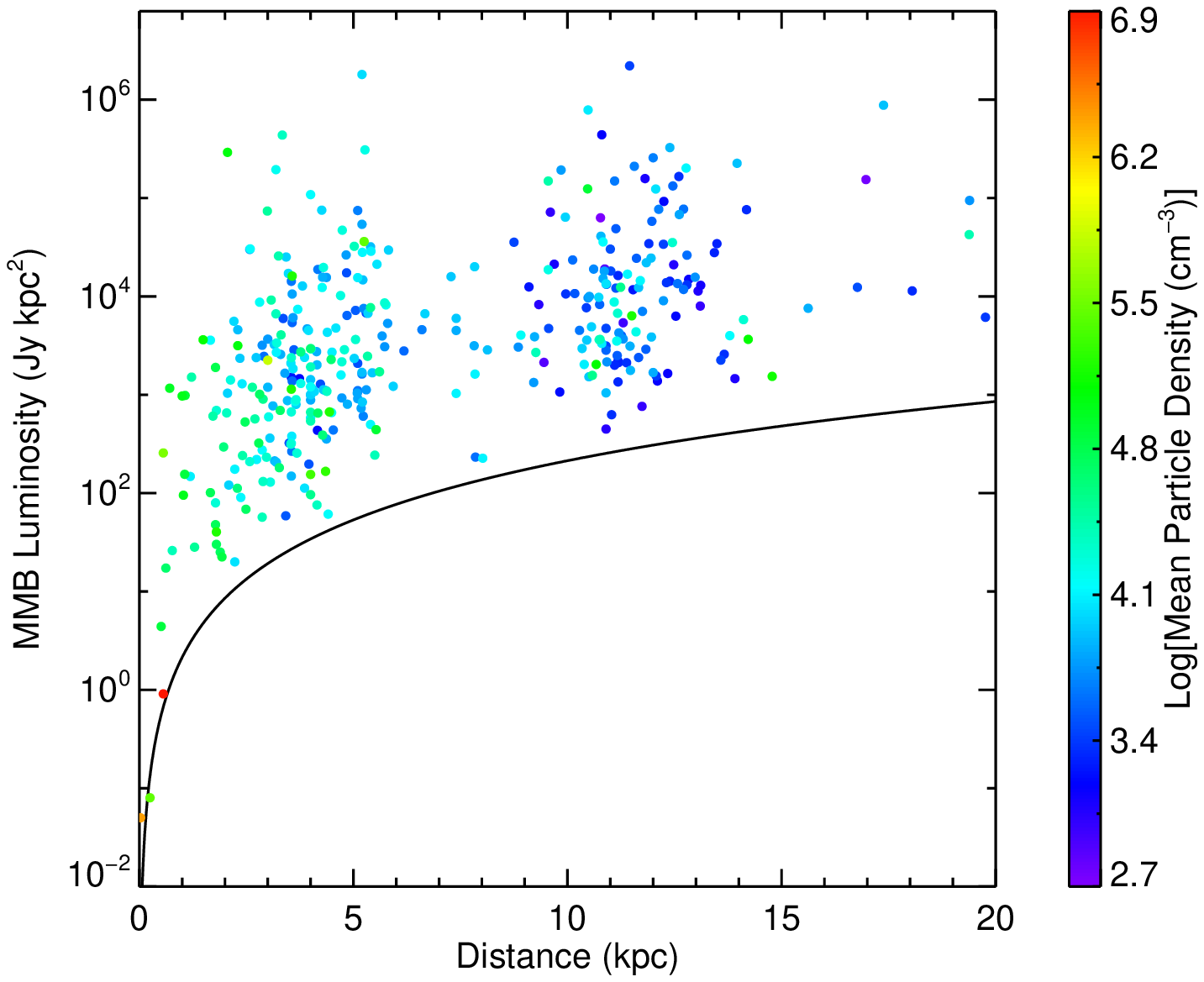}

\caption{\label{fig:mmb_luminosity}  In the left panel we present the methanol maser luminosity distribution of the ATLASGAL-MMB associations. The vertical blue dashed line shows the MMB luminosity completeness limit. The bin size used is 0.25\,dex. In the right panel we show the source luminosity as a function of heliocentric distance. The solid black line indicated the MMB surveys luminosity sensitivity limit. The colours used to plot each source show the clump-averaged volume density; see color bar to the right of the plot for values.} 

\end{center}
\end{figure*}

\subsubsection{Clump mass function}

In Fig.\,\ref{fig:mass_radius_dndm_histogram} we present the differential mass distribution for the ATLASGAL-MMB associated clumps. On this plot the dashed vertical blue line indicates the completeness limit and the solid red line shows the results of a linear least-squares fit to the bins above the peak in the mass distribution shown in the left panel of Fig.\,\ref{fig:clump_mass} ($\sim$3,000\,\msun). This line provides a reasonable fit to all of the bins above the peak mass. It does not fit the two bins just above the completeness limit, which may suggest that a second power law is required to account for these mass bins. The derived exponent ($\alpha$ where ${\rm{d}}N/{\rm{d}}M \propto M^{\alpha}$) of the fit to the high-mass tail for the ATLASGAL-MMB associated sources is $-2.0\pm0.1$, which is similar to values ($-2.0$\ to $-2.3$) derived by \citet{williams2004}, \citet{reid2005}, \citet{beltran2006} from studies of more evolved stages and \citealt{tackenberg2012} who targeted a sample of starless clumps identified from ATLASGAL data. This similarity between the clump mass function over the different stages of massive star formation would suggest that the CMF does not change significantly as the embedded star formation evolves.

\subsection{Methanol maser luminosity}

In the left panel of Fig.\,\ref{fig:mmb_luminosity} we present the 6.7\,GHz luminosity distribution of the ATLASGAL-MMB associated masers for which we have derived a distance. These ``luminosities'' have been calculated assuming the emission is isotropic using:

\begin{equation}
L_{\rm{MMB}} \, = \, 4\pi D^2 S_\nu
\end{equation}

\noindent where $D$ is the heliocentric distance in kpc and $S_\nu$ is the methanol maser peak flux density in Jy. $L_{\rm{MMB}}$ thus has units of Jy\,kpc$^2$. The distribution shown in the left panal of Fig.\,\ref{fig:mmb_luminosity} peaks at $\sim$3,000\,Jy\,kpc$^2$, but is skewed to higher luminosities with a mean value of $3\times10^4$\,Jy\,kpc$^2$. In the right panel of Fig.\,\ref{fig:mmb_luminosity} we show the methanol maser luminosities as a function of heliocentric distance. We can see from this plot that the MMB survey is effectively complete across the Galaxy ($\sim$20\,kpc) to all methanol masers with luminosities greater than $\sim$1,000\,Jy\,kpc$^2$. This completeness limit is shown by the dashed vertical line plotted on the luminosity distribution shown in the left panel of this figure. 

\subsubsection{Luminosity-volume density correlation}
\label{lum_vol_correlation}
There has been a number of recent methanol and water-maser and dust-clump studies that have reported a trend towards lower mean volume densities with increasing maser luminosity, which has been interpreted as the result of the evolution of the embedded star formation (i.e., \citealt{breen2011_water,breen2011_methanol,breen2012}).

\begin{figure}
\begin{center}
\includegraphics[width=0.49\textwidth, trim= 0 0 0 0]{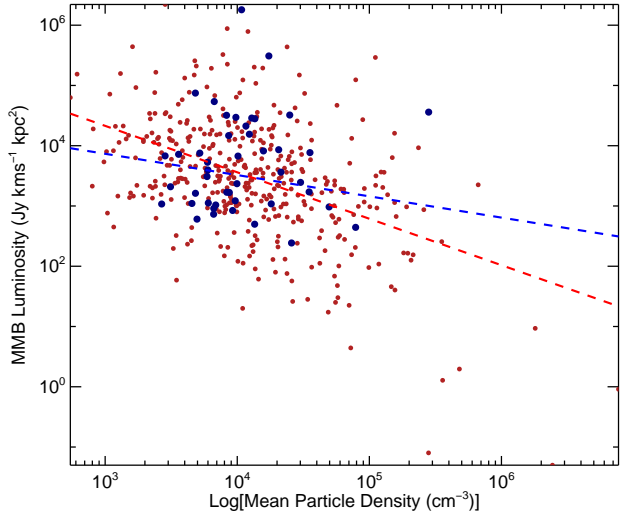}

\caption{\label{fig:maser_lum_volume_density}  The isotropic methanol-maser luminosity as a function of clump-averaged volume density. The red lines show the result of a linear least-squares fit to the data.} 

\end{center}
\end{figure}

To test this trend for our sample of methanol masers, we plot the maser luminosity as a function of clump-averaged volume density in Fig.\,\ref{fig:maser_lum_volume_density} for the whole sample of 374 ATLASGAL-MMB associations for which we have sufficient data to derive the maser luminosity and volume density.\footnote{Beam-averaged volume densities would be significantly larger. However, we use clump-averaged values here in order to be consistent with the analysis of \citealt{breen2011_methanol}.} We more or less replicate the correlation coefficient and least-squares fit gradient reported by \citet{breen2011_methanol} using the whole sample. Correlation coefficients are $-$0.43 and $-$0.46 and gradients $-0.78\pm0.09$ and $-0.85\pm0.16$ for the fit presented in Fig.\,\ref{fig:maser_lum_volume_density} and Fig.\,2 of \citet{breen2011_methanol}, respectively. However, from a casual inspection of the mean volume densities (see colour bar of the right panel of Fig.\,\ref{fig:mmb_luminosity}) it is clear that there is a significant distance dependency. It would appear that the poorer sensitivity to lower luminosity masers, and the increase in the spatial volume being sampled at larger distances, results in a decrease in the mean volume-densities of clumps and an increase in maser luminosities.

 To test this further, we have performed a partial Spearman correlation test of the mean volume-densities and maser luminosities to remove their mutual dependence on the distance (e.g., \citealt{yates1986}), of the form $r_{\rm AB,C}$, where
\begin{equation}
r_{\rm AB,C} \; = \; \frac {r_{\rm AB} -  r_{\rm AC} r_{\rm BC}}
{[(1-r^2_{\rm AC})(1-r^2_{\rm BC})]^{1/2}},
\end{equation}
\noindent where A, B and C are the maser luminosity, mean particle density and distance respectively, and  $r_{\rm AB}$, $r_{\rm AC}$ and $r_{\rm BC}$ are the Spearman rank correlation coefficients for each pair of parameters. The significance of the partial rank correlation coefficients is estimated using  \mbox{$r_{\rm AB,C} [(N-3)/
(1-r_{\rm AB,C}^2)]^{1/2}$} assuming it is distributed as Student's t statistic (see \citealt{collins1998} for more details).

This returned a partial correlation coefficient value of $-$0.06 and a $p$-value $\sim$0.6 and we are therefore unable to exclude the null hypothesis that the sample is drawn from a population where  rho\,=\,0. We would conclude that there is no intrinsic correlation between maser luminosity and clump-averaged volume density. We additionally performed tests on distance selected subsamples and obtained the same result. The scales traced by the ATLASGAL observations are much larger than those of either an HMC or an UCHII region, which by definition are $< 0.1$\,pc. We do not see evidence for radical changes in the density distribution of the envelope on pc scales with evolutionary state (the distribution of Y values being similar). Hence, the density distribution of the envelope does not evolve appreciably over the relevant timescale. For the clump-averaged volume density, the volume density of the envelope dominates over the much smaller core, and so we would not expect volume densities derived from single dish observations to change appreciably either.

Our conclusion that there is no correlation between clump density and methanol maser luminosity is supported by a recent study by \citet{cyganowski2013} who also failed to find a correlation between water maser luminosities and clump densities towards a sample of extended green objects (EGOs; \citealt{cyganowski2008}). This is not surprising, since for the masers to work, the density must be within a narrow range, which has little to do with the density sampled in the ATLASGAL beam. Since the H$_2$ number densities form a central part of the evolutionary arguments put forward by Breen et al. the results presented here cast significant doubt on some of their conclusions.

\subsection{Maser luminosity-clump mass: completeness}
\label{sect:completeness}

\begin{figure}
\begin{center}
\includegraphics[width=0.49\textwidth, trim= 0 0 0 0]{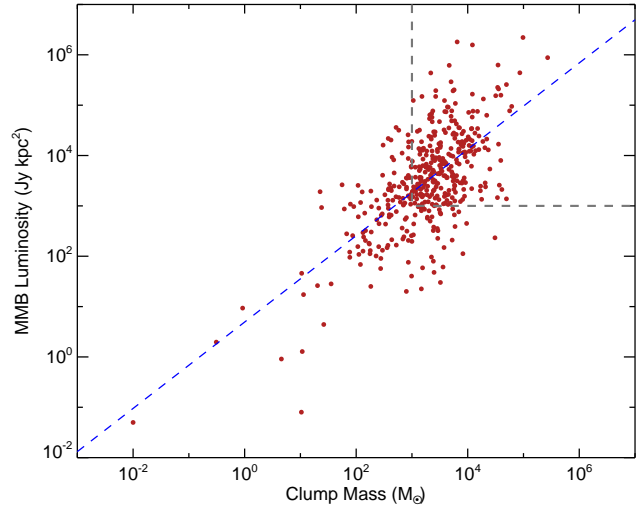}

\caption{\label{fig:maser_mass_plot} The relationship between clump mass and luminosity of the embedded maser. The region in the top right corner of the plot outlined by the dashed box shows the part of the parameter space we are complete to both methanol maser luminosity and clump mass. The dashed blue lines show the results of a linear fit to the data. Using a partial correlation function to remove the dependency of both of these parameters on distance$^2$ we obtain a coefficient value of 0.37.} 

\end{center}
\end{figure}

In Fig.\,\ref{fig:maser_mass_plot} we present a plot of the maser luminosity as a function of clump mass for the 442 ATLASGAL-MMB associations for which we have a distance. The upper right region outlined by the grey dashed line of this figure indicates the region of parameter space that we are complete across the Galaxy to both methanol maser luminosity and clump mass; this region contains 280 sources. 

Applying a partial correlation function to the clump mass and maser luminosity to remove the dependence of these two parameters on distance (\citealt{collins1998}) we obtain a partial correlation coefficient value of 0.37 with a $p$-value $\ll$ 0.01, suggesting that there is a weak correlation between the maser luminosity and clump mass. We fit these data with a linear least squared function the result of which is shown in Fig.\,\ref{fig:maser_mass_plot} as a blue dashed line. The parameters of the fit are: Log($L_{\rm{MMB}})=(0.857\pm0.046)\,M_{\rm{clump}}+(0.692\pm0.155)$. Within the errors the fit to the data has a gradient close to one and so the relationship between mass and maser luminosity is relatively linear. As discussed in Sect.\,4.3 the most massive clumps are likely to be forming more massive stars (assuming a Kroupa IMF and a SFE of 30\,per\,cent) and therefore the weak correlation between the clump mass and maser luminosity may be related to this. This suggests that higher (isotropic) maser luminosity is related to higher stellar luminosity in some way, perhaps via the pumping mechanism or maybe the larger clump/core just provides a longer maser amplification column.

\section{Discussion}

There have been a number of studies that have tried to firmly establish a connection between the presence of a methanol maser and ongoing massive star formation and these have been relatively successful. Most have searched for methanol masers towards low luminosity protostellar sources and, when no masers were detected, have concluded that methanol masers are exclusively associated with high-mass protostars (e.g., \citealt{minier2003,bourke2005}). However, most of these surveys have focused on small samples, with poorly defined selection criteria, and often use IRAS fluxes to determine the luminosity of the embedded source and so the luminosities may have been overestimated. So, although these studies have set a lower limit to the luminosity of the associated protostar, it is unclear whether the results obtained are applicable to the whole methanol maser population.

In this subsection we will draw on the results presented in the previous sections to test this hypothesis. The ATLASGAL-MMB sample presented in this paper includes 94\,per\,cent of the MMB sources in the overlap region of the two surveys, and 90\,per\,cent of the entire MMB published catalogue. Moreover, since the number of methanol masers in the whole Galaxy is not expected to exceed a few thousand, our sample is likely to incorporate a large fraction of the whole Galactic population. Therefore any statistical results drawn from this sample will be a fair reflection of the properties of the general population.

\subsection{Empirical mass-size relationship for massive star formation}
\label{mass-size-relationship}

\subsubsection{Criterion for massive star formation}

In two papers \citet{kauffmann2010a, kauffmann2010b} investigated the mass-radius relationship of nearby ($<$500\,pc) molecular cloud complexes (i.e., Ophiuchus, Perseus, Taurus and the Pipe Nebula) and found that clouds that were devoid of any high-mass star formation generally obeyed the following empirical relationship:

\begin{equation}
m(r) \le 580\,{\rm{M}}_\odot\,(R_{\rm{eff}}/{\rm{pc}})^{1.33}
\end{equation}

\noindent where $R_{\rm{eff}}$ is the effective radius as defined by \citet{rosolowsky2010}.\footnote{ Note that when deriving this relationship \citet{kauffmann2010b} reduced the dust opacities of \citet{ossenkopf1994} by a factor of 1.5. This reduced value for the opacities has not been applied when determining the clump masses presented here and therefore we have rescaled the value given by \citet{kauffmann2010b} of 870 by this factor to the value of 580 given in Eqn.\,5 (cf. \citealt{dunham2011}).} Comparing this mass-size relation with samples of known high-mass star forming regions such as those of \citet{beuther2002}, \citet{hill2005}, \citet{motte2007} and \citet{Mueller2002}, \citet{kauffmann2010b} found that all of these regions occupied the opposing side of the parameter space (i.e., where $m(r) \ge 580$\,\msun\ $(R_{\rm{eff}}/{\rm{pc}})^{1.33}$. This led them to suggest that the relation may approximate a requirement for massive star formation, where only more massive clumps have the potential to form massive stars. However, \citet{kauffmann2010b} states that larger samples of massive star forming clouds are required to strengthen this hypothesis.

\begin{figure}
\begin{center}
\includegraphics[width=0.49\textwidth, trim= 0 0 0 0]{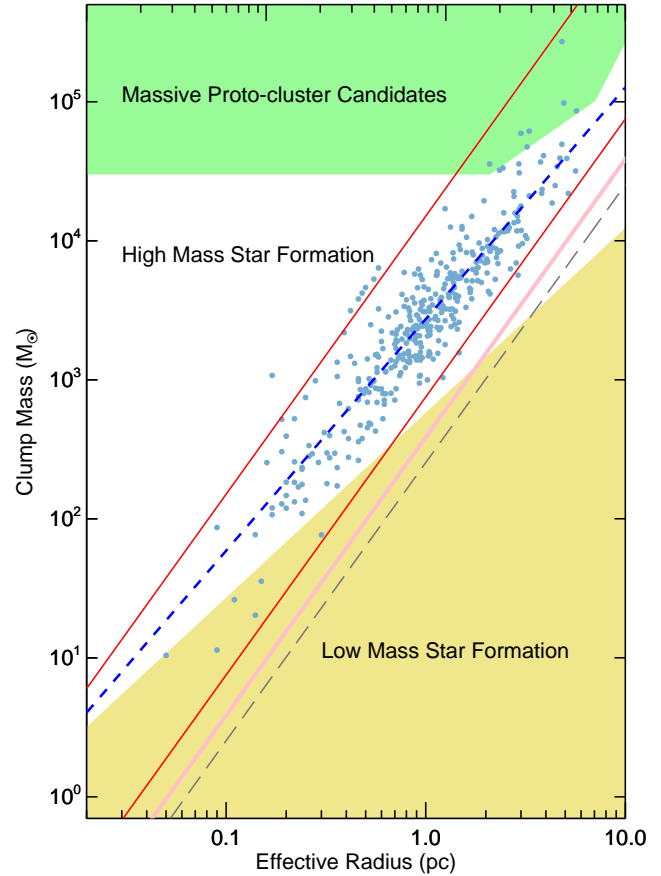}

\caption{\label{fig:mass_radius_distribution} The mass-size relationship of the ATLASGAL-MMB associated clumps. The yellow shaded region shows the part of the parameter space found to be devoid of massive star formation that satisfies the relationship $m(r) \le 580$\,\msun\, $(R_{\rm{eff}}/{\rm{pc}})^{1.33}$ (cf. \citealt{kauffmann2010c}).$^3$ The green shaded region indicates the region of parameter space where the young massive cluster progenitors are expected to be found (i.e., \citealt{bressert2012}). The dashed blue line shows the result of a linear least-squares fit to the resolved dust sources. The grey dashed line shows the sensitivity of the ATLASGAL survey and the upper and lower solid red line shows the surface densities of 1\,g\,cm$^{-2}$ and 0.05\,g\,cm$^{-2}$, respectively. The diagonal pink band fills the gas surface density ($\Sigma(\rm{gas})$) parameter space between 116-129\,\msun\,pc$^{-2}$ suggested by \citet{lada2010} and \citet{heiderman2010}, respectively, to be the threshold for ``efficient'' star formation.}

\end{center}
\end{figure}

In Fig.\,\ref{fig:mass_radius_distribution}, we present the mass-size relationship for the ATLASGAL-MMB associated sources. This sample consists of 375 clumps that have distance estimates and are spatially resolved in the APEX beam. Of these, we find that 363 have masses larger than the limiting mass for their size, as determined by \citet{kauffmann2010b} for massive star formation. This corresponds to $\sim$97\,per\,cent of the sample and, although this in itself does not confirm that the embedded source is a high-mass protostar, it does at least suggest that these clumps have the potential to form one. This also supports our earlier statement (in Sect.\,4.3) that the less massive clumps (i.e., less than 1,000\,\msun) also have the potential to form massive stars, as long as they are relatively compact.

Only 6 ATLASGAL-MMB associated sources are in the part of the parameter space that was found to be devoid of massive stars. All of these sources have been placed at the near distance and if the wrong distance has been assigned then this would explain their location in the mass-radius plot. Checking the confidence flag given by \citet{green2011b} we find that three of these sources have been given a flag of $b$, indicating their distance assignments are less reliable, and may explain why these sources fail to satisfy \citet{kauffmann2010b} mass-radius requirement for massive star formation.

Turning our attention back to Fig.\,\ref{fig:mass_radius_distribution} we see that the ATLASGAL-MMB data form a fairly continuous distribution over almost four orders of magnitude in mass and two orders of magnitude in radius. The dashed blue line overlaid on this plot shows the result of a linear least-squares fit to the data. This fit provides a good description of the mass-size relationship for these objects (Log($M_{\rm{clump}}) = 3.4\pm0.013 + (1.67\pm0.036)\times {\rm{Log}}(R_{\rm{eff}}$)). The upper and lower red diagonal lines indicate constant surface densities, $\Sigma({\rm{gas}})$, of 1\,g\,cm$^{-2}$ and 0.05\,g\,cm$^{-2}$, respectively.  These two lines provide fairly reliable empirical upper and lower bounds for the clump surface densities required for massive star formation. Furthermore, the lower bound of 0.05\,g\,cm$^{-2}$ provides a better constraint than Kauffmann et al.\ for the high-mass end of the distribution (i.e., $R_{\rm{eff}} > $ $\sim$0.5\,pc or $M_{\rm{clump}} > $ $\sim$500\,\msun). 

The thick pink line shows the threshold derived by \citet{lada2010} and \citet{heiderman2010} (116 and 129\,\msun\,pc$^{-2}$, respectively; hereafter LH threshold) for ``efficient'' star formation. Above this threshold the observed star formation in nearby molecular clouds (d $\le 500$\,pc) is linearly proportional to cloud mass and negligible below. This line corresponds to an extinction threshold of A$_K\simeq0.9$\,mag or visual extinction, A$_V$, of $\simeq8$\,mag. As discussed in the previous paragraph, the lower mass-radius envelope of the ATLASGAL-MMB associations is well modeled by $\Sigma({\rm{gas}})=0.05$\,g\,cm$^{-2}$, which is approximately twice the LH threshold. However, we note that the star formation associated with the nearby molecular clouds used to determine the LH threshold is likely to be predominantly low-mass, while the threshold determined from the ATLASGAL-MMB associations may be a requirement for efficient formation of intermediate- and high-mass stars. This result suggests that a volume density threshold may apply (e.g., \citealt{parmentier2011}), however, with the data presented here it is not possible to determine the density distribution or the volume density at the scales involved with star formation.

Given that this sample includes a large range of physical sizes (cores, clumps and clouds) and evolutionary stages (HMC, HMPO and \uchii\ regions)  it is somewhat surprising to find such a strong correlation. However,  as mentioned in Sections 4.2 and 4.3, the clumps appear to have a scale-free envelope structure that is not significantly changed as the embedded YSOs evolve towards the main sequence.

\subsubsection{Precursors to young massive clusters}
 
\begin{table*}

\begin{center}\caption{Derived parameters for massive proto-clusters. Masses have been estimated assuming a dust temperature of 20\,K for all sources except G002.53+00.016 for which a range of temperatures between 19 and 27\,K estimated from SED fits to individual pixels across the clump (see \citealt{longmore2012a} for details).}
\label{tbl:MPC_derived_para}
\begin{minipage}{\linewidth}
\begin{tabular}{l...c}
\hline \hline
  \multicolumn{1}{c}{ATLASGAL Name}&  \multicolumn{1}{c}{Distance}&	\multicolumn{1}{c}{Effective Radius}  &	\multicolumn{1}{c}{Clump Mass} & \multicolumn{1}{c}{Reference}\\
  \multicolumn{1}{c}{}&  \multicolumn{1}{c}{(kpc)}&	\multicolumn{1}{c}{(pc)}  &	\multicolumn{1}{c}{(Log(\msun))} &\\
 \hline
AGAL010.472+00.027	&	11.0	&	3.0	&	4.77 & 1	\\
AGAL328.236$-$00.547	&	11.4	&	4.9	&	4.99& 1	\\
AGAL329.029$-$00.206	&	11.7	&	3.2	&	4.68& 1	\\
AGAL345.504+00.347	&	10.8	&	5.7	&	4.93& 1	\\
AGAL350.111+00.089	&	11.4	&	2.1	&	4.55& 1	\\
AGAL351.774$-$00.537	&	17.4	&	4.8	&	5.43& 1	\\
AGAL352.622$-$01.077	&	19.4	&	3.3	&	4.79& 1	\\
\hline
G000.253+00.016 & 8.4 & 2.8 & 5.10  & 2 \\
G010.472+00.026 & 10.8 & 2.1 & 4.58 & 3 \\
G043.169+00.009 & 11.4 & 2.2&5.08 & 3\\
G049.489$-$00.370$^a$& 5.4 & 1.6 & 4.68 & 3\\
G049.489$-$00.386$^a$& 5.4 & 1.6 & 4.72&3\\
\hline\\
\end{tabular}\\
References: (1) this work, (2) \citet{longmore2012a}, (3) \citet{ginsburg2012} \\
Notes: $^a$These two sources are considered as a single MPC in the discussion presented by  \citet{ginsburg2012}.

\end{minipage}

\end{center}
\end{table*}

The green shaded area in the upper left part of Fig.\,\ref{fig:mass_radius_distribution} indicates the region of the mass-radius parameter space in which young massive proto-cluster (MPC) candidates are thought to be found (see \citealt{bressert2012} for details).  MPCs are massive clumps with sufficient mass that, assuming a fairly typical SFE (i.e., $\sim$30\,per\,cent), have the potential to be the progenitors of future young massive clusters (YMCs) that have masses of $>$10$^4$\,\msun\ such as the Arches and Quintuplet clusters (\citealt{portegies_zwart2010}). We consider both starless and star-forming clumps to be MPCs as anything with a gas envelope is a possible precursor. In an analysis of the BGPS data, \citet{ginsburg2012} identified only three MPC candidates in a longitude range of $\ell = 6$-90\degr\ from a sample of $\sim$6,000 dust clumps. Using their detection statistics, \citet{ginsburg2012} go on to estimate the number of MPCs in the Galaxy to be $\leq 10\pm6$, however, they state that a similar study of the southern Galactic plane is needed.

Given that the lifetime of the starless massive clumps is relatively short ($\sim$0.5\,Myr; \citealt{tackenberg2012,ginsburg2012}) it is probably safe to assume that the ATLASGAL-MMB sample probably includes all of the likely MPC candidates found in the $20\degr > \ell > 280\degr$ of the Galaxy. Applying the \citet{bressert2012} criteria to our ATLASGAL-MMB sample, we have identified 7 young massive proto-cluster candidates, one of which (i.e., AGAL010.472+00.027) is located in the overlapping regions analysed by \citet{ginsburg2012} who also identified this source as an MPC candidate. The ATLASGAL clump names and their derived parameters are tabulated in Table\,\ref{tbl:MPC_derived_para} along with the MPC candidates identified by \citet{ginsburg2012} and \citet{longmore2012a}.

We note that the detection rate of young massive proto-cluster candidates identified from the ATLASGAL-MMB sample is twice that identified from the BGPS (\citealt{ginsburg2012}). Both studies covered a similar area of the Galactic disk (approximately 30\,per\,cent each between galactocentric radii of 1-15\,kpc), however, ATLASGAL covers a much broader range of Galactic latitude ($|b|<1.5\degr$ and $|b|<0.5\degr$ for ATLASGAL and the BGPS, respectively) and we note that three of the MPC candidates identified here have latitudes greater than 0.5\degr. The difference in MPC candidate detection rates between the two surveys is likely a result of the difference in latitude coverage.  It is therefore also likely that \citet{ginsburg2012} have underestimated the number of MPC candidates in the first quadrant by a factor of two. Combining the number of MPCs identified and the volume coverage we estimate the total number of MPC candidates in the Galaxy is $\leq 20\pm6$.

The number of MPC candidates is a factor of two larger than the number of YMCs known in the Galaxy, but similar to the number of embedded clusters in the Galaxy estimated by \citet{longmore2012a} (i.e., $\sim$25). However, this is almost an order of magnitude greater than the number of MPCs expected assuming a YMC lifetime of 10\,Myr and a formation time of $\sim$1\,Myr. This would suggest that either only a small number of the MPC candidates will evolve into YMCs, which would require a SFE lower than the assumed 30\,per\,cent, or there are many more YMCs that have as yet not been identified.

\subsection{Galactic distribution}

\begin{figure}
\begin{center}
\includegraphics[width=0.49\textwidth, trim= 0 0 0 0]{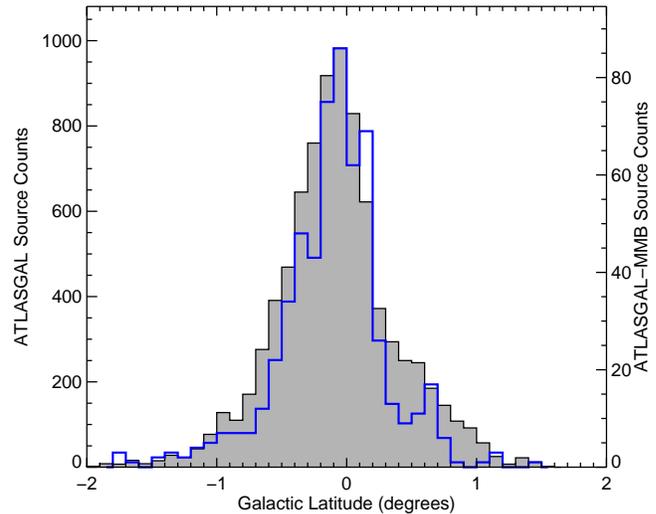}

\caption{\label{fig:gal_lat} Galactic latitude distribution of ATLASGAL sources (grey filled histogram) and ATLASGAL-MMB associated sources (blue histogram). The bin size used is 0.1\degr.} 

\end{center}
\end{figure}

\begin{figure*}
\begin{center}
\includegraphics[width=0.98\textwidth, trim= 0 0 0 0]{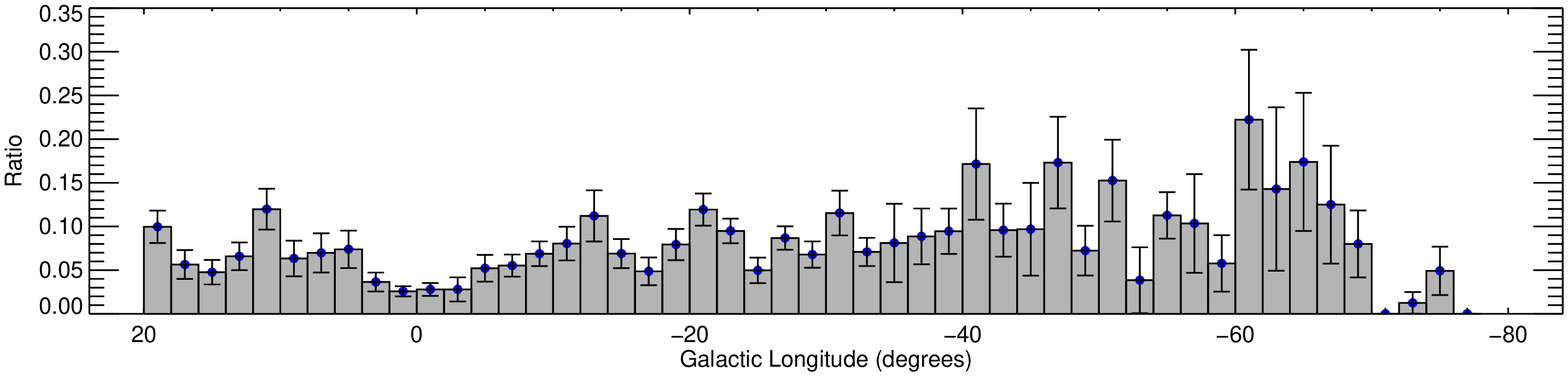}\\
\includegraphics[width=0.98\textwidth, trim= 0 0 0 0]{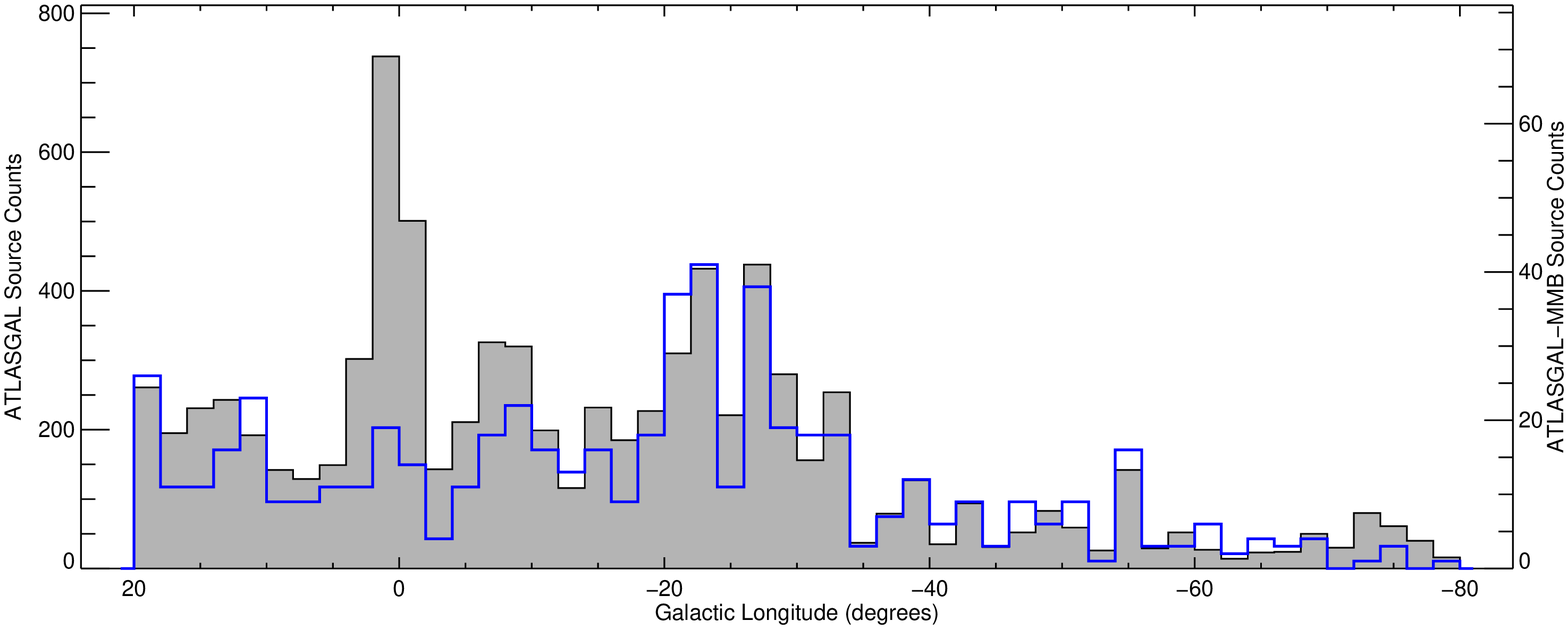}\\

\caption{\label{fig:gal_long} Galactic longitude distribution of ATLASGAL sources (grey filled histogram) and ATLASGAL-MMB associated sources (blue histogram) are presented in the lower panel, while in the upper panel we present the fraction of ATLASGAL sources associated with a methanol maser. The bin size used is 2\degr.} 

\end{center}
\end{figure*}

In the previous subsection we have shown that the clumps identified from their association with methanol masers are likely to be involved with the formation of the next generation of massive stars. Massive star formation has been found to be almost exclusively associated with the spiral arms of nearby analogues of the Milky Way (\citealt{kennicutt2005}) and therefore the Galactic distribution of this sample of ATLASGAL-MMB associated clumps may provide some insight into where in the Galaxy massive star formation is taking place and its relation to the spiral arms.

Figs.\,\ref{fig:gal_lat} and \ref{fig:gal_long} we show the distribution in Galactic latitude and longitude of ATLASGAL sources and ATLASGAL-MMB associations in our sample, in the range $20^{\circ} \ge \ell \ge 280^{\circ}$. We find no significant difference between the ATLASGAL-MMB sample's latitude distributions and the overall ATLASGAL source distribution; which has been previously commented on by \citet{beuther2012} and \citet{contreras2013} and so will not be discussed further here. The distribution of source counts at 2-degree resolution is shown in the lower panel of Figs.\,\ref{fig:gal_long} and the fraction of ATLASGAL sources with MMB associations in the upper panel of this figure.  While the total number of sources is dependent on both the local source density and the line-of-sight distribution, the ratio is independent of these and shows the incidence rate of methanol masers in dense clumps.

\begin{figure*}
\begin{center}
\includegraphics[width=0.90\textwidth, trim= 0 0 0 0]{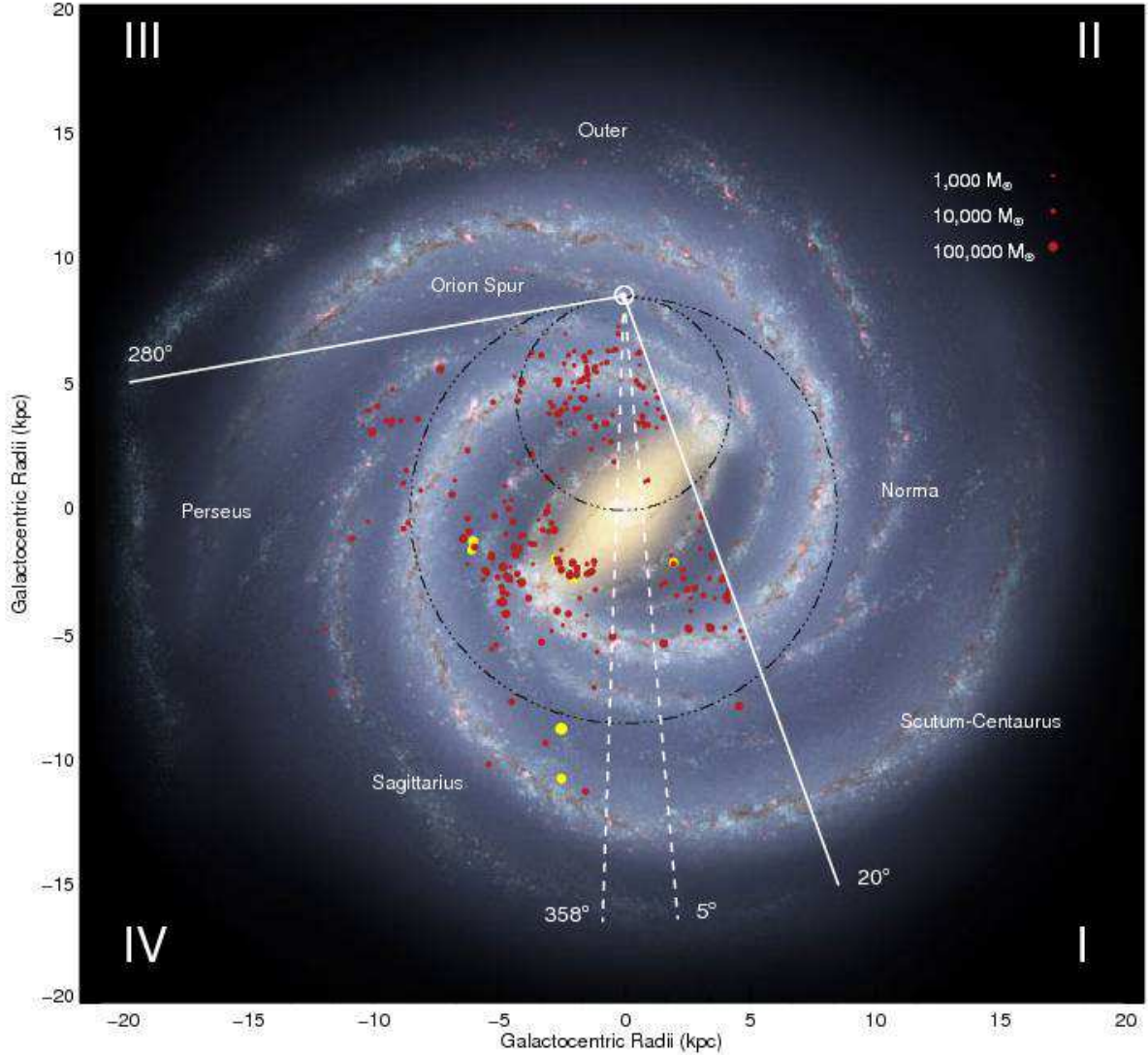}

\caption{\label{fig:galactic_mass_radius_distribution}Galactic distribution of the ATLASGAL-MMB sources with known distances and clump masses $\geq$ 10$^3$\,\msun and $L_{\rm{MMB}}\geq$ 1,000\,Jy\,km\,s$^{-1}$\,kpc$^2$). We show the kinematic positions of our sample as red-in-black (yellow-in-black for the PMC candidates) circles, the sizes of which give an indication of their respective dust mass. In the upper right corner we give the masses for a sample of clumps. We have superimposed the ATLASGAL-MMB source distribution over a sketch of how the Galaxy is thought to appear if viewed face-on. This image has been produced by Robert Hurt of the Spitzer Science Center in consultation with Robert Benjamin and attempts to synthesise many of the key elements of Galactic structure using the best data currently available (Courtesy NASA/JPL-Caltech; see text for more details). The position of the Sun is shown by the small circle above the image centre. The Roman numerals in the corners refer to the Galactic quadrants and the two white solid lines originating from the location of the Sun enclose the region of the Galactic Plane overlapped by the MMB and ATLASGAL surveys, while the two white dashed lines indicate the region towards the Galactic centre excluded by \citet{green2011b} as kinematic distance in this region are unreliable. The dot-dashed circles represent the locus of tangent points and the solar circle.} 

\end{center}
\end{figure*} 

The distribution of ATLASGAL sources as a function of Galactic longitude in Fig.\,\ref{fig:gal_long} (lower panel) reflects Galactic structure, with a strong, narrow peak towards the Galactic Centre and a broader maximum centred around $\ell\,\sim\,330$, which corresponds to a group of sources in the Scutum-Centaurus arm and possibly along the Norma arm tangent, which is close to the same direction (see \citealt{beuther2012} and \citealt{contreras2013} for further discussion). Since methanol masers are associated with the early stages of massive star formation (\citealt{minier2003}), maser source counts can be taken as a measure of the massive star-formation rate (SFR).  If the total maser counts trace the massive SFR on this spatial scale, then this fraction traces an analogue of the star-formation efficiency (SFE), i.e. the rate at which dense, submillimetre-traced clumps are producing massive YSOs, within the timescale appropriate to the methanol-maser stage of evolution ($\sim$2.5-$4.5\times 10^4$\,yr; \citealt{walt2005}).  The MMB fraction dips significantly at longitudes $|\,\ell\,|\,<$\,4\degr, and perhaps within 10\degr, suggesting that the SFE is significantly reduced near the Galactic centre. Outside this narrow zone, there is little evidence of any significant changes in SFE on these scales, associated with other features of Galactic structure within the survey area.  In particular, the ratio is more or less constant across both the peak in source counts seen at $\ell \simeq 330^{\circ}$ mentioned above, and the clear drop in counts at larger longitudes. Similar results were reported by \citet{beuther2012} from a comparison of GLIMPSE source counts to the ATLASGAL longitude distribution. \citet{moore2012} report increases in the large-scale SFE associated with some spiral-arm structures, while \citet{eden2012} found no significant variations associated with major features of Galactic structure in the fraction of molecular cloud mass in the form of dense clumps.

The lower SFE towards the Galactic centre inferred from the ratio of ATLASGAL-MMB associated clumps is supported by a recent study presented by \citet{longmore2012b}. These authors used the integrated emission of the inversion transition of ammonia (1,1) from HOPS (\citealt{walsh2011, purcell2012}) and the 70-500\,\mum\ data from the Hi-GAL survey (\citealt{molinari2010a}) to trace the Galactic distribution of dense gas and compared this to star formation tracers such as the water, methanol masers and \uchii\ regions identified by HOPS, MMB and the Green Bank Telescope HII region Discovery Survey (HRDS; \citealt{bania2010}), respectively, to derive the star formation rate per unit gas mass. The overall distribution of the integrated ammonia emission is very similar to that of the ATLASGAL source distribution showing a very strong peak towards the Galactic centre (cf. Fig.\,3  \citealt{purcell2012}). Although \citet{longmore2012b} found the ammonia and 70-500\,\mum\ dust emission to be highly concentrated towards the Galactic centre they found the distribution of masers and \hii\ regions to be relatively uniform across the Galaxy. Measuring the SFR, \citet{longmore2012b} found it to be an order of magnitude lower than what would be expected given the region's surface and volume densities.

We are unable  to independently estimate the absolute SFR from the MMB and ATLASGAL source counts.  However, we can state that we find the fraction of clumps associated with the formation of massive stars is a factor of 3-4 lower in the Galactic centre compared with that found for the rest of the Galaxy. The lower SFE seen towards the Galactic centre is likely a result of the extreme environmental conditions found in this region of the Galaxy (\citealt{immer2012}).  

In Fig.\,\ref{fig:galactic_mass_radius_distribution} we present the 2-D Galactic distribution of ATLASGAL-MMB associations. In this figure we only include ATLASGAL-MMB associations that have masses and maser luminosities for which the sample is complete across the Galaxy (i.e., $M \ga$\,1,000\,\msun\ and $L_{\rm{MMB}}\ga$ 1,000\,Jy\,km\,s$^{-1}$\,kpc$^2$). We find that the positions of the ATLASGAL-MMB associations are in reasonable agreement with the main structural features of the Galaxy, as determined by other tracers  (taking into account the kinematic distance uncertainty due to peculiar motions is of order $\pm$1\,kpc; \citealt{reid2009}). However, we also note that the degree of correlation is not uniform, with high source densities coincident with the near section of the Scutum-Centaurus arm and the southern end of the Galactic Bar. There is also a smaller number of sources that are distributed along the length of the Sagittarius arm.  However, one of the most interesting and surprising features of Figs.\,\ref{fig:gal_long} and \ref{fig:galactic_mass_radius_distribution} is the lack of any clear enhancement in ATLASGAL sources or ATLASGAL-MMB associations corresponding to the Scutum-Centaurus arm tangent at $\ell\,\sim\,310\degr$-320\degr. The low number of clumps with or without masers associated with the southern Scutum-Centaurus arm implies that this arm is not strongly forming stars of any type, and not just a lack of high-mass star formation. This is more surprising when we consider the distribution of CO emission, which is found to peak between $\ell =310-312$\degr\ (see Figs.\,6 and 7 presented in \citealt{bronfman1988}). This would suggest that this arm is associated with significant amounts of molecular material, but that the clump forming efficiency is relatively low for some reason. 

\begin{figure}
\begin{center}
\includegraphics[width=0.49\textwidth, trim= 0 0 0 0]{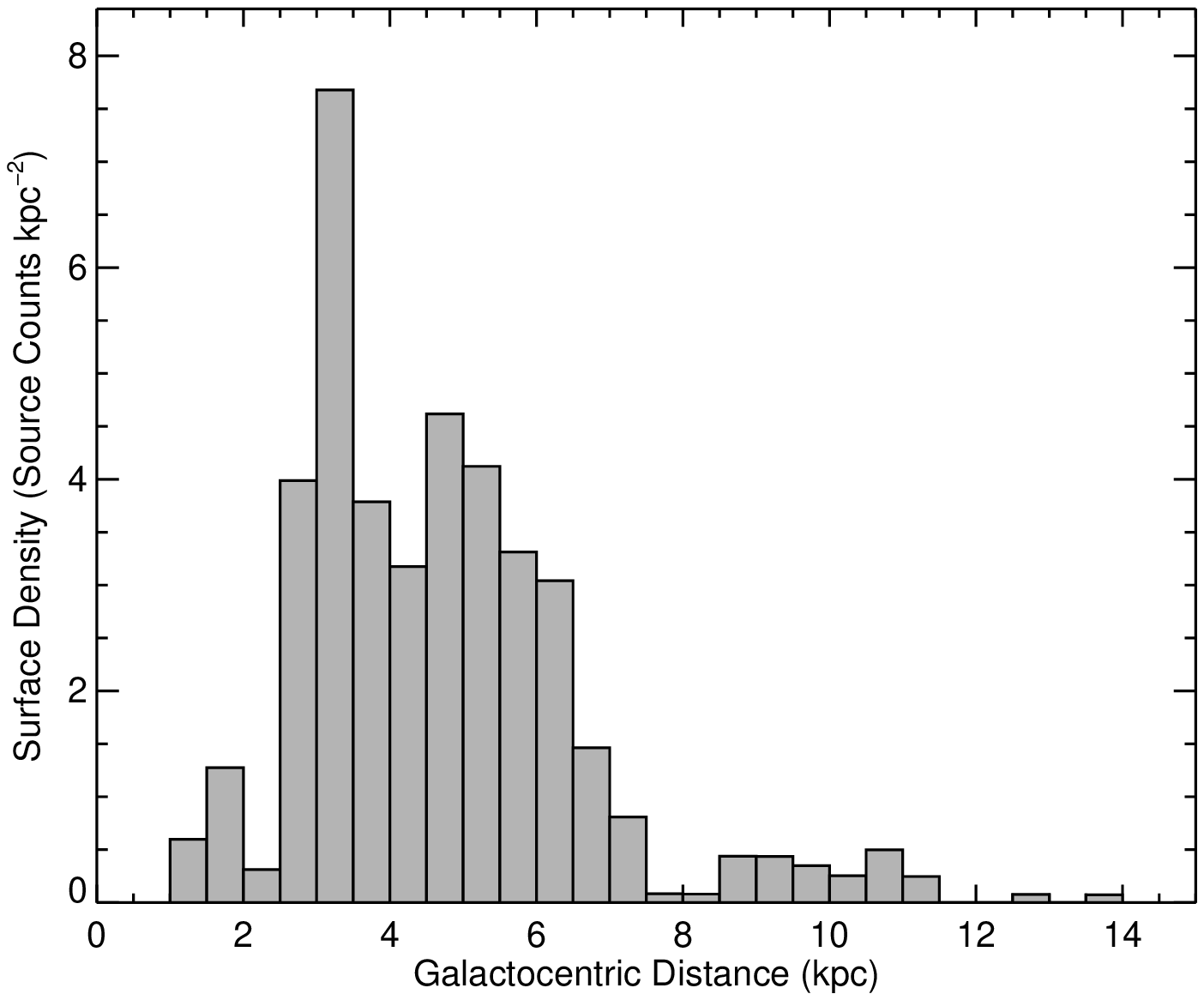}

\caption{ \label{galactocentric_distribution} Surface density of ATLASGAL-MMB associated sources as a function of galactocentric distance. As with Fig.\,\ref{fig:galactic_mass_radius_distribution} we only include ATLASGAL-MMB associations that have masses and maser luminosities for which the sample is complete across the Galaxy (i.e., $M \ga$\,1,000\,\msun\ and $L_{\rm{MMB}}\ga$ 1,000\,Jy\,km\,s$^{-1}$\,kpc$^2$). The bin size is 0.5\,kpc.} 

\end{center}
\end{figure}

Another interesting feature seen in Fig.\,\ref{fig:galactic_mass_radius_distribution} is the relatively low number of ATLASGAL-MMB asscociated clumps located at large Galactic radii. Since we have only plotted clumps with masses above 1,000\,\msun\ this low number of sources found outside the solar circle is not due to a lack of sensitivity. In Fig.\,\ref{galactocentric_distribution} we present the ATLASGAL-MMB source surface density as a function of Galactocentric radius. The three peaks in the Galactic radial distribution at $\sim$3, 5 and 10\,kpc correspond to the far 3-kpc spiral arm, a combination of the near section of the Scutum-Centaurus arm and the southern end of the Galactic Bar, and the Sagittarius spiral arm, respectively. There is a strong enhancement in the massive star-formation rate in the far 3-kpc arm with an otherwise relatively constant surface density between 3 and 7\,kpc, however, the surface density drops significantly at large radii.   

There is clearly a significant difference between the surface density inside and outside of the solar circle. The co-rotation radius is approximately the same as the solar circle (i.e, R$_{\rm{GC}}=$ 8.5\,kpc), which is where the spiral arms --- in principle --- are less important because the ISM is no longer being shocked as it runs into an arm. So one explanation could be that the spiral arms play an important role in creating the conditions required for massive star formation within the inner Galaxy (e.g., efficiently forming molecular clouds from the gas entering the spiral arm) compared to the outer Galaxy. The co-rotation radius is also where the metallicity has been found to drop sharply (e.g., \citealt{lepine2011}) and therefore metallicity may also play a role. 

Since the ATLASGAL-MMB sample is primarily tracing massive star formation it is unclear whether this difference in the surface density of star formation between the inner and outer Galaxy is restricted to the massive SFR or if it is also found in the low- and intermediate-mass star formation; if it is a feature of the massive SFR then it could have implications for the IMF. However, it is also possible that the galactocentric massive star formation surface density is simply reflecting the underlying distribution of molecular and atomic gas in the Galaxy. This is something we will revisit when distances and masses are available for a larger fraction of the ATLASGAL compact source catalogue.

The surface density distribution presented in Fig.\,\ref{galactocentric_distribution} has been determined using only the ATLASGAL-MMB associations that are above the clumps mass and maser luminosity completeness limit. By multiplying the surface density in each of the bins in this plot by the area of the bin annulas we estimate the total number of massive star forming clumps with masses $>$ 1000\,\msun\ in the Galaxy to be $\sim$560. The sample presented here therefore represents approximately 50\,per\,cent of the whole Galactic population of massive star forming clumps that are associated with a methanol maser. 

We can take this analysis a step further and estimate the contribution these star forming clumps make to the Galactic SFR. If we take the median clump mass of $\sim$3000\,\msun\ and a SFE of 30\,per\,cent then combined we estimate these clumps will produce clusters with a total stellar mass of $\sim5\times10^5$\,\msun. The formation time for the most massive stars in these cluster was calculated by \citet{molinari2008} to be $\sim$1.5$\times10^5$\,yr and \citet{davies2011} estimated the combined statistical lifetime of the MYSO and \uchii\ region stages to be several $10^5$\,yr, and therefore a reasonable lower limit for the cluster formation time is likely to be $\sim$0.5\,Myr. However, most of the total cluster stellar mass comes from lower-mass stars, which take longer to form and will increase the cluster formation time to perhaps 1\,Myr. Using these two cluster formation times as an upper and lower limit we estimate that these methanol maser associated clumps have combined SFR of between 0.5-1\,\msun\,yr$^{-1}$, and therefore could be responsible for up to half the current Galactic SFR $\sim$2\,\msun\,yr$^{-1}$ (\citealt{davies2011}).   

\subsection{Luminosity-mass correlations}

In Sect.\,\ref{sect:completeness} we found a weak correlation between the clump mass and maser luminosity and speculated that this may be related to higher luminosities of the embedded young stars. In this section we will explore this possible link. 

\begin{figure}
\begin{center}
\includegraphics[width=0.49\textwidth, trim= 0 0 0 0]{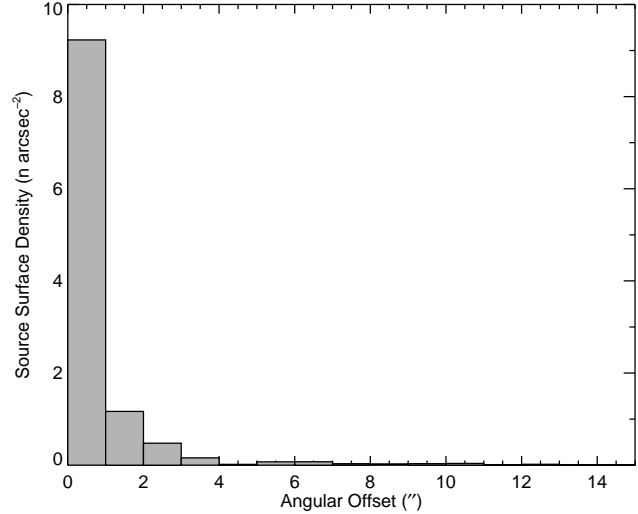}

\caption{ \label{rms_surface_density} Surface density of RMS-MMB associations as a function of separation. We have truncated the $x$-axis of this plot at 15\arcsec. The bin size is 1\arcsec.} 

\end{center}
\end{figure}

In a recent paper \citet{gallaway2013} reported the positional correlation of MMBs with other samples of young high mass stars such as the RMS (\citealt{urquhart2008}) and with EGOs which are thought to trace shocked gas associated with the outflows of MYSOs (e.g., \citealt{cyganowski2008,cyganowski2009}). They identified 82 RMS sources (\uchii\ regions and/or YSOs) within 2\arcsec\ of an MMB source and found that all of these have luminosities consistent with the presence of an embedded high-mass star. We have repeated the spatial correlation described by Gallaway et al., however, we relaxed the matching radius to 20\arcsec\ in order to determine the standard deviation of the offsets and set the association radius accordingly; this identifies 137 possible matches. In Fig.\,\ref{rms_surface_density} we present a plot of the surface density as a function of angular separation between the matched RMS and MMB sources. The distribution is peaked at offsets between 0 and 1\arcsec\ and rapidly falls off to approximately zero for separations greater than 4\arcsec. We consider ATLASGAL-MMB sources to be associated with an RMS source if the offset is less than 4\arcsec; this identifies 102 reliable matches, which corresponds to approximately 20\,per\,cent of the MMB sample in the $20\degr > \ell > 10\degr$ and $350 \degr > \ell >280\degr$ region (the RMS survey excluded sources within 10\degr\ of the Galactic centre due to the increased background confusion and larger distance uncertainties).

\begin{figure}
\begin{center}
\includegraphics[width=0.49\textwidth, trim= 0 0 0 0]{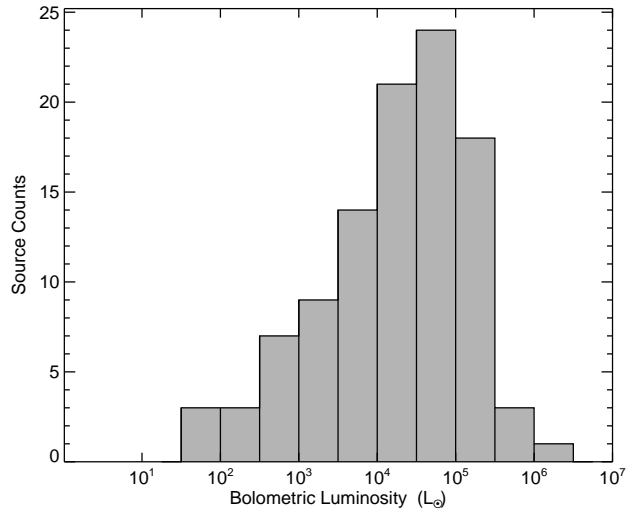}

\caption{ \label{fig:rms_luminosity} Bolometric luminosity distribution of the RMS sources associated with a methanol maser. The bin size is 0.5\,dex.} 

\end{center}
\end{figure}

Bolometric luminosities have been estimated for the RMS sources (\citealt{mottram2010,mottram2011b}) and thus allow us to investigate the luminosity distribution of these RMS-MMB associations. The fluxes used to estimate these luminosities are effectively clump-average values and are therefore a measure of the total cluster luminosity. However, if we assume that the luminosity is emitted from zero age main sequence (ZAMS) stars then the cluster luminosity is dominated by the most massive star in the cluster and so can be used to probe the correlation between it and the methanol maser. Since many of these masers are associated with \uchii\ regions the ZAMS assumption is reasonable.

\begin{figure}
\begin{center}

\includegraphics[width=0.49\textwidth, trim= 0 0 0 0]{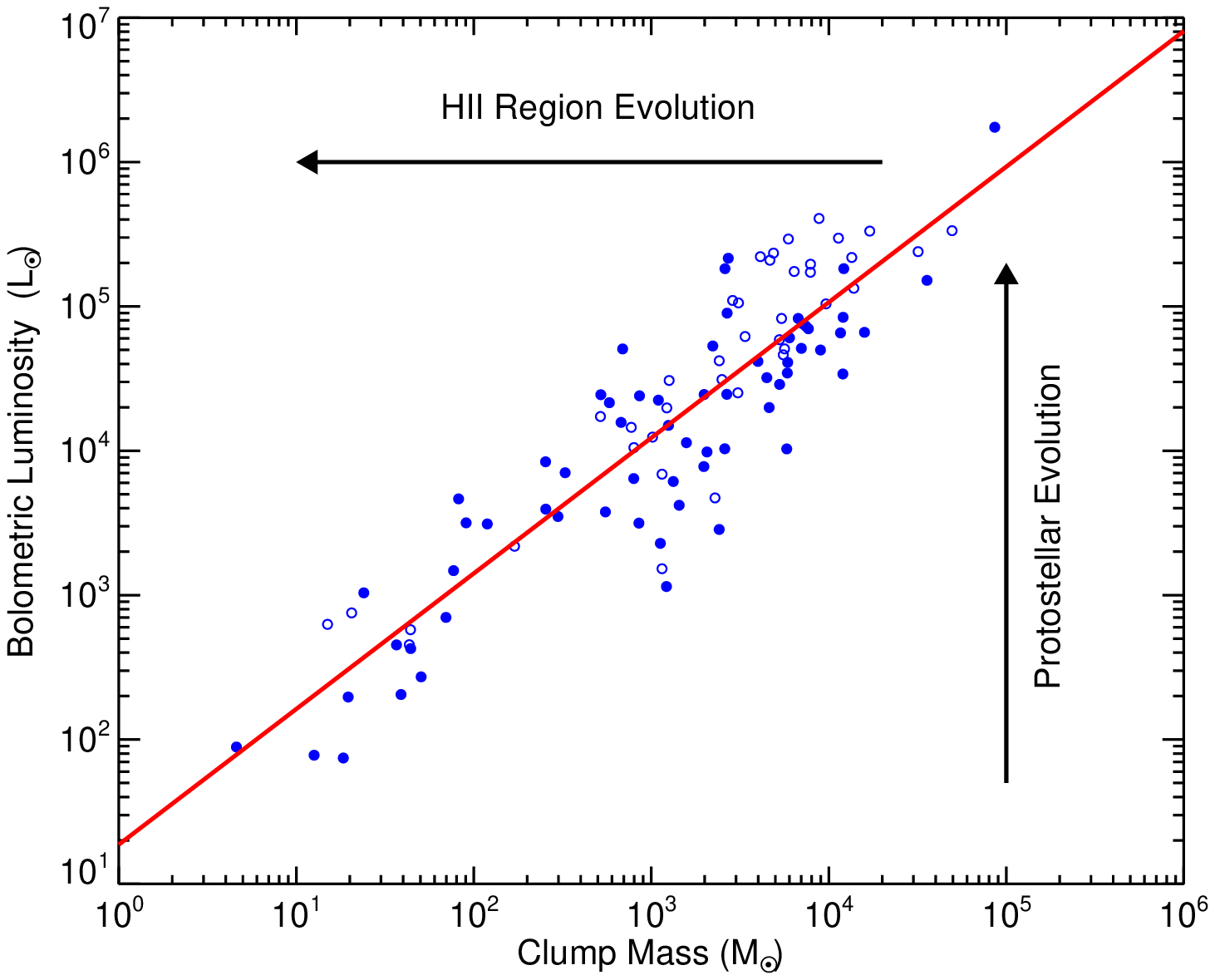}
\includegraphics[width=0.49\textwidth, trim= 0 0 0 0]{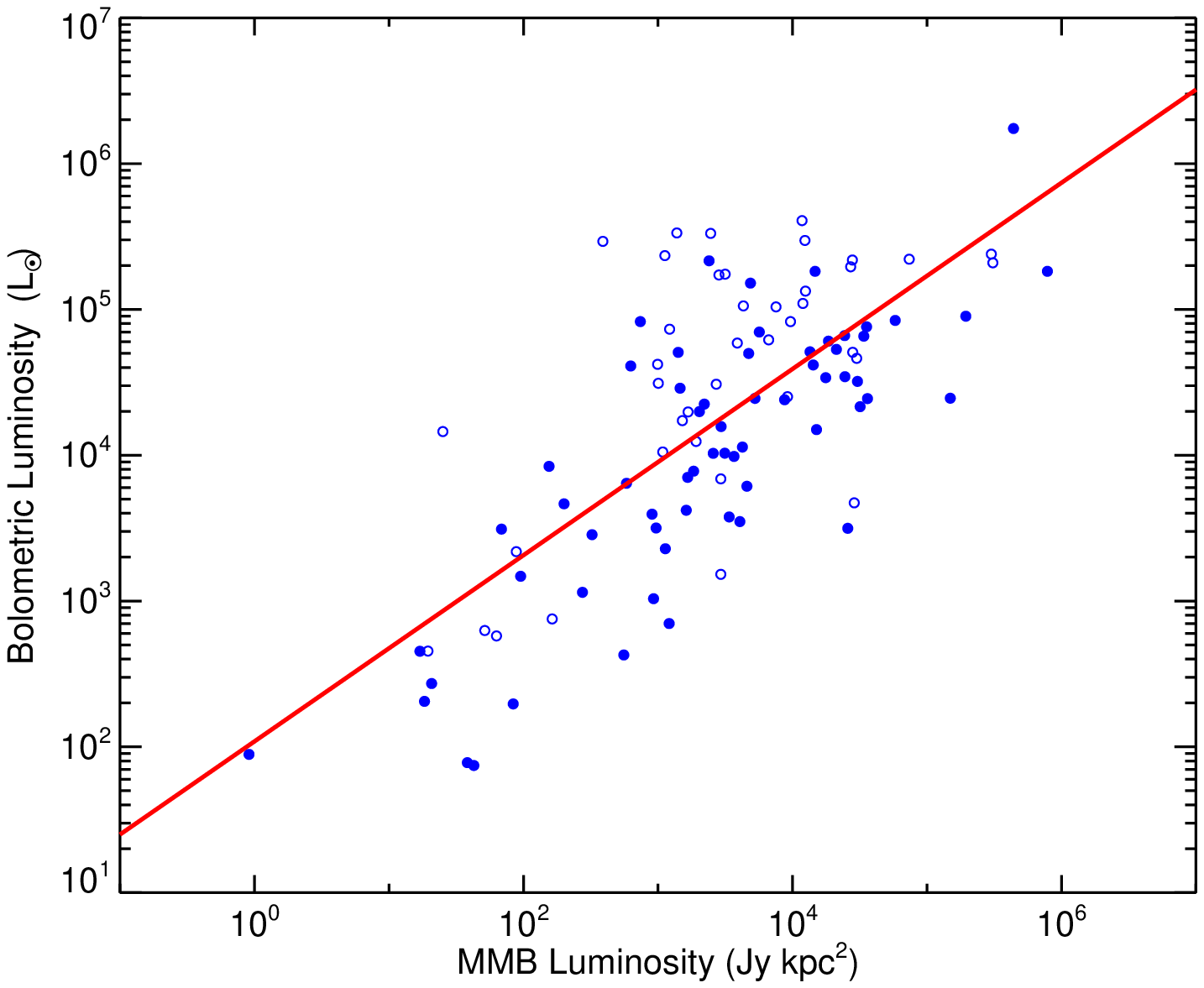}

\caption{ \label{rms_luminosity_mass} The bolometric luminosity plotted against clump mass (upper panel) and methanol maser luminosity (lower panel). Sources classified as YSOs and UCHII regions are shown as filled and open circles, respectively. The red line show the result of a linear least-squares log-log fit to the whole sample.} 

\end{center}
\end{figure}

The luminosities range from $\sim$100 to 10$^6$\,\lsun\ with a peak in the distribution at 10$^{4.5-5.0}$\,\lsun\ (see Fig.\,\ref{fig:rms_luminosity}). The RMS survey is complete for embedded young massive stars with luminosities a few times 10$^4$\,\lsun\ (\citealt{davies2011}) and so the fall off in the distribution for luminosities lower than this value is due to incompleteness. The minimum luminosity of a high-mass star is $\sim$1,000\,\lsun\ (corresponding to a star of spectral type B3; \citealt{boehm1981,meynet2003}) and so the fact that most of the RMS-MMB sample  ($\sim$90\,per\,cent) have luminosities above this supports their association with high-mass star formation. 

While the majority of the MMB sources appear to be associated with massive stars we do find a handful of sources with luminosities towards the lower end of the distribution (i.e., 100--500\,\lsun). Errors in the kinematic distance solution could possibly account for a few of these but it is unlikely to explain them all. It therefore seems likely that these methanol masers are associated with intermediate-mass stars with masses of several solar masses. This is consistent with the findings of \citet{minier2003} who were able to put a lower limit of $\sim$3\,\msun\ for strong maser emission. It is possible that these are intermediate mass protostars that are still in the process of accreting mass on their way to becoming a high-mass protostar. This is consistent with the mass-size relationship discussed in the previous subsection, however, as we will show in the next few paragraphs, there is a strong correlation between clump mass and bolometric luminosity and therefore these low luminosity sources also tend to be associated with the lower mass clumps and would require a high SFE ($>$50\,per\,cent) to form a massive star.

In the upper and lower panels of Fig.\,\ref{rms_luminosity_mass} we plot the bolometric luminosity as a function of clump mass and methanol maser luminosity for these ATLASGAL-MMB-RMS associated sources, respectively. In these plots we indicate the evolutionary type as classified by the RMS team, however, a KS test is unable to reject the null hypothesis that the two source types are drawn from the same populations and therefore we do not distinguish between them here. Again calculating the partial coefficients to remove the dependence of these parameters on distance we find a strong correlation between the bolometric luminosity and clump mass ($r$=0.78 and $p$-value $\ll 0.01$) and a weaker correlation between the bolometric and maser luminosities ($r$=0.42 and $p$-value $\ll 0.01$). The linear log-log fit to the bolometric luminosity and clump mass gives a slope of $0.94\pm0.04$ and so not only are these two parameters strongly correlated but their relationship is very close to linear. This value is slightly shallower than the slope of $\sim$1.3 reported by \citet{molinari2008} from their analysis of a sample of 42 regions of massive star formation. However, we note that they use IRAS fluxes to determine source luminosities and this will tend to overestimate fluxes for the more distant sources, which are typically also the most massive sources. This would lead to a steepening of the luminosity-mass slope when compared to the RMS luminosities, which used the higher resolution 70\,\mum\ MIPSGAL band to compute luminosities and therefore should be more reliable.

\begin{figure*}
\begin{center}
\includegraphics[width=0.33\textwidth, trim= 0 0 0 0]{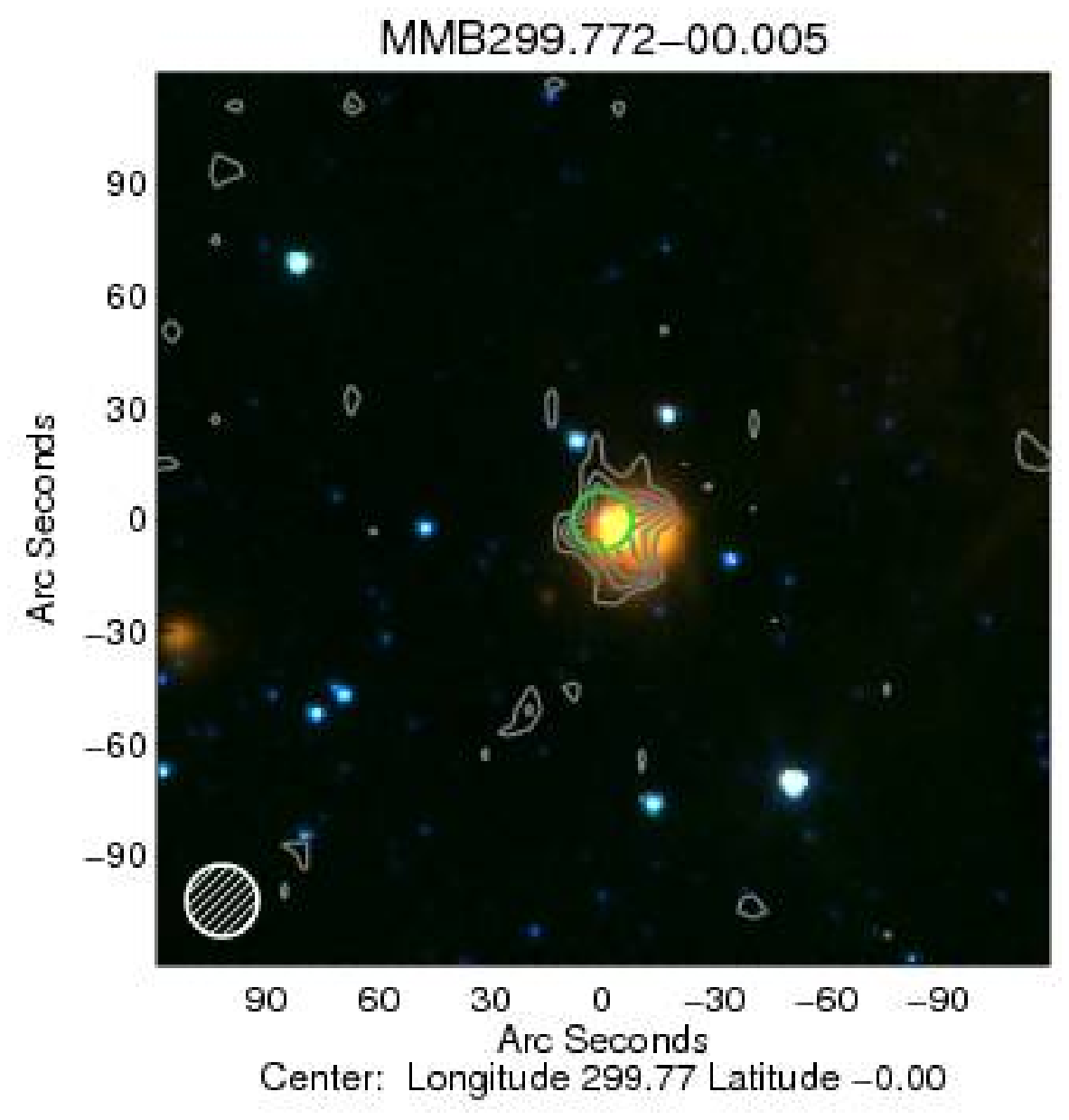}
\includegraphics[width=0.33\textwidth, trim= 0 0 0 0]{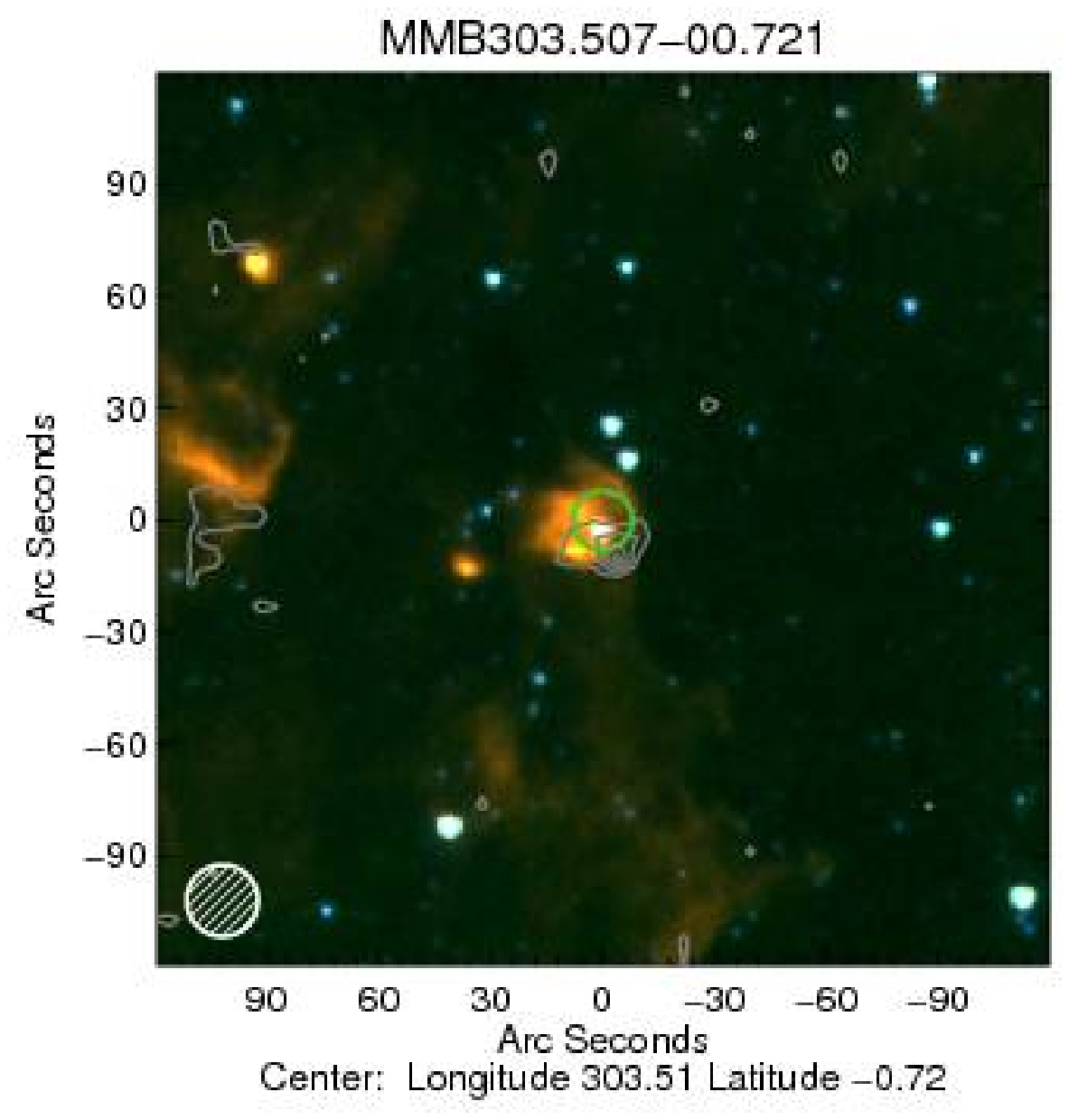}
\includegraphics[width=0.33\textwidth, trim= 0 0 0 0]{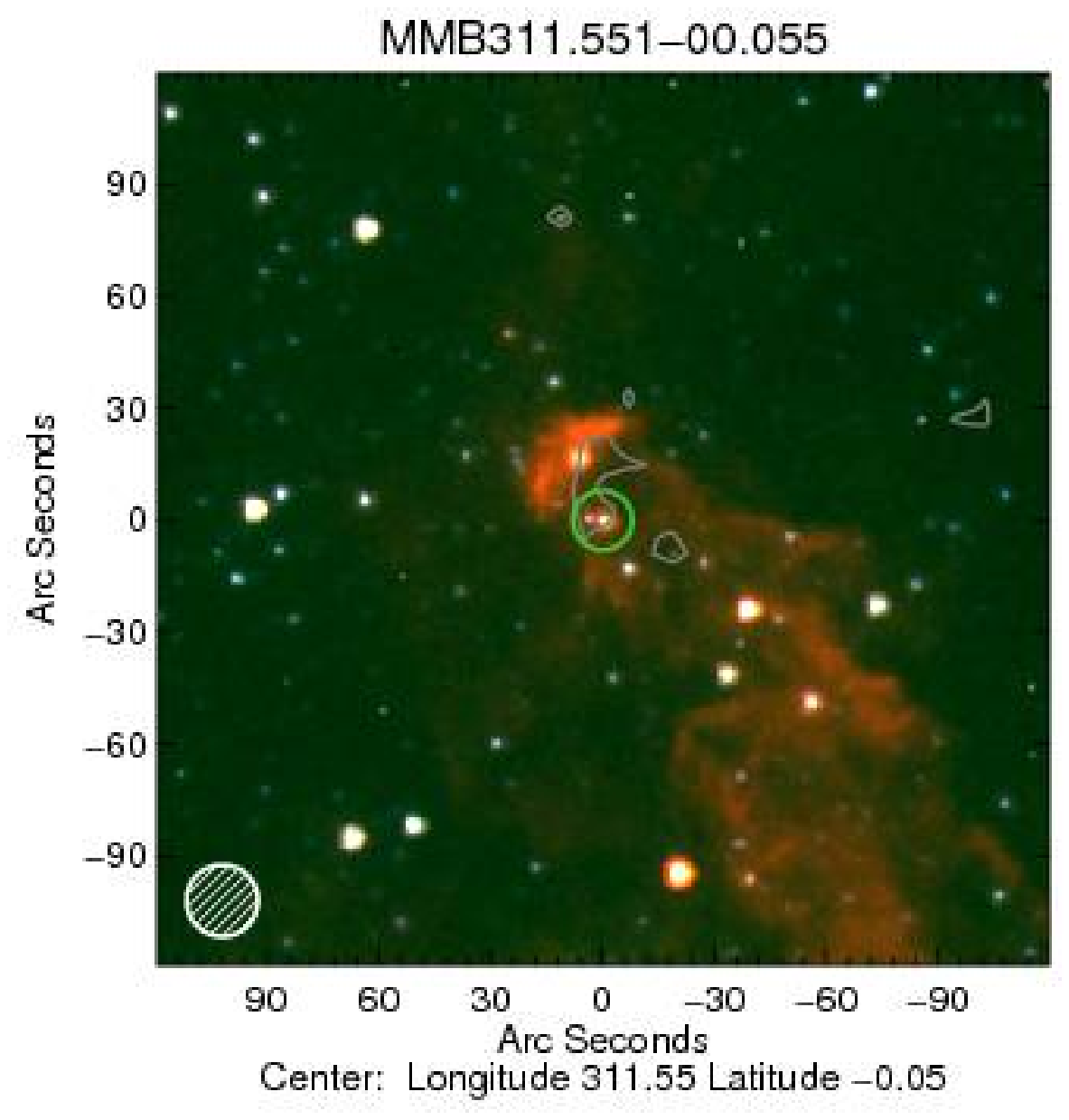}\\
\includegraphics[width=0.33\textwidth, trim= 0 0 0 0]{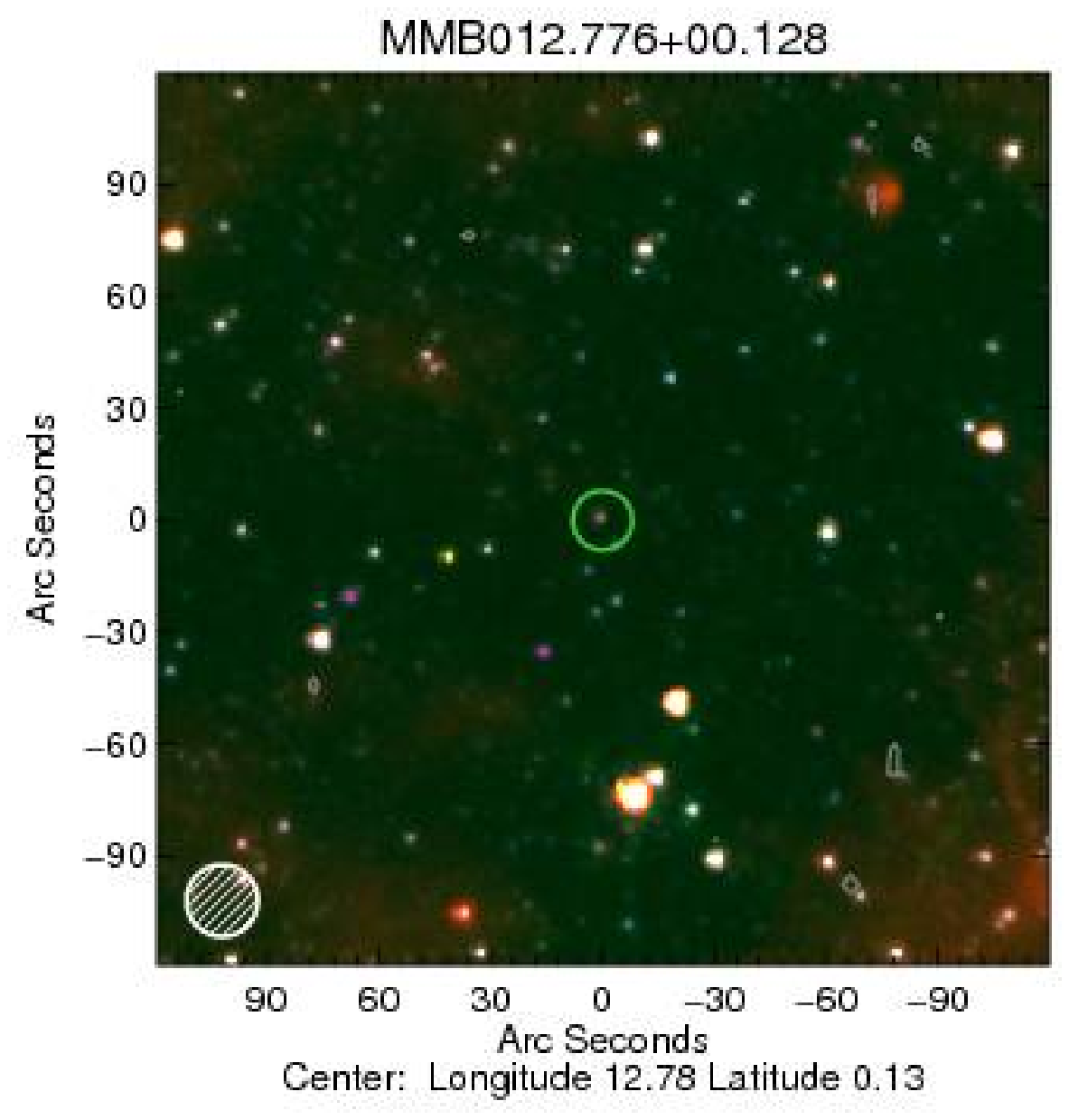}
\includegraphics[width=0.33\textwidth, trim= 0 0 0 0]{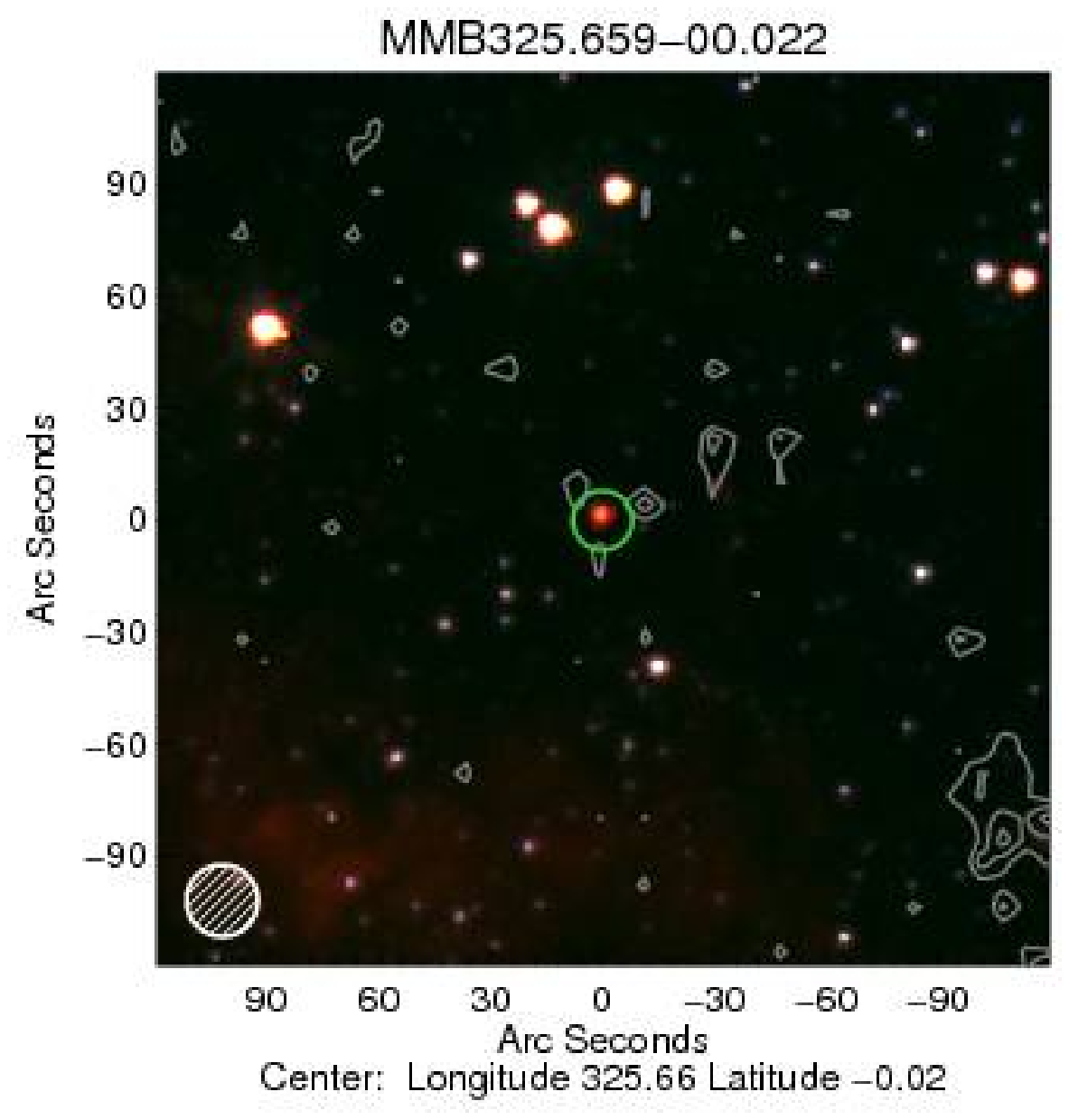}
\includegraphics[width=0.33\textwidth, trim= 0 0 0 0]{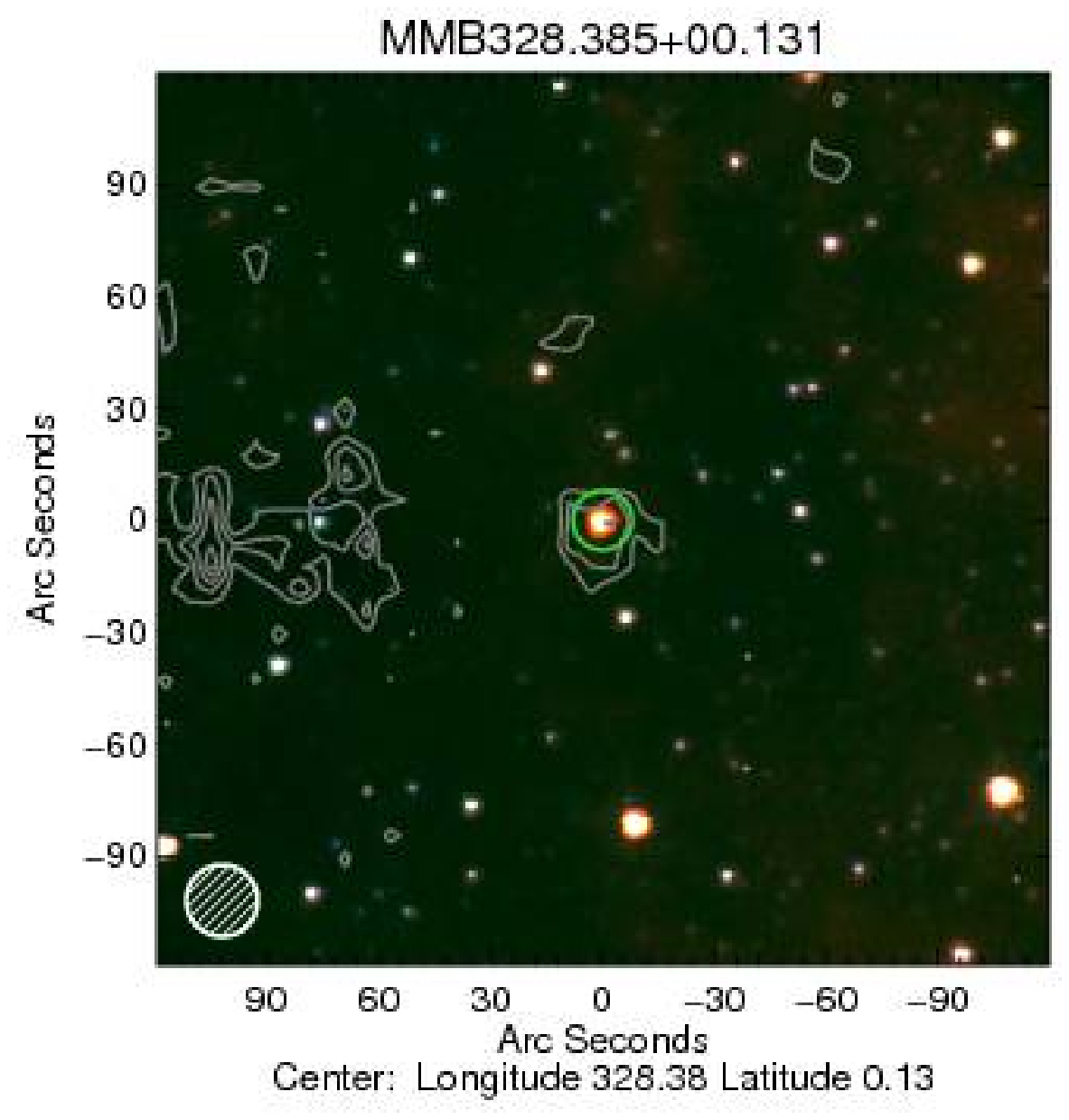}

\caption{\label{fig:irac_images_unassociated_mmb} Examples of the local mid-infrared environment found towards the unassociated methanol masers. Images and symbols are as described in the caption of Fig.\,\ref{fig:irac_images_associated_atlasgal}. The contour levels start at 1$\sigma$ and increase in steps of 1$\sigma$ to emphasis any weak compact or diffuse emission in the vicinity of the methanol maser. In the upper panels we present images of MMB sources that show extended 8\,$\mu$m emission commonly associated with star forming regions, while in the lower panels we present images of MMB sources that appear to be isolated and therefore possible evolved stars.} 

\end{center}
\end{figure*}

Assuming that the clump mass remains relatively constant through the early embedded stages of massive star formation while the luminosity increases as both the accretion rate and the mass of the protostar increase, the source should move vertically upwards in the mass-luminosity plot. Once the \hii\ region has formed and starts to disrupt its host clump the source's luminosity will be constant (luminosity of the ZAMS star ionizing the \hii\ region) and its mass will begin to decrease resulting in it moving horizontally to the left. Since there is no significant difference between the luminosity-mass distributions of MYSO and \uchii\ region stages we can assume they have similar ages; at both of these stages core hydrogen burning has begun, however, the \uchii\ regions are slightly more evolved and have started to ionise their surroundings. The fit to the data (solid red line in the upper panel of Fig.\,\ref{rms_luminosity_mass}) is a crude approximation of the transition between protostellar and \hii\ region evolution. 

As Gallaway et al. point out, the MMB sources are likely to be associated with a broad range of evolutionary states from HMCs to \uchii\ regions with their luminosity increasing through accretion as they develop. As previously mentioned only $\sim$20\,per\,cent of ATLASGAL-MMB associated sources are associated with the later stages (i.e., YSO and \uchii) and therefore the majority are younger and less evolved embedded objects whose luminosity is not yet sufficient to dominate that of its associated cluster. All of these sources will be located below the red line with more evolved \hii\ regions dominating the upper left part of the parameter space. 

We have found the mass and bolometric luminosity to be weakly correlated with the methanol maser luminosity. The mass and bolometric luminosity are strongly correlated with each other (i.e., $r=0.78$) and have very similar correlation values when compared to the maser luminosity (i.e., $r=0.42$ and 0.44, respectively), which might make the underlying cause of the correlation hard to distinguish. However, since the maser arises from a population inversion pumped by far-infrared and submillimetre photons emitted from the embedded YSO, it is the intrinsic luminosity of the YSO that is more likely the fundamental cause of these correlations. The weaker correlation between the bolometric and maser luminosities is probably related to the fact that the former is a measure of the total luminosity of the embedded proto-cluster, whereas the latter is more likely to be driven by a single cluster member,  and no doubt maser variability also plays a role.

A similar correlation between bolometric luminosity and water maser luminosity was reported by \citet{urquhart2011b} from a single dish survey of MYSOs and \uchii\ regions identified by the RMS survey. Having found a correlation between the methanol and water maser luminosities for these two advanced stages we would conclude that the strength of the maser emission is dominated by the energy output of the central source and not driven by source evolution as suggested by Breen et al. (see discussion in Sect.\,\ref{lum_vol_correlation}). One caveat to this is that both models of massive star formation (i.e., monolithic collapse and competitive accretion) predict that massive stars gain their mass through accretion from a circumstellar disk, and therefore their maser luminosity is likely to increase proportionally with their mass and luminosity. However, the large range of possible final masses of OB stars renders the maser luminosity insensitive to the current evolutionary stage of a particular stars in the same way that the bolometric luminosity is insensitive to evolutionary state of the embedded object.

\subsection{The nature of the unassociated MMB sources}
\label{sect:non-atlas-mmb}

\setlength{\tabcolsep}{3pt}

\begin{table*}

  \begin{center}\caption{\label{tbl:ATLASGAL_dark_sources} MMB sources without a counterpart in the ATLASGAL catalogue.}
\begin{minipage}{\linewidth}
\begin{tabular}{c.....}
  \hline \hline
    \multicolumn{1}{l}{MMB Name$^{\rm{a}}$}
  &  \multicolumn{1}{c}{MMB Peak Flux} &\multicolumn{1}{c}{870\,\mum\ Upper Limit} &  \multicolumn{1}{c}{Distance} &\multicolumn{1}{c}{$z$}&\multicolumn{1}{c}{Mass Upper Limit} \\
     \multicolumn{1}{c}{}
  &  \multicolumn{1}{c}{(Jy)} &\multicolumn{1}{c}{(Jy\,beam$^{-1}$)} &  \multicolumn{1}{c}{(kpc)} &\multicolumn{1}{c}{(pc)}&\multicolumn{1}{c}{(\msun\,beam$^{-1}$)} \\

  \hline
MMB005.677$-$00.027$^\dagger$	&	0.79	&	0.23	&	\multicolumn{1}{c}{$\cdots$}	&	\multicolumn{1}{c}{$\cdots$}	&	\multicolumn{1}{c}{$\cdots$}	\\
MMB006.881+00.093$^\dagger$	&	3.12	&	0.37	&	17.8	&	29.0	&	650.28	\\
MMB012.776+00.128$^\dagger$	&	0.84	&	0.21	&	13.0	&	29.0	&	197.67	\\
MMB013.696$-$00.156$^\dagger$	&	1.90	&	0.23	&	10.9	&	-29.7	&	147.46	\\
MMB014.521+00.155$^\dagger$	&	1.40	&	0.20	&	5.5	&	14.9	&	33.52	\\
MMB015.607$-$00.255$^\dagger$	&	0.43	&	0.22	&	11.4	&	-50.8	&	161.13	\\
MMB016.976$-$00.005$^\dagger$	&	0.64	&	0.20	&	15.5	&	-1.4	&	270.72	\\
MMB018.440+00.045$^\dagger$	&	1.85	&	0.21	&	11.6	&	9.1	&	152.98	\\
MMB019.614+00.011$^\ddagger$	&	3.99	&	0.22	&	13.1	&	2.5	&	210.32	\\
MMB292.468+00.168	&	4.40	&	0.72	&	7.9	&	23.3	&	248.31	\\
MMB299.772$-$00.005$^\dagger$	&	15.65	&	0.47	&	\multicolumn{1}{c}{$\cdots$}	&	\multicolumn{1}{c}{$\cdots$}	&	\multicolumn{1}{c}{$\cdots$}	\\
MMB303.507$-$00.721$^\dagger$	&	2.06	&	0.28	&	10.9	&	-136.5	&	179.64	\\
MMB303.846$-$00.363$^\dagger$	&	7.40	&	0.27	&	11.9	&	-75.3	&	210.62	\\
MMB303.869+00.194$^\ddagger$	&	0.90	&	0.27	&	\multicolumn{1}{c}{$\cdots$}	&	\multicolumn{1}{c}{$\cdots$}	&	\multicolumn{1}{c}{$\cdots$}	\\
MMB307.132$-$00.476	&	1.20	&	0.34	&	\multicolumn{1}{c}{$\cdots$}	&	\multicolumn{1}{c}{$\cdots$}	&	\multicolumn{1}{c}{$\cdots$}	\\
MMB307.133$-$00.477$^\dagger$	&	2.36	&	0.28	&	\multicolumn{1}{c}{$\cdots$}	&	\multicolumn{1}{c}{$\cdots$}	&	\multicolumn{1}{c}{$\cdots$}	\\
MMB308.715$-$00.216$^\dagger$	&	1.04	&	0.26	&	\multicolumn{1}{c}{$\cdots$}	&	\multicolumn{1}{c}{$\cdots$}	&	\multicolumn{1}{c}{$\cdots$}	\\
MMB311.551$-$00.055$^\dagger$	&	1.00	&	0.30	&	5.6	&	-5.4	&	53.33	\\
MMB311.729$-$00.735$^\dagger$	&	0.46	&	0.31	&	14.3	&	-183.6	&	346.68	\\
MMB312.501$-$00.084$^\dagger$	&	1.19	&	0.36	&	13.6	&	-19.9	&	363.05	\\
MMB312.698+00.126$^\dagger$	&	1.65	&	0.30	&	14.4	&	31.6	&	338.24	\\
MMB312.702$-$00.087$^\dagger$	&	0.81	&	0.27	&	5.8	&	-8.8	&	50.34	\\
MMB313.774$-$00.863$^\dagger$	&	14.30	&	2.16	&	3.3	&	-49.4	&	128.08	\\
MMB316.484$-$00.310$^\dagger$	&	0.72	&	0.33	&	\multicolumn{1}{c}{$\cdots$}	&	\multicolumn{1}{c}{$\cdots$}	&	\multicolumn{1}{c}{$\cdots$}	\\
MMB324.789$-$00.378$^\dagger$	&	1.15	&	0.24	&	15.2	&	-100.0	&	301.03	\\
MMB325.659$-$00.022$^\dagger$	&	0.57	&	0.18	&	17.4	&	-6.7	&	293.79	\\
MMB327.282$-$00.469$^\dagger$	&	5.40	&	0.26	&	14.5	&	-118.4	&	296.61	\\
MMB327.863+00.098$^\ddagger$	&	1.58	&	0.25	&	\multicolumn{1}{c}{$\cdots$}	&	\multicolumn{1}{c}{$\cdots$}	&	\multicolumn{1}{c}{$\cdots$}	\\
MMB328.385+00.131$^\dagger$	&	1.60	&	0.32	&	18.0	&	41.2	&	579.17	\\
MMB329.526+00.216$^\ddagger$	&	1.81	&	0.30	&	\multicolumn{1}{c}{$\cdots$}	&	\multicolumn{1}{c}{$\cdots$}	&	\multicolumn{1}{c}{$\cdots$}	\\
MMB330.998+00.093$^\dagger$	&	0.70	&	0.20	&	12.8	&	20.7	&	177.95	\\
MMB331.900$-$01.186$^\star$	&	2.50	&	0.83	&	11.8	&	-244.5	&	636.07	\\
MMB332.854+00.817$^\dagger$	&	1.10	&	0.25	&	11.8	&	167.5	&	192.35	\\
MMB332.960+00.135$^\dagger$	&	2.00	&	0.16	&	3.7	&	8.6	&	11.95	\\
MMB334.933$-$00.307$^\ddagger$	&	3.30	&	0.19	&	9.4	&	-50.4	&	91.48	\\
MMB337.517$-$00.348$^\ddagger$	&	1.50	&	0.20	&	17.1	&	-103.7	&	326.59	\\
MMB345.205+00.317$^\ddagger$	&	0.80	&	0.16	&	11.5	&	63.6	&	117.06	\\
MMB345.949$-$00.268$^\dagger$	&	1.53	&	0.29	&	13.9	&	-65.1	&	306.05	\\
MMB348.723$-$00.078$^\dagger$	&	2.58	&	0.22	&	11.2	&	-15.2	&	149.49	\\
MMB350.470+00.029$^\dagger$	&	1.44	&	0.16	&	1.2	&	0.6	&	1.34	\\
MMB350.776+00.138$^\dagger$	&	0.65	&	0.17	&	11.4	&	27.5	&	119.29	\\
MMB355.545$-$00.103$^\dagger$	&	1.22	&	0.55	&	11.5	&	-20.7	&	404.14	\\
MMB356.054$-$00.095$^\ddagger$	&	0.52	&	0.19	&	\multicolumn{1}{c}{$\cdots$}	&	\multicolumn{1}{c}{$\cdots$}	&	\multicolumn{1}{c}{$\cdots$}	\\
\hline
\end{tabular}\\
$^{\rm{a}}$ Sources with a superscript have been searched for mid-infrared emission by \citet{gallaway2013}:  $\dagger$ and $\ddagger$ indicate infrared bright and infrared dark sources, respectively, and $\star$ identifies the sources they were unable to classify.\\ 

\end{minipage}
\end{center}
\end{table*}

\setlength{\tabcolsep}{3pt}

In Sect.\,\ref{sect:unmatched_mmb} we identified 43 methanol masers that were not matched to an ATLASGAL source. In this section we will examine the available evidence to try and investigate the nature of these sources.

We begin by inspecting the ATLASGAL maps at the locations of the \unmatchedmmb\ MMB sources. In many cases, the position of the maser is coincident with either low-surface brightness, diffuse submillimetre emission, which would have been filtered out by the background subtraction used in the source-extraction process, or weak compact emission that fell below the detection threshold used in the source extraction. We have measured the peak 870\,$\mu$m flux at the position of the MMB source and estimated the 3$\sigma$ noise from the standard deviation of nearby emission-free regions in the map, and have used the higher of these two values as the upper limit to the submillimetre flux. We have used these values with their assigned distances (as discussed in Sect.\,\ref{sect:distance}) to estimate an upper limit for their masses using Eqn.\,1 and again assuming a dust temperature of 20\,K. These results are summarised in Table\,\ref{tbl:ATLASGAL_dark_sources}.

As previously mentioned, it is widely accepted that methanol masers are \emph{almost} exclusively associated with high-mass star-forming regions. This is supported by many of the findings presented in this paper. One explanations is that these unassociated MMB sources are  located at larger distances and that their submillimetre emission simply falls below the ATLASGAL detection sensitivity. Indeed, the median distance for the unassociated maser sample is $\sim$12\,kpc, which is much larger than the median value of $\sim$5\,kpc found for the ATLASGAL-MMB associations and the KS test showed the two distributions to be significantly different (see Fig.\,\ref{fig:atlas_mmb_distance_hist} for comparison of distance distributions and Sect.\,\ref{sect:distance} for discussion). Looking at the estimated upper limits for the mass we find that they are all significantly lower than the 1,000\,\msun\ assumed to be the minimum required for massive star formation. However, as shown in Sect.\,\ref{mass-size-relationship} it is at least feasible for less massive clumps to form massive stars, but in the majority of cases these sources would need to be significantly smaller than the beam. The only caveat is that for all of the other ATLASGAL-MMB associations we found a scale-free envelope, but these sources would need to be relatively discrete and isolated clumps that are not embedded in a larger structure. Alternatively, it is also possible that these methanol masers are associated with embedded sources that will go on to form intermediate-mass stars, in which case the low values obtained for the upper limits to the masses may not be all that important.

Assuming this hypothesis is correct then, even though the dust emission is too weak to be detected, we might expect to see associated mid-infrared emission from diffuse nebulosity and evidence of extinction from dark lanes of dust often seen towards sites of star formation. {bf Approximately 80\,per\,cent of these sources are associated with mid-infrared emission (these are indicated by superscripts given by the MMB name in Table\,\ref{tbl:ATLASGAL_dark_sources}), which is similar to the proportion found by \citealt{gallaway2013} for the whole MMB catalogue (i.e., 83\,per\,cent). We present a sample of these false-colour mid-infrared images in Fig.\,\ref{fig:irac_images_unassociated_mmb} again created by combining data extracted from the GLIMPSE archive.}

The upper middle panel of this figure shows the mid-infrared image of the MMB source MMB303.507$-$00.721, which is located at a distance of $\sim$11\,kpc. This source is found to have extended 8\,$\mu$m emission commonly associated with star-forming regions and there is evidence of compact dust emission to the south-west of the mid-infrared emission. From a visual inspection of the mid-infrared emission we estimate that in  approximately a third of cases the emission is consistent with this maser being associated with a more distant star-formation site, however, this is probably a lower limit.

So for a significant fraction of these unassociated methanol masers it is likely that the lack of dust emission can be explained by them being more distant and their associated dust emission falling below the ATLASGAL detection threshold. However, there is also a small number of sources for which this explanation is not satisfactory.  There are three sources that are located at relatively near distances (i.e., MMB292.468+00.168, MMB311.551$-$00.055 and MMB312.702$-$00.087) where we would expect to have detected their dust emission. There are two more that are located at the far distance, but where the far distance allocation puts them much farther from the Galactic mid-plane than expected for star forming regions (i.e., MMB311.729$-$00.735 and MMB331.900$-$01.186 that have $z$ distance of $-$183.6 and $-$244.5\,pc, respectively), which casts some doubt on the distance allocation if indeed these are star forming. However, there is another intriguing possibility which is that these masers may arise in the circumstellar shells associated with evolved stars (e.g., \citealt{walsh2003}).

A search of the SIMBAD database revealed that only ten of these methanol masers were previously known: one identified in the IRAS point-source catalogue; two detected in the BGPS, and so it is likely that their dust emission falls below our detection threshold but is detected by the BGPS due to their superior low surface-brightness sensitivity; and 7 are included in \citet{robitaille2008} intrinsically red GLIMPSE source catalogue, six of which they classify as YSOs, and one (MMB328.385+00.131) they classified as a possible asymptotic giant branch star. In the lower right panel of Fig.\,\ref{fig:irac_images_unassociated_mmb}, we present the mid-infrared image of the MMB source MMB328.385+00.131 that shows only a single point source coincident with the position of the methanol maser. This point source is isolated in the image and the lack of any extended 8\,$\mu$m emission or extinction feature, that are commonly associated with star forming regions, along with the classification made by \citet{robitaille2008} from its mid-infrared colours, would suggest this maser might be associated with an evolved star. However, this could simply be due to a chance alignment along the same line of sight and so this association requires further investigation to test its reliability.

Currently, there is not enough complementary data available for all of these sources to be able to properly evaluate these two explanations.  The Hi-GAL survey of the inner Galactic plane at 70-500\,$\mu$m (\citealt{molinari2010a}) will provide a way to definitively test these two possibilities (e.g., \citealt{anderson2012}) and will be discussed in a future publication. 

\section{Summary and conclusions}

The ATLASGAL survey (\citealt{schuller2009}; $280\degr < \ell < 60\degr$) has identified the Galactic distribution of dust through its thermal emission at 870\,\mum\ and is complete to all massive clumps above 1,000\,\msun\ to the far side of the inner Galaxy ($\sim$20\,kpc). In total the ATLASGAL survey has identified some 12,000 compact sources, many of which have the potential to form the next generation of massive stars. Methanol masers have been found to be associated with high mass star formation and we have therefore taken advantage of the availability of an unbiased catalogue of these objects compiled by the methanol multibeam (MMB; \citealt{green2009}) survey team to identify a large sample of high-mass star forming clumps.

Cross-matching these two surveys we have identified 577 ATLASGAL-MMB associated clumps within the overlapping region of both surveys (i.e, $280\degr <  \ell < 20\degr$ and $|b| < 1.5\degr$) with two or more methanol masers being detected towards 44 clumps. We find $\sim$90\,per\,cent of the matches are within 12\arcsec\ ($\sim$3$\sigma$) of the peak of the submillimetre emission revealing a strong correlation with column density and the location of a methanol maser within the clumps. We fail to identify any significant 870\,\mum\ emission towards 43 MMB sources.

Assuming a dust temperature of 20\,K and using distances provided by \citet{green2011b}, and derived here, we are able to estimate the clump masses and radii, column and volume densities, and methanol maser luminosities for almost 500 of the ATLASGAL-MMB associations. We find we are complete across the Galaxy to all dust clumps with masses larger than 1,000\,\msun\ hosting a methanol maser with luminosities $>$1,000\,Jy\,kpc$^{2}$. We use these parameters to investigate the link between methanol masers and massive star forming clumps and as a probe of Galactic structure. Our main findings are as follows:

\begin{enumerate}

\item The clump radii cover a range from 0.1 to several parsecs with the larger clumps generally found at larger distances. With a median aspect ratio of 1.4 the ATLASGAL-MMB associations are fairly spherical centrally condensed structures, however, with a median $Y$-factor of $\sim$5 a significant amount of their mass is located outside the central region. The position of the methanol masers is strongly correlated with the peak column density at the centre of the clumps. We find no correlation between the aspect ratio and $Y$-factor with distance, which suggests that the envelope structures of these massive star forming clumps are scale-free.  The apparently simple clump structure (with masers at the central col density peak and scale-free radial structure) suggests the formation of one central stellar cluster per clump.  Formation of multiple clusters might be expected to be accompanied by more complex clump structure.
 
\item We are complete to all dust clumps harbouring a methanol maser with masses over 1,000\,\msun\ across the inner Galactic disk (i.e., $\sim$20\,kpc). The median clump dust mass ($\sim$3,000\,\msun) is significantly larger than the completeness level, which confirms that methanol masers are preferentially associated with massive clumps. Furthermore, assuming a Kroupa IMF and a star formation efficiency of 30\,per\,cent we find that 72\,per\,cent of these clumps (i.e., $M_{\rm{clump}} >$ 1,000\,\msun) are in the process of forming clusters hosting one or more 20\,\msun\ star(s). Although 28\,per\,cent of the clumps have masses lower than 1,000\,\msun\ we find these are also more compact objects ($\simeq$0.3\,pc) that are likely to form either single stars or small multiple systems of bound stars that are also very likely to include a massive star. This is supported by  the empirical mass-radius criterion for massive star formation (\citealt{kauffmann2010b}) that shows that 97\,per\,cent of ATLASGAL-MMB associations have masses and sizes that are consistent with all other known massive star forming clumps, including the lower mass clumps. We conclude that the vast majority of clumps associated with methanol masers are in the process of forming high-mass stars. 

\item Inspecting the mass-radius relation for the ATLASGAL-MMB associations we find that a surface density of 0.05\,g\,cm$^{-2}$ provides a better estimate of the lower envelope of the distribution for masses and radii greater than 500\,\msun and 0.5\,pc, respectively, than the criterion given by  \citet{kauffmann2010b}. This surface density threshold corresponds to a A$_K\sim2$\,mag or visual extinction, A$_V$, of $\simeq16$\,mag, which is approximately twice the required threshold determined by \citet{lada2010} and \citet{heiderman2010} for ``efficient'' low-mass star formation. This would suggest that there is a clear surface density threshold required for clumps before star formation can begin but a higher threshold is required to for more massive star formation.

\item Testing the evolutionary trend reported in the literature (i.e. \citealt{breen2011_methanol,breen2012}) between the methanol maser luminosity and clump-averaged volume density we fail to find any correlation. Although we are able to reproduce the results of these previous studies we find that both parameters have a strong dependence on distance and that once this is removed the correlation between them drops to zero. However, we do find a the bolometric and methanol maser luminosities are correlated with each other. 

\item We have identified seven clumps that have masses large enough to be classified as massive protocluster (MPC) candidates which are expected to form the next generation of young massive clusters (YMCs) such as the present day Archers and Quintuplet clusters. Using the Galactic plane coverage of this study and the number of MPC candidates detected we estimate the Galactic population to be $\le$20$\pm$6. This value is twice as many as previously estimated and similar to the number of currently known YMCs, which would suggest that only a few of these MPC candidates will successfully convert the 30\,per\,cent of their mass into stars required to form a YMC. 

\item The Galactic distribution reveals the Galactic centre region having a significantly lower star formation efficiency (SFE), than the rest of the Galaxy covered by this survey, which is broadly flat even towards the spiral arm tangents. The lower SFE is probably a reflection of the much more extreme environment found in the central region of the Milky Way. Interestingly we find no enhancement in either the ATLASGAL or ATLASGAL-MMB source counts in the direction of the Scutum-Centaurus arm tangent, from which we conclude that this arm is not actively forming stars of any type.

\item The galactocentric distribution reveals very significant differences between the surface density of the massive star formation rate between the inner and outer Galaxy. We briefly speculate on possible explanations, however, it is clear further work is required before the reasons behind this difference can be properly understood. Using the surface density distribution with the completeness levels applied we estimate the total Galactic population to be $\sim$560, which means that the sample presented here represents approximately 50\,per\,cent of the whole population. We estimate the star formation associated with these methanol maser associated clumps may contribute up to 50\,per\,cent of the Galactic star formation rate.

\item Bolometric luminosities are available from the literature for $\sim$100 clumps and these range between $\sim$100 to 10$^6$\,\lsun\ with the distribution peaking at $\sim$10$^5$\,\lsun. This confirms the association between methanol masers and massive young stars for 90\,per\,cent of this sample of clumps, but also reveals that there are some masers associated with intermediate-mass stars. For lower luminosity clumps (i.e., $M_{\rm{clump}}<$1,000\,\lsun) these may be intermediate-mass protostars that are still accreting mass and will eventually go on to form a high-mass star, however, these tend to be associated with the lower mass clumps and would therefore require a SFE $>$50\,per\,cent. It therefore may be the case that a small number of these clumps are destined to only form intermediate-mass stars.

\item We investigated the available evidence for the 43 methanol masers towards which no 870\,\mum\ emission has been detected and concluded that while perhaps 50\,per\,cent may be more distant star forming regions where the dust emission has simply fallen below the ATLASGAL surveys sensitivity, the nature of the other masers is yet to be determined. 

\end{enumerate}

This is the first of a series of three papers planned to use the ATLASGAL survey to conduct a detailed and comprehensive investigation of high-mass star formation. The main aim of these papers is to use the unbiased nature of the dust emission mapped by ATLASGAL over the inner Galactic plane to connect the results derived from different high-mass star formation tracers. In subsequent papers we will investigate the dust properties of an unbiased sample of ultra-compact HII regions identified from the CORNISH survey and a complete sample of massive YSOs identified by the RMS survey.

\section*{Acknowledgments}

The ATLASGAL project is a collaboration between the Max-Planck-Gesellschaft, the European Southern Observatory (ESO) and the Universidad de Chile. 
This research has made use of the SIMBAD database operated at CDS, Strasbourg, France. This work was partially funded by the ERC Advanced Investigator Grant
  GLOSTAR (247078) and was partially carried out within the Collaborative
  Research Council 956, sub-project A6, funded by the Deutsche
  Forschungsgemeinschaft (DFG). L. B. acknowledges support from CONICYT project Basal PFB-06. JSU would like to dedicate this work to the memory of J.\,M.\,Urquhart.

\bibliography{mmb}

\bibliographystyle{mn2e_new}

\end{document}